\documentclass[a4paper,fleqn,usenatbib]{mnras}
\usepackage{newtxtext,newtxmath}

\usepackage[T1]{fontenc}
\usepackage{ae,aecompl}

\usepackage{graphicx}	
\usepackage{amsmath}	
\usepackage{amssymb}	

\title[First Billion Years of Cosmic Metal Evolution]{Metal-enriched Galaxies in the First $\sim 1$ Billion Years: Evidence of a Smooth Metallicity Evolution at $z \sim 5$}

\author[S. Poudel et al.]{
Suraj Poudel,$^{1}$\thanks{E-mail: spoudel@email.sc.edu}
Varsha P. Kulkarni,$^{1}$
Frances H. Cashman,$^{1}$
Brenda Frye,$^{2}$
\newauthor
 C\'eline P\'eroux,$^{3,4}$
Hadi Rahmani,$^{4,5}$
and Samuel Quiret$^{4}$
\\
$^{1}$University of South Carolina, Dept. of Physics and Astronomy, Columbia, SC 29208, USA\\
$^{2}$University of Arizona, Dept. of Astronomy and Steward Observatory, Tucson, AZ 85721, USA\\
$^{3}$European Southern Observatory, Karl-Schwarzschild-Strasse 2, 85748 Garching bei Munchen, Germany\\
$^{4}$Laboratoire d'Astrophysique de Marseille, UMR 7326, F-13388, Marseille, France\\
$^{5}$Observatoire de Paris, 61 Avenue de l'Observatoire, 75014 Paris, France
}

\date{Accepted XXX. Received YYY; in original form ZZZ}

\pubyear{2018}

\begin{document}
\label{firstpage}
\pagerange{\pageref{firstpage}--\pageref{lastpage}}
\maketitle

\begin{abstract}
We present seven new abundance measurements of the elements O, C and Si at $z>4.5$, doubling the existing sample of weakly depleted elements in gas-rich galaxies, in order to constrain the first $\sim$1 billion years of cosmic metal evolution. These measurements are based on quasar spectra of damped Lyman-alpha absorbers (DLAs) and sub-DLAs obtained with the Magellan Inamori Kyocera Echelle (MIKE) and Magellan Echellette (MagE) spectrographs on Magellan-South, and the X-Shooter spectrograph on the Very Large Telescope. We combine these new measurements with those drawn from the literature to estimate the $N_{\rm H\, I}$-weighted binned mean metallicity of $-1.51\pm0.18$ at $z=4.8$. This metallicity value is in excellent agreement with the prediction from lower redshift DLAs, supporting the interpretation that the metallicity evolution is smooth at $z\sim5$, rather than showing a sudden decline at $z>4.7$. Furthermore, the metallicity evolution trends for the DLAs and sub-DLAs are similar within our uncertainties. We also find that the [C/O] ratios for $z\sim5$ DLAs are consistent with those of the very metal-poor DLAs. Additionally, using [C/O] and [Si/O] to constrain the nucleosynthesis models, we estimate that the probability distributions of the progenitor star masses for three relatively metal-poor DLAs are centered around 12 M$_{\odot}$ to 17 M$_{\odot}$. Finally, the $z\sim 5$ absorbers show a different metallicity-velocity dispersion relation than lower redshift DLAs, suggesting that they may be tracing a different population of galaxies.
\end{abstract}




\section{Introduction}
Metal abundance measurements throughout the cosmic ages track the history of galaxy formation and evolution from the initial pristine stars and galaxies to the present day metal-rich galaxies.  At $z\sim5$, the cosmic stellar mass density is expected to have doubled every 300 Myr \citep*[e.g.][]{Gonzalez et al. 2011}. As a result, due to the finite age of the universe, metals observed in absorption at this epoch are only a few hundred million years old and are produced either by core-collapse supernovae of the massive metal-poor stars or by pair-instability supernovae. Therefore, the first 1 billion years of the cosmic metal evolution are influenced by the nucleosynthetic signatures from the early stars, and measuring abundances during the $z \sim5$ epoch can constrain the nature of the pristine Population III stars and the metal-poor Population II stars. The initial mass function of Population III stars is still a subject of debate, some studies suggest masses as high as 100 M$_{\odot}$ \citep*[e.g.][]{Bromm et al. 1999, Nakamura 2001, Abel et al. 2002}, where other studies suggest somewhat smaller values. For example, fragmentation due to turbulence can result in smaller Population III IMF (Initial Mass Function) \citep[]{Clark et al. 2011}, and multiple smaller Population III stars can be formed in a given minihalo \citep*[e.g.][]{Stacy 2013, Hirano et al. 2014, Stacy et al. 2016}. Measurements of relative abundances such as [C/O] in chemically young systems can put a strong constraint on the mass of the progenitor Population III stars \citep[]{Cooke et al. 2017}. Finally, abundance measurements can be compared with the results from different cosmic metal evolution models with the inclusion or exclusion of the Population III stars \citep*[e.g.][]{Maio 2015, Kulkarni et al. 2013} to explore the early enrichment history of star formation and enrichment in galaxies.\\

 While probing the chemical enrichment of galaxies from continuum or line emission is difficult at higher redshift, measurement of absorption lines in the spectra of background objects such as quasars provides a powerful method to measure chemical abundances \citep*[e.g.][]{Pettini et al. 1994, Prochaska et al.  2003a, Kulkarni et al. 2005}. Also, unlike emission-based detection, the absorption-based method can select gas-rich galaxies independent of their brightness \citep[]{Wolfe et al. 2005}. Particularly interesting absorption systems are the DLAs (Damped Lyman-alpha Absorbers) and sub-DLAs (sub-Damped Lyman-alpha Absorbers) in background quasar spectra, which have high neutral hydrogen column densities (log $N_{\rm H\, I} \ge 20.3$ and 19.0 $\le$ log $N_{\rm H\, I} < 20.3$, respectively). DLAs and sub-DLAs together dominate the neutral gas mass density of the universe at high redshift \citep*[e.g.][]{Wolfe et al. 2005, Zafar et al. 2013, Berg et al. 2019}, and provide the neutral gas reservoir for star formation \citep*[e.g.][]{Nagamine et al. 2004a, Nagamine et al. 2004b, Wolfe 2006}. The presence of damping wings in the Lyman-alpha lines for DLAs and sub-DLAs allows precise measurements of neutral hydrogen column densities. Furthermore, DLAs and sub-DLAs are also powerful probes of metals in the interstellar medium and the circumgalactic medium of distant galaxies. Finally, unlike the Lyman alpha forest and Lyman limit systems, DLAs are primarily neutral and insensitive to the ionization corrections \citep*[e.g.][]{ Erb et al. 2006, Aguirre et al. 2008, Lehner et al. 2013}.\\  
From a number of previous observations, DLA metallicity is observed to decrease gradually from redshift $z=0$ to $z=4$ with a rate of about 0.2 dex per redshift \citep*[e.g.][]{ Kulkarni 2002, Prochaska et al. 2003a, Kulkarni et al. 2005, Kulkarni et al. 2007, Rafelski et al. 2012, Jorgenson et al.  2013, Som et al. 2013, Som et al. 2015, Quiret et al. 2016}. However, the DLA metallicity evolution at high redshift is still unclear, primarily due to inadequate samples having robust measurements of dust-free metallicity. Some previous studies \citep*[e.g.][]{Rafelski et al. 2012, Rafelski et al. 2014} reported a sudden decline of DLA metallicity at $z>4.7$, suggesting a sudden change in the chemical enrichment process. However, such a sudden drop would be expected to be associated with a sudden change in the star formation history of galaxies, which is not observed. Indeed, the observed star formation history of galaxies based on UV and IR observations, shows no sudden change at $4 \lesssim z \lesssim 8 $ \citep[]{Madau 2014}. Recent studies \citep*[e.g.][]{Morrison et al. 2016, De Cia et al. 2018, Poudel et al. 2018} have shown that the sudden drop reported at $z>4.7$  can be explained in terms of  dust depletion, which cannot be ignored even at $z\sim5$. As observed in the Milky Way (MW) Interstellar Medium (ISM), elements with a higher condensation temperature (e.g, Si, Fe etc) are subject to more depletion into dust grains than elements with a relatively lower condensation temperature (S, O, Zn etc). Even in DLAs the depletion is more severe for refractory elements like Si, Fe, than for volatile elements O, S \citep[e.g.][]{De Cia et al. 2016}. \\

The results from a small sample of seven systems \citet{Poudel et al. 2018} with dust-free abundance measurements suggests that the N$_{\rm H\,I}$-weighted mean metallicity at $z \sim5$ absorbers agrees with the prediction from $z < 4.5$ DLAs within $<0.5 \sigma $. Furthermore, the N$_{\rm H\, I}$-weighted mean metallicity at $z \sim 5$ from \citet{Poudel et al. 2018} is in excellent agreement with results of recent simulations from \citet{Finlator et al. 2018}.\\

Evolution of metallicity as a function of redshift is a powerful tracer of the cosmic star formation history. It is therefore important to determine the metallicity evolution of DLAs/sub-DLAs accurately. Given the potentially interesting implications of a sudden drop in metallicity at z$\sim$5 and the conflict such a drop would produce with existing chemical evolution models, it is especially important to check with a larger sample whether or not such a sudden drop in metallicity actually exists. With this goal, we have been expanding measurements of S and/or O in other DLAs at  $z \ge 4.5$. Here we present results of VLT and Magellan observations of 7 more absorbers in this redshift range. Inclusion of these new measurements doubles the current high-$z$ sample of volatile elements, and thus offers a better understanding of DLA metallicity evolution at $z \sim 5$. We also examine the relative element abundances and their variation with velocity of the absorbing gas.\\ 

This paper is organized as follows: section~\ref{sec:obs} presents our observations and details of data reduction. Section~\ref{sec:method} presents the measurements of the metal lines and determination of the element abundances. Section~\ref{sec:result} describes the results for the individual absorbers. Section~\ref{sec:discussion} presents a discussion of our results. Finally, section~\ref{sec:conclusion} summarizes our conclusions.
 

\section{Observations and data reduction}
\label{sec:obs}
Our sample consists of seven absorbers with neutral hydrogen column density of log N$_{\rm H\, I}$ = 19.65 to 20.75, at redshifts 4.59 to 5.05 along the sight lines to four quasars. These absorbers were chosen from the $z > 4.5$ SDSS (Sloan Digital Sky Survey) absorbers with log $N_{\rm H\,I} \ge 20.0$ listed in \citet{Noterdaeme et al. 2012}. To check the validity of the systems, the SDSS spectra were checked to make sure that at least 1 metal line of any low ion (e.g. C II, Si II, O I) was detected at the same redshift as the H I DLA redshift listed in \citet{Noterdaeme et al. 2012} (or within $\pm 1500$ km s$^{-1}$ of that redshift), regardless of the strength of that metal line. In other words, the selection of the absorbers was driven by the redshift, the H I column density, and the mere existence of at least 1 other low-ionization metal line, but was independent of the strength of the metal line. \\

Two of the quasars J1557+1018 and J1253+1046 probing three absorbers were observed with MIKE in April 2017 on the Magellan Clay telescope at Las Campanas observatory in Chile (PI: Frye, 2017A). Observations were carried out with a 1" slit, and reached a spectral resolution of $\sim$22,000 with the red arm and $\sim$28,000 with the blue arm, giving a combined wavelength coverage of 3500 to 9400 {\AA}. The quasar J1233+0622 was observed with X-Shooter in 2014, a medium resolution spectrograph on VLT (Very Large Telescope) with a spectral coverage of 3000 to 25,000 {\AA}. While the near infrared arm of X-Shooter provides resolution of 5300, the visible arm of the spectrograph in which most of our lines fall, provides a resolution of 8900 with a 0.9" slit. Finally, J0306+1853 was observed with Magellan Echellette (MagE) in January 2014 using a 1" slit with a resolution of $\sim$4100 over 6000-10,000 {\AA}. The details of the observations are summarized in Table~\ref{tab:sum_obs}.\\

The MIKE data were reduced using the MIKE pipeline reduction code written in IDL  and developed by S. Burles, J. X. Prochaska and R. Bernstein. The reduction package performs all the necessary steps for bias subtraction, flat fielding and sky-subtraction. Before and after each science exposure, comparison exposures using the Th-Ar lamp were taken for wavelength calibration purposes. After the calibration and the heliocentric velocity corrections, the pipeline extracts each individual echelle order from all the exposures. Multiple exposures were obtained and combined to facilitate the rejection of cosmic rays.\\ 

X-Shooter observations of J1233+0622 were reduced using the X-Shooter Common Pipeline Library \citep[]{Goldoni et al. 2006} release 6.5.1\footnote{http://www.eso.org/sci/facilities/paranal/instruments/xshooter/} to obtain the flat fielded, wavelength calibrated and sky-subtracted 2D spectra from the 2-D raw images. The final 1-D spectrum was extracted using spectral point spread function (SPSF) subtraction as described in \citet{Rahmani 2016}. The details of the reduction procedure are described in \citet{Poudel et al. 2018}.\\ 

The MagE data were reduced using a custom-built pipeline and flux-calibrated using techniques described in \citet{Becker et al. 2015}.\\ 

For all objects, the final 1-D spectra were continuum-normalized and multiple exposures were combined thereafter using IRAF\footnote{IRAF is distributed by the National Optical Astronomy Observatory, which is operated by the Association of Universities for Research in Astronomy, Inc., under cooperative agreement with the National Science Foundation.}.

\section{Voigt profile fitting and abundance measurements}
\label{sec:method}
We used the Voigt profile fitting program, VPFIT\footnote{https://www.ast.cam.ac.uk/rfc/vpfit.html} v. 10.0 for the determination of column densities of metal lines and neutral hydrogen. VPFIT minimizes the $\chi ^2$ residual between the data and theoretical Voigt profiles convolved with the instrumental line spread profile through multiple iterations. VPFIT makes it possible to perform simultaneous multi-component Voigt profile fitting of multiple lines of multiple ions, while allowing the Doppler $b$ parameters and redshifts of the corresponding components to be tied together. While the line blending can be an issue with lower resolution spectrographs, three out of the seven absorbers in our sample have a relatively higher resolution of ~13.7 km s$^{-1}$ which can resolve the blending of metal lines. For many of the systems we were also able to use multiple lines to minimize the error caused by possible saturation effects \citep[e.g.][]{Penprase et al. 2010}.  The fitted column densities and associated uncertainties determined with VPFIT were confirmed by comparing the calculated and observed line profiles (in the core as well as the wings) and verifying that the profiles agreed within the noise level present. Another potential issue with high-redshift systems is the severe blending of metal lines with hydrogen lines in the dense Lyman forest. However, all of the lines used for the column density determinations were outside the Lyman-alpha forest and were not affected by the forest lines. For the sake of comparison, we have also included apparent optical depth (AOD) column densities \citep[e.g.][]{Savage 1991} in addition to the column densities derived from the Voigt profile fitting for all of our systems. In most cases, the column densities estimated from the AOD method are consistent with those estimated from the Voigt profile fitting method within $\sim1 \sigma$. In cases with only one saturated line each for O I and/or C II, the AOD method cannot give reliable lower limits to the column densities. Limits from a linear curve of growth are also not applicable for such saturated lines. In such cases, we adopt the lower limits obtained from Voigt profile fitting.\\
 
Another challenge affecting high-redshift systems is the continuum normalization blueward of the Lyman-alpha emission. This is partly due to severe blending with the Lyman-alpha forest and partly because of the relatively low signal-to-noise ratio for the faint quasars. The typical resolution of 13 - 55 km s$^{-1}$  used for most of our absorbers, was adequate to resolve the forest for determining the neutral hydrogen column density. For each sightline, we selected the entire available spectrum blueward of the Lyman-alpha emission for the continuum normalization. In most cases, the spectra covered at least the first three Lyman-series lines. Regions of the spectra unaffected by any absorption were selected to fit the continuum. A spline polynomial, typically of order 5, was fitted to the selected continuum region using the CONTINUUM task in IRAF. H~ I column densities were estimated by fitting Voigt profiles to the Lyman series lines that were usable. For most of the systems, the H I column densities were measured by fitting Voigt profiles to the Lyman-alpha lines taking the core and the wings of the profiles into consideration \citep[e.g.][]{Poudel et al. 2018}. In some cases, we were able to use both the Lyman-beta and the Lyman-alpha lines to estimate the H I column densities. While the Lyman-alpha lines for the absorbers at  z$=$4.589, 4.600 along the sightline to the quasar J1253+1046 were blended with each other, we were able to make a joint fit of Lyman-alpha and Lyman-beta lines of these two systems together to estimate the H I column densities for both the absorbers. Moreover, both the Lyman-alpha and Lyman-beta lines were used to estimate the H I column density for the absorber at z$=$4.987 towards J0306+1853. In several cases, Lyman-series lines beyond Lyman-alpha were not useful, partly due to severe blending with the dense Lyman-alpha forest at these high redshifts, and partly due to the high noise level (see Fig. \ref{fig:lyman_series1} and Fig. \ref{fig:lyman_series2} in Appendix B). To estimate the uncertainties in the H I column densities, we examined the range of values for which the fitted profiles are consistent with the observed data within the noise level \citep[e.g.][]{Poudel et al. 2018}. We note that the uncertainties thus estimated are more conservative than estimates based on the continuum placement errors alone. (The uncertainties  in the H I column densities contributed by error in continuum placement were estimated by shifting the continuum up and down by the 1$\sigma$ uncertainty, renormalizing the continuum at those levels, and then refitting the Voigt profiles in each case. The uncertainties in the H I column densities resulting from continuum placement errors thus derived were typically $\lesssim$0.1 dex)\\

We have not made ionization corrections to any of the abundances since observations of higher ionization lines are not available to place constraints on the ionization parameter. In any case, the ionization corrections are expected to be negligible for DLAs and $\lesssim0.2$ dex for sub-DLAs \citep*[e.g.][]{Dess et al. 2003, Meiring et al. 2009, Cooke et al. 2011, Som et al. 2015}. \\

All abundances reported in this paper are relative to the solar abundances taken from \citet{Asplund et al. 2009}.  The rest-frame wavelengths and oscillator strengths used for the Voigt profile fitting were taken from \citet{Morton 2004} and \citet{Cashman et al. 2017}.



\begin{table*}
	\centering
	\caption{Summary of targets and observations}
	\label{tab:sum_obs}
	\begin{tabular}{ccccccccc} 
		\hline
		\hline
		\parbox[t]{1.6in}{QSO\par RA(J2000), Dec(J2000)\strut}  & m$_i$ & z$_{em}$ & z$_{abs}$ & log $N_{\rm H\, I}$ & Instrument & \parbox[t]{0.5in}{Exposure  \par Time (s)\strut} & \parbox[t]{0.7in}{Wavelength \par Coverage({\AA})\strut} & \parbox[t]{0.8in}{Spectral\par Resolution (R)\strut}\\  
		\hline
		\hline
\parbox[t]{1.6in}{J0306+1853\par RA: 03:06:42.51; Dec: +18:53:15.8\strut} & 17.96 & 5.363 & 4.987 &  20.60$\pm$0.15 & Magellan MagE & $1800\times 3$ & 6000-10,000 & 4100\\
	\hline	
\parbox[t]{1.6in}{J1233+0622\par RA: 12:33:33.47; Dec: +06:22:34.3\strut} & 20.09 & 5.311 & \parbox[t]{0.3in}{4.859 \par 5.050 \strut} & \parbox[t]{0.5in}{20.75$\pm$0.15 \par 20.10$\pm$0.15 \strut} & VLT X-Shooter & $3600$ & 3000-25,000 & \parbox[t]{0.7in}{VIS: 8800\par NIR: 5300\strut}\\
	\hline	
\parbox[t]{1.6in}{J1253+1046\par RA: 12:53:53.35; Dec: +10:46:03.1\strut} & 19.40 & 4.925 & \parbox[t]{0.3in}{4.589 \par 4.600 \par 4.793 \strut} & \parbox[t]{0.5in}{19.75$\pm$0.15 \par 20.35$\pm$0.15 \par 19.65$\pm$0.10 \strut} & Magellan MIKE & $2700 \times 3$ & 3500-9,400 & \parbox[t]{0.7in}{Red: 22,000\par Blue: 28,000\strut}\\
	\hline	
\parbox[t]{1.6in}{J1557+1018\par RA: 15:57:00.17; Dec: +10:18:41.8\strut} & 19.98 & 4.765 & 4.627 &  20.75$\pm$0.15 & Magellan MIKE & $2700 \times 5$ & 3500-9,400 & \parbox[t]{0.7in}{Red: 22,000\par Blue: 28,000\strut}\\
		
		\hline
	\end{tabular}
\end{table*}

\begin{figure*}
\centering
  \begin{tabular}{@{}cc@{}}

   \includegraphics[width=.37\textwidth]{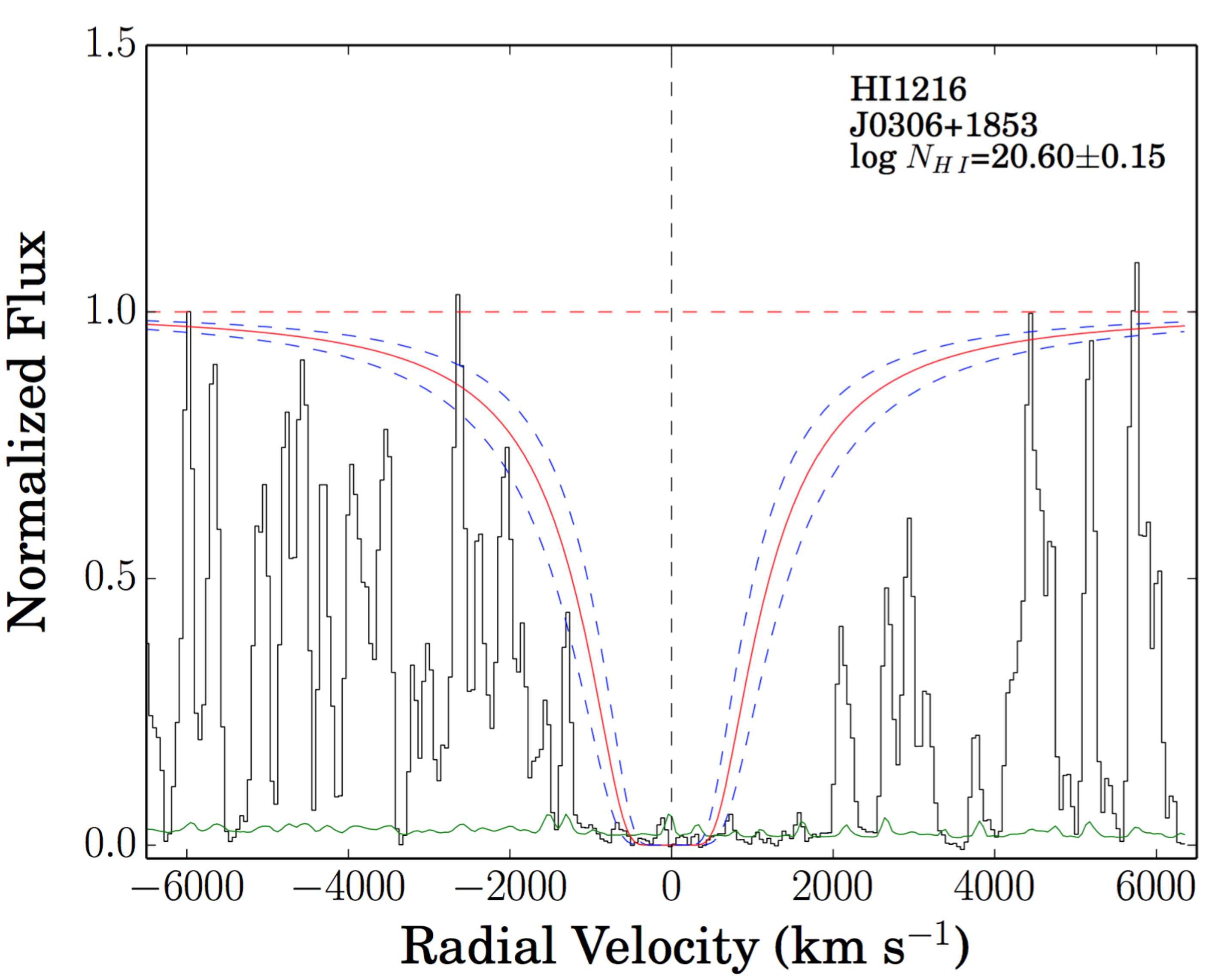} &
   \includegraphics[width=.37\textwidth]{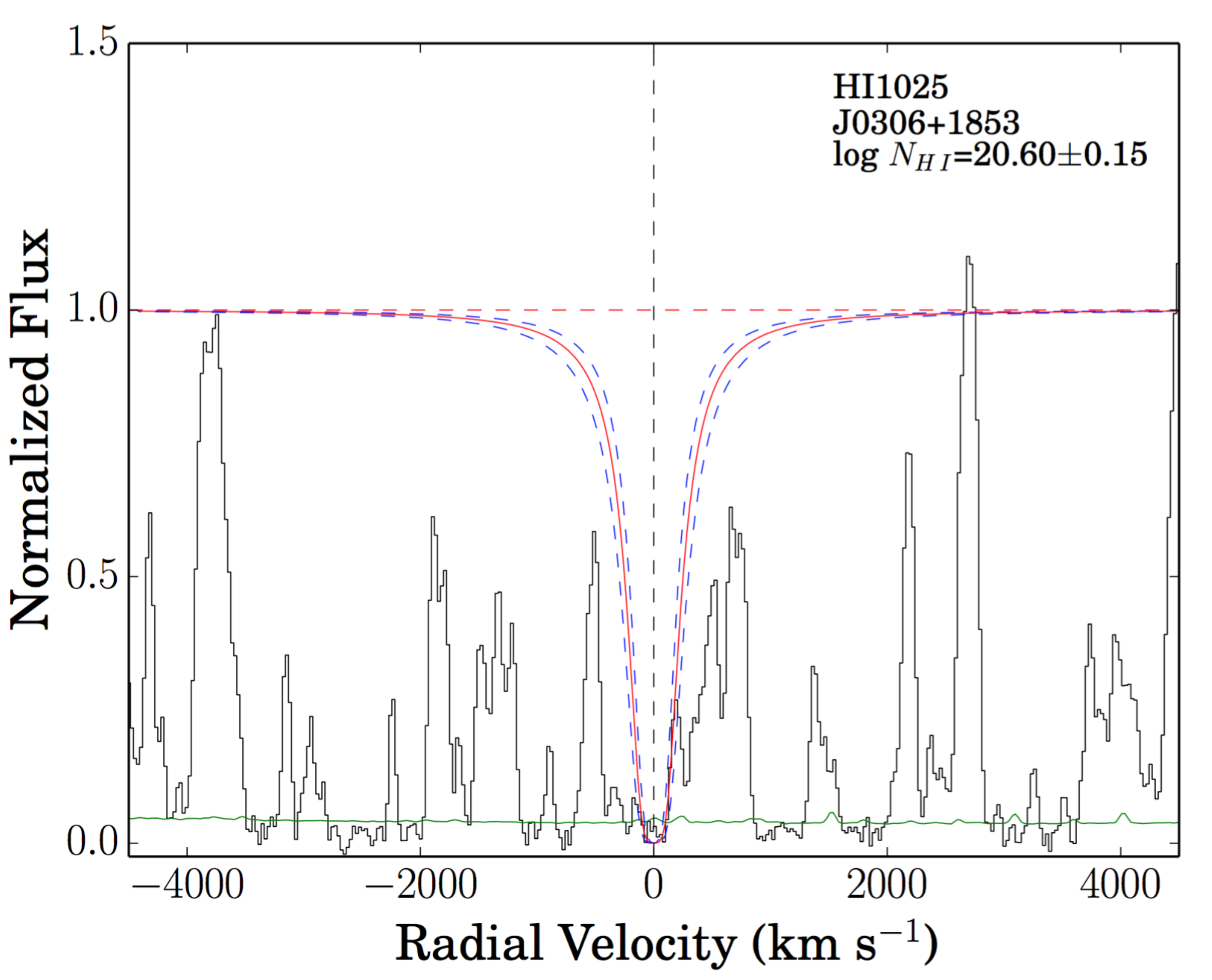}\\
    \includegraphics[width=.37\textwidth]{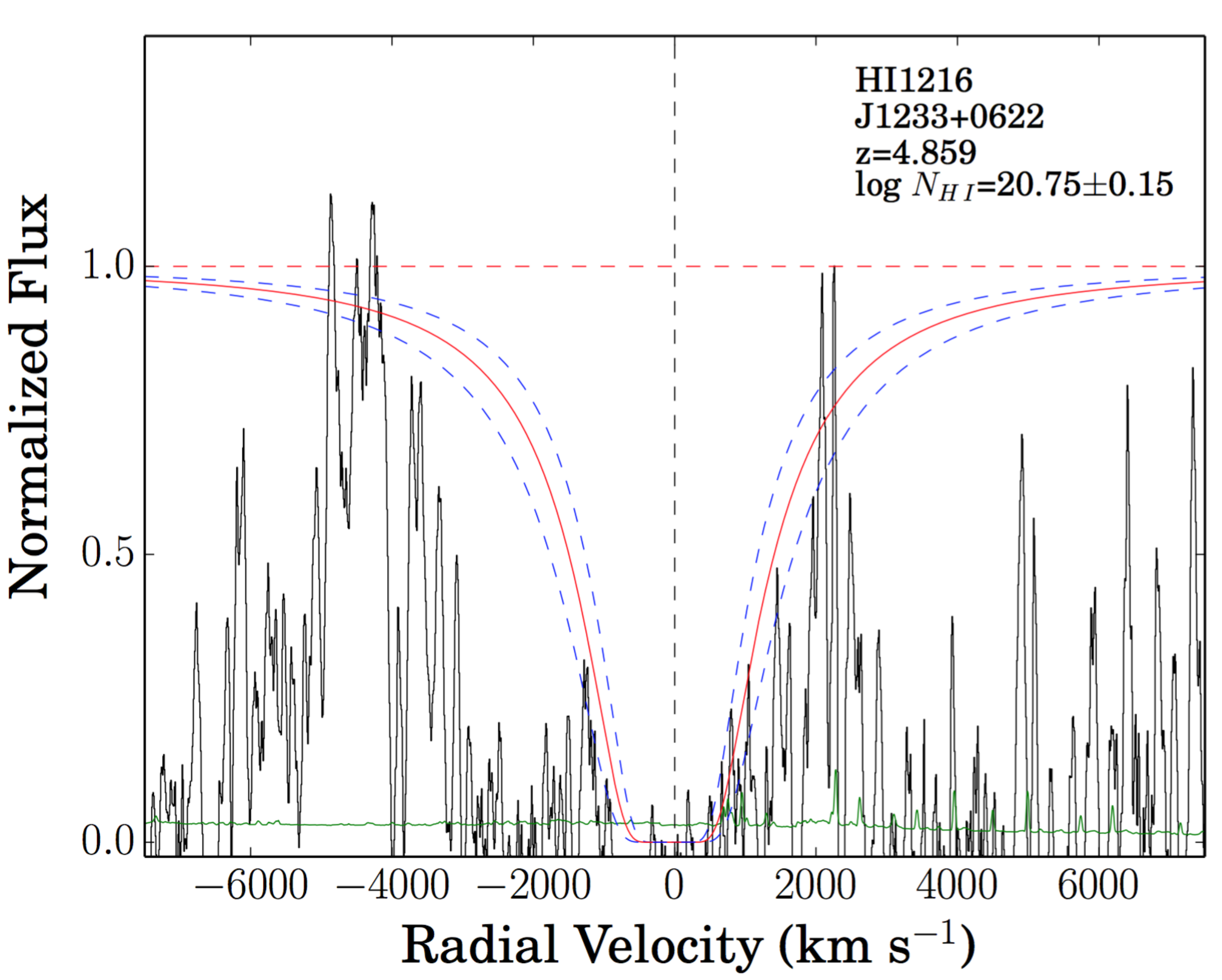} & 
     \includegraphics[width=.37\textwidth]{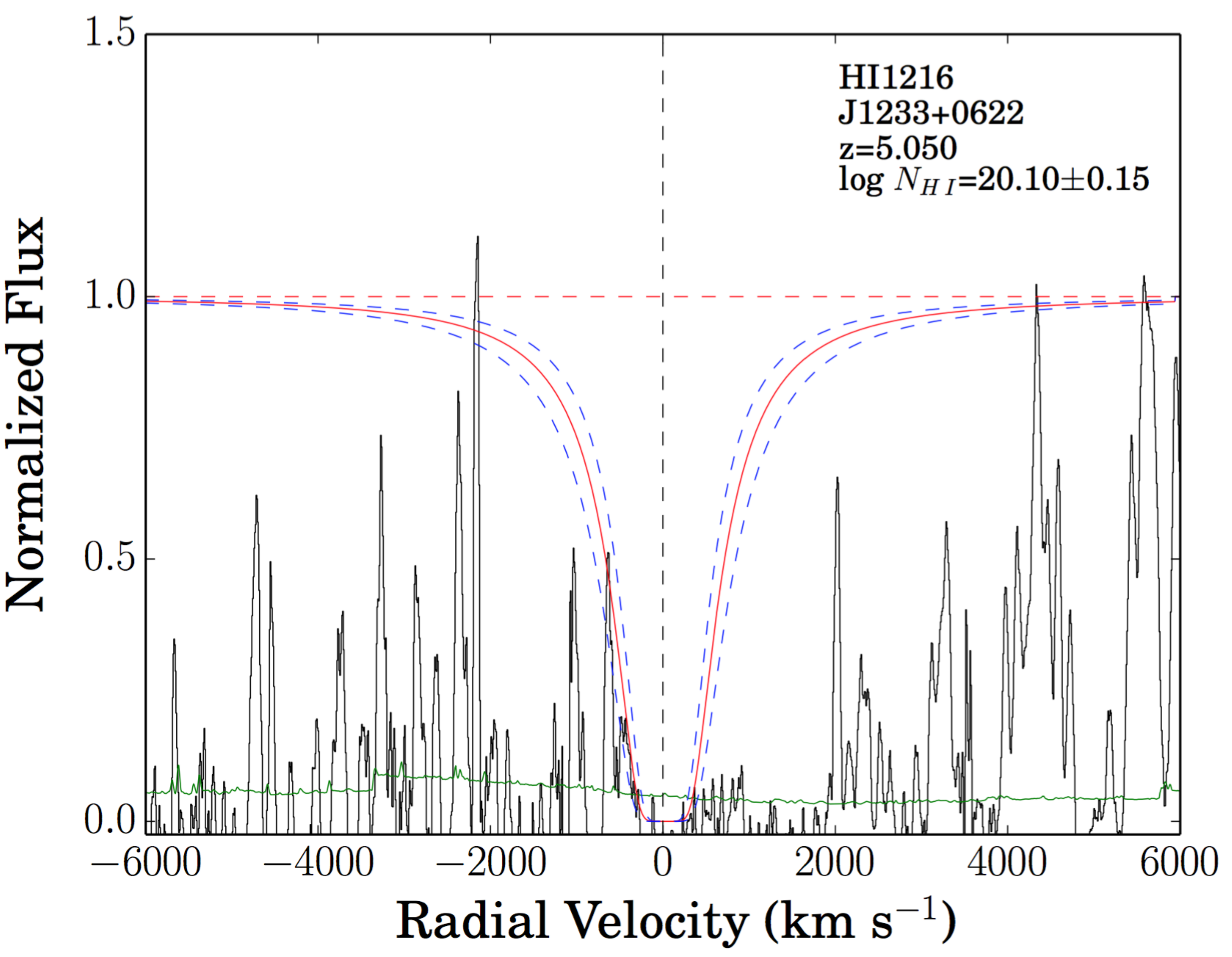} \\ 
    \includegraphics[width=.37\textwidth]{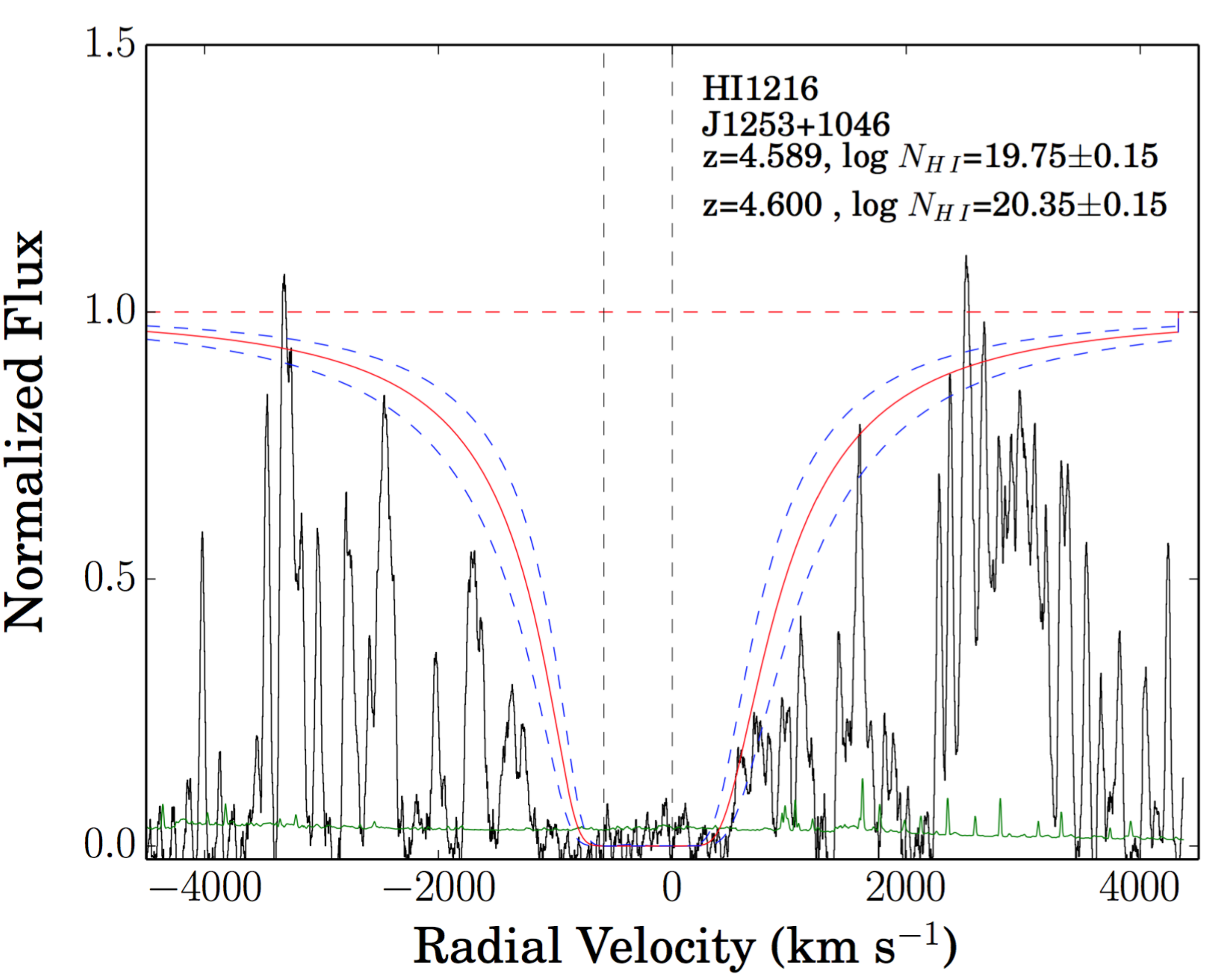} &
    \includegraphics[width=.37\textwidth]{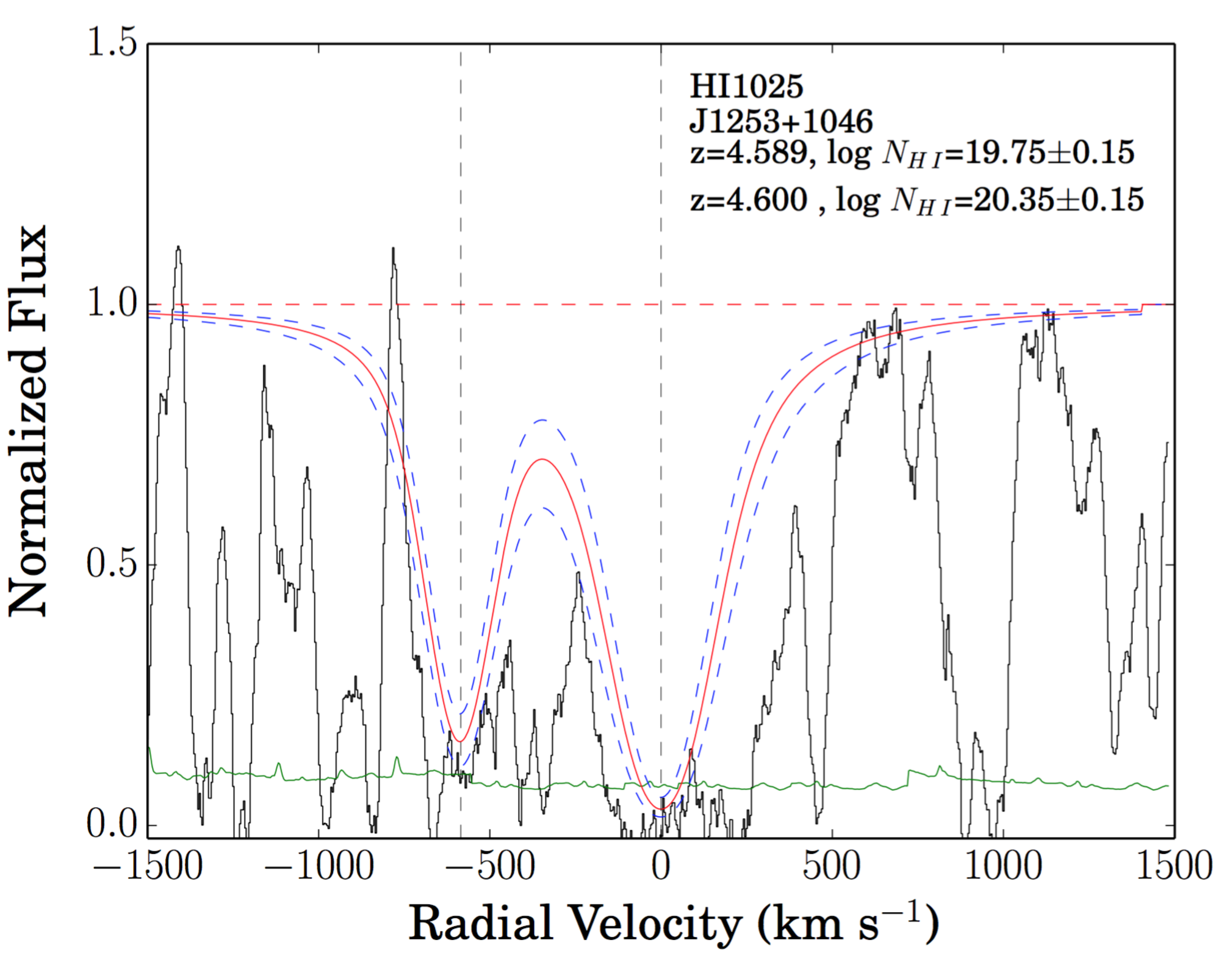} \\
    \includegraphics[width=.37\textwidth]{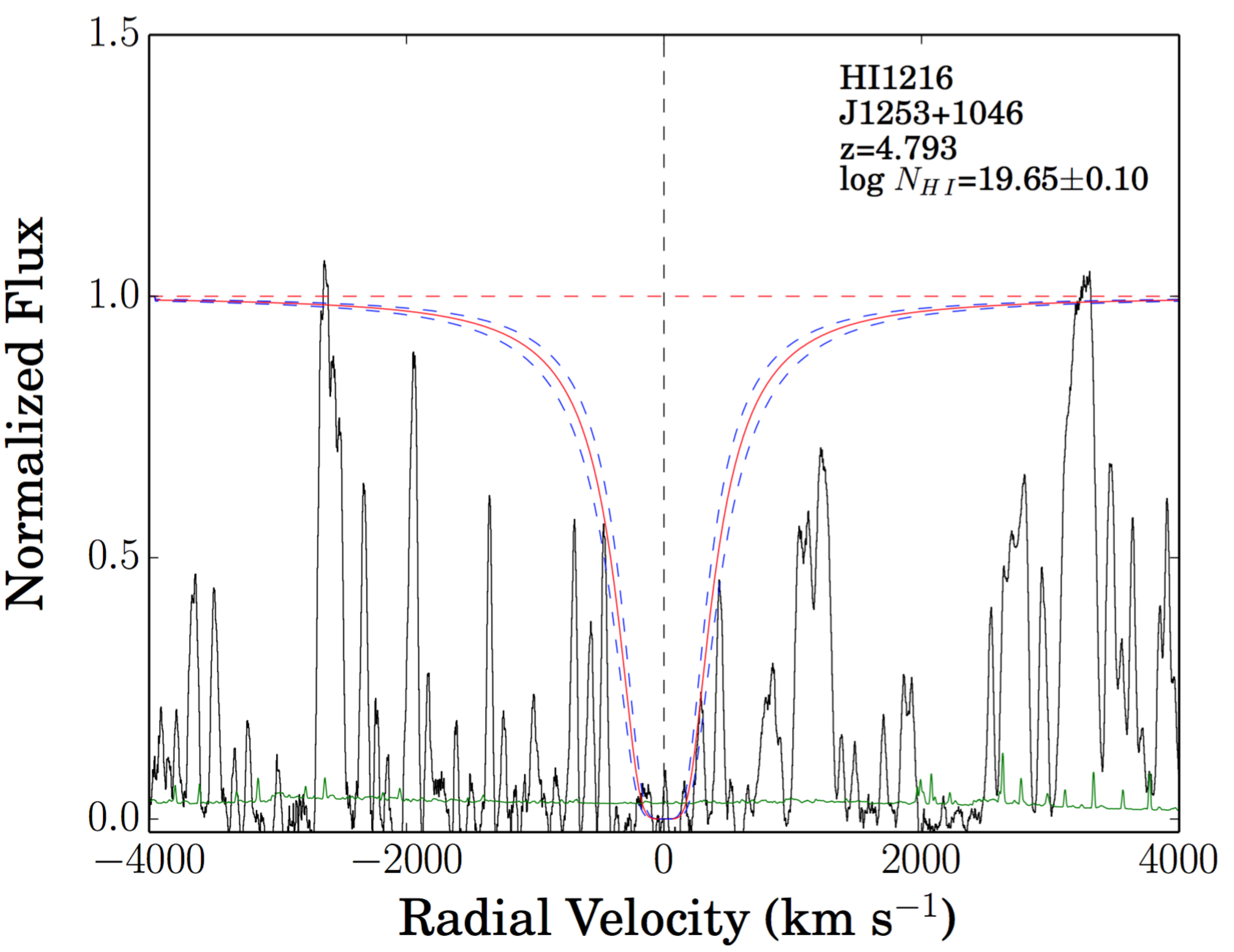}&
   \includegraphics[width=.37\textwidth]{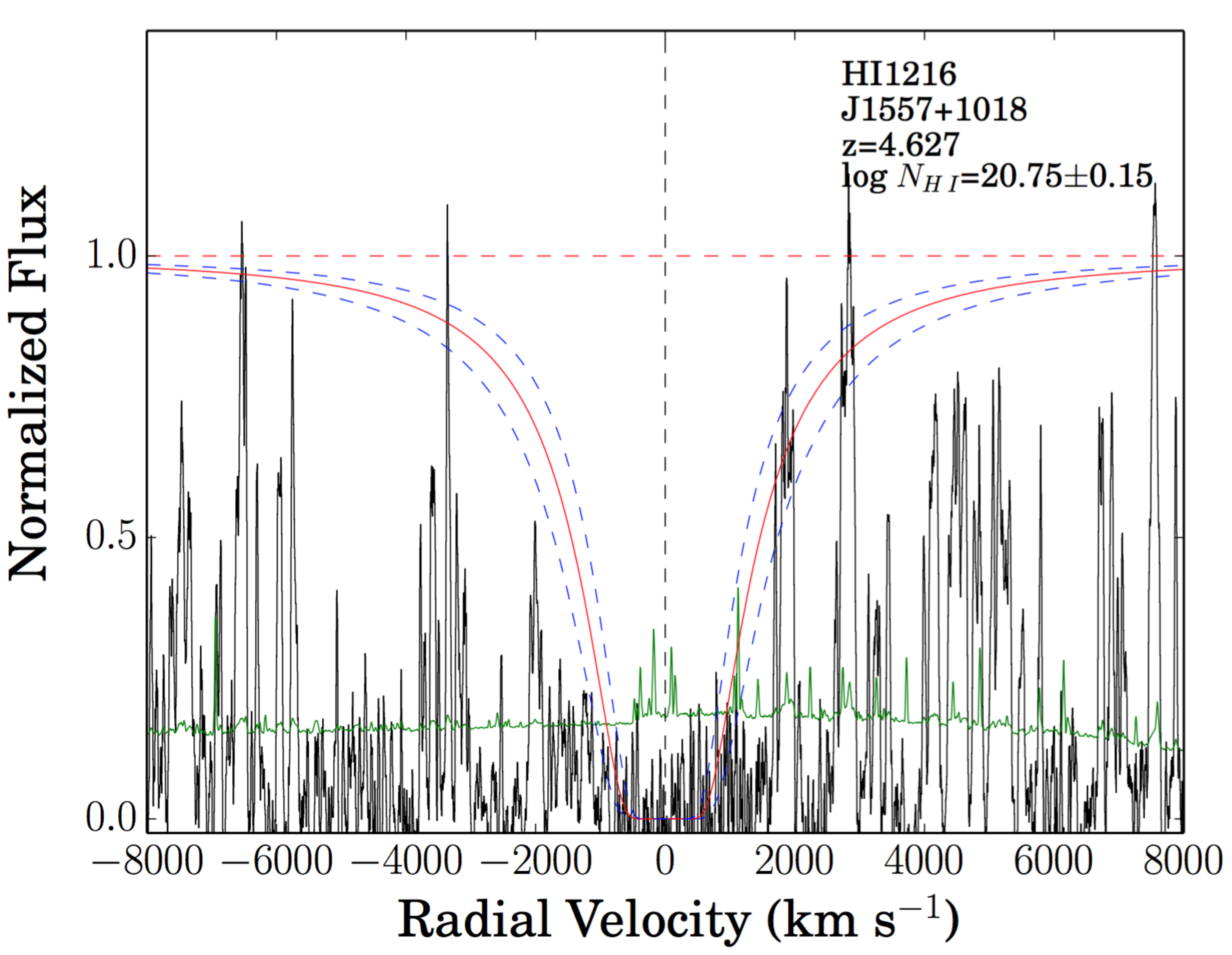}
  \end{tabular}
  \caption{Plots showing the estimation of neutral hydrogen column densities for all of the absorbers. In each case, the observed continuum normalized flux is shown in black and the best fitted profile is shown in red. The blue dashed curves above and below the fitted profile represent the uncertainty in the column density determination. The $1\sigma$ error in the normalized flux is shown in green at the bottom of each panel. The centre of the profile is fixed at the redshift determined during metal line fitting and is shown as a black dashed line. The placement of continuum is shown by the horizontal dashed red line. Lyman-alpha and Lyman-beta were used to determine hydrogen column density for an absorber at z$=$4.987 toward J0306+1853 and the two absorbers at z$=$4.589, 4.600 toward J1253+1046, while only Lyman-alpha was used for the other absorbers (see text for further details). The estimated log $N_{\rm H\,I}$ values and redshifts are mentioned in the top right corner of each panel.}
  \label{fig:lyman}
\end{figure*}

\section{Results for individual absorbers}
\label{sec:result}
In this section, we report the results for the column densities derived from Voigt profile fitting, the inferred absolute and relative abundances, and the gas kinematics determined from velocity dispersion measurements for individual absorbers. For each system, the H I lines are shown in Fig.~ \ref{fig:lyman}.

\subsection{Absorber at $z=4.987$ along the sight line to J0306+1853}
\begin{figure}
\begin{tabular}{l}
\hspace*{-0.27in}
\includegraphics[height=9.5cm, width=10cm]{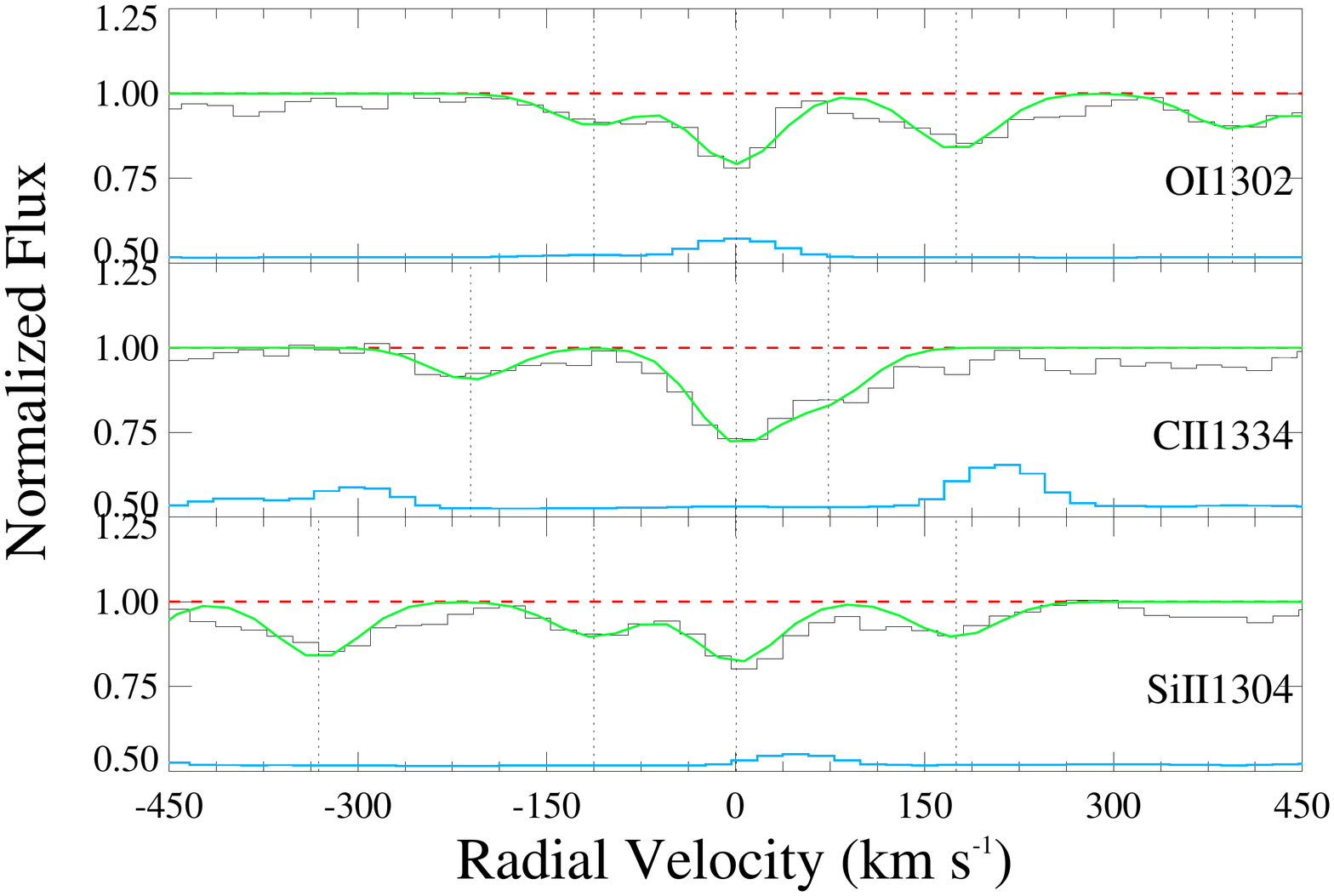}
\end{tabular}
\caption{Velocity plots for metal lines for the absorber at $z=4.987$ in the sight line to J0306+1853 from the MagE spectrograph. In each panel, the observed data are shown in black and the fitted profiles are shown in green. The blue line at the bottom of each panel shows the $1\sigma$ error in the normalized flux and is shifted by +0.5. The placement of continuum is shown by the horizontal dashed red line. The vertical dotted lines show the different velocity components included in the profile fits. It is to be noted that the rightmost vertical dotted line in O I $\lambda$1302 is attributed to Si II absorption and the leftmost vertical dotted line in Si II $\lambda$1304 is attributed to O I absorption.}
    \label{fig:metals_mage}
\end{figure}

This absorber was observed with lower spectral resolution than the other absorbers and was initially reported in \citet{Wang et al. 2015}. We reanalyzed this system by doing Voigt profile fitting for O I $\lambda$1302, Si II $\lambda$1304, and C II $\lambda$1334 and calculated the corresponding element abundances. Fig. \ref{fig:metals_mage} shows the Voigt profile fits for the metal lines. This system is a metal-poor DLA with log $N_{\rm H\, I} = 20.60\pm0.15$ and [O/H]$=-2.69\pm0.17$. Our estimate of $N_{\rm H\, I}$ is consistent with the estimate log $N_{\rm H\,I} = 20.50^{+0.10}_{-0.12}$ of \citet{Wang et al. 2015}. \citet{Wang et al. 2015} had estimated log $N_{\rm Si\,II} =14.8$ using the apparent optical method for Si II $\lambda$1304 and Si II $\lambda$1527, and thereby estimated [Si/H] = $-1.3 \pm 0.1$. The Si II $\lambda$1526 line is, however, blended significantly. We therefore did Voigt profile fitting for only the Si II $\lambda$1304 line and got a lower Si II column density log $N_{\rm Si\,II} = 14.22$. The apparent optical depth method applied to this line also gives a similar value for the Si II column density. \\

By comparison, the O I $\lambda$1302 line is weak, implying that Si is strongly enhanced relative to O, with [Si/O] = $0.79 \pm 0.09$.  Such a high [Si/O] value is surprising, given that Si and O are both alpha-elements. Thus, [Si/O] or [Si/S] is usually either zero (in the absence of dust depletion) or negative (in the presence of dust depletion, since Si is more severely depleted than the more volatile elements O or S). Given the relative weakness of O I $\lambda$1302, the high [Si/O] value is not expected to be an artifact of saturation. Nevertheless, higher resolution observations of this sight line would help to determine the column densities more accurately.

\subsection{Absorber at $z=4.859$ along the sight line to J1233+0622}
\begin{figure}
\begin{tabular}{l}
\hspace*{-0.27in}
\includegraphics[height=9.5cm, width=10cm]{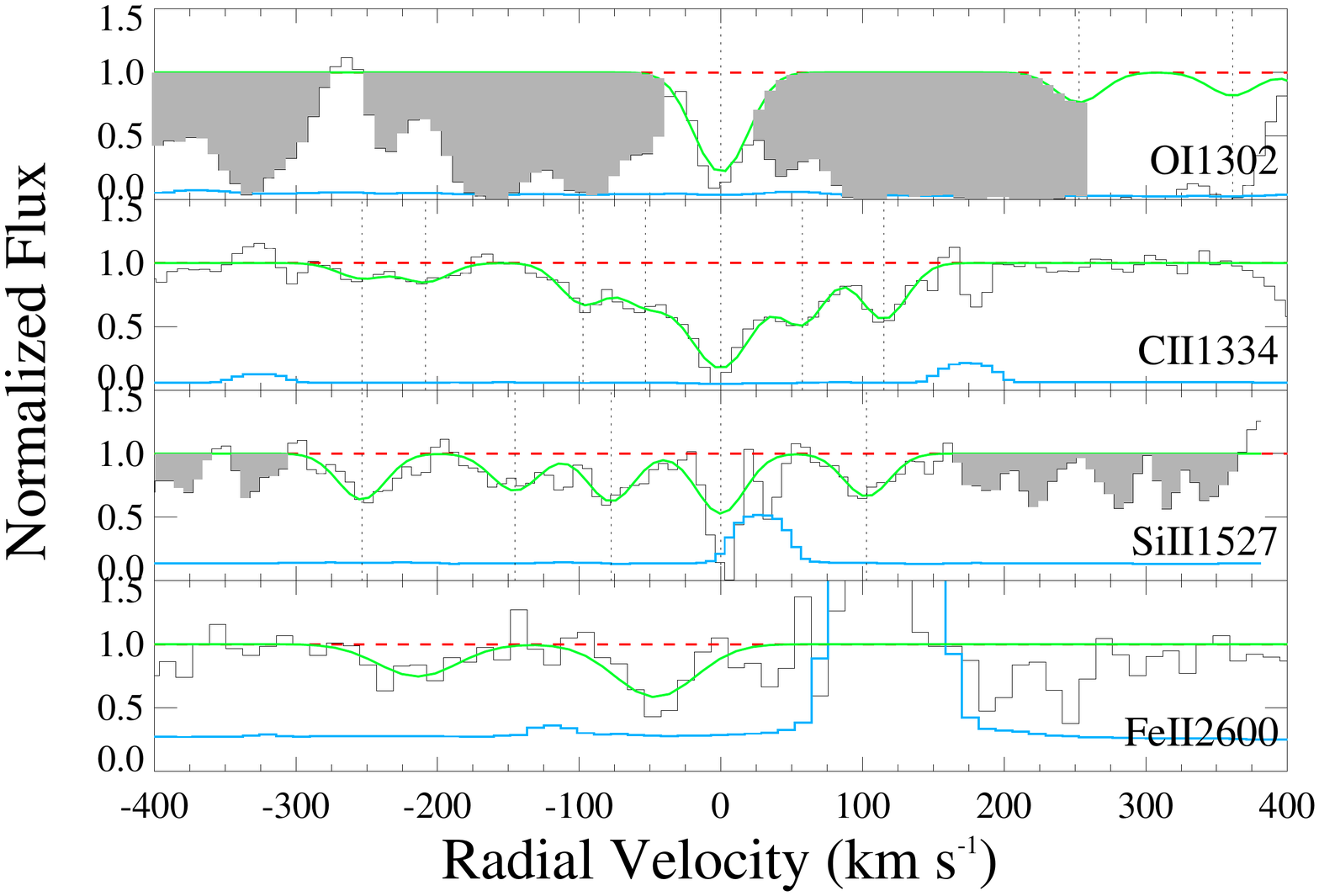}
\end{tabular}
\caption{ Velocity plots for metal lines for the absorber at $z=4.859$ in the sight line to J1233+0622 from the X-Shooter spectrograph. In each panel, the observed data are shown in black and the fitted profiles are shown in green. The blue line at the bottom of each panel shows the $1\sigma$ error in the normalized flux and the placement of continuum is shown by the horizontal dashed red line. The vertical dotted lines show the different velocity components included in the profile fits. The shaded regions are the telluric and unrelated absorption features. }
    \label{fig:metals_485}
\end{figure}

A DLA at $z =4. 859$ along the sight line to J1233+0622 has been detected with log $N_{\rm H\, I} = 20.75\pm0.15$. The Voigt profile fits for the metal lines are shown in Fig. \ref{fig:metals_485}. While C II $\lambda$1334, Fe II $\lambda$2600, and Si II $\lambda$1527 are detected for this system, the high- and low-velocity components for the O I $\lambda$1302 line are blended with telluric absorption. This absorber is spread over a large radial velocity range with at least seven components detected in C II $\lambda$1334 with X-Shooter resolution. This absorber shows some variations among different velocity components. For example, three of the components seen in Si II $\lambda$1527 do not seem to be perfectly matched with those in C II $\lambda$1334. We believe there is no possibility of blending in Si II $\lambda$1527, since it lies outside the Lyman-alpha forest, and the only other known absorber in this sightline does not have suitable lines that could cause blending. We also do not find any obvious problems with data reduction. The differences between the Si II $\lambda$  1527 and C II $\lambda$ 1334 lines may thus arise partly due to a slight offset in the wavelength calibrations of the visible and near-IR parts of X-Shooter spectra and partly from the differences in ionization, depletion, or intrinsic nucleosynthetic ratios. Such differences in nucleosynthesis/depletion have been reported before in other high-redshift absorbers even with the higher spectral resolution of Keck HIRES  \citep*[e.g.][]{Morrison et al. 2016}. Moreover, we are able to estimate the lower limit of the O I $\lambda$1302 column density by overfitting the profile for the central main component, fixing the redshift and Doppler b parameter determined in C II $\lambda$1334. Taking only this central component (excluding  the other components that are heavily blended with telluric absorption features), we place a lower limit on [O/H] $> -2.14$. For C II $\lambda$1334, as shown in Fig. \ref{fig:saturation}, we have analyzed the effect of changing the column density for the strongest component in steps of 0.1 dex. The difference between the profile and the data becomes large, considering the 1$\sigma$ uncertainty in the normalized flux, if the C II column density is increased by $> 0.1$dex compared to the adopted value of 14.42. This is consistent with the uncertainty of 0.07 dex obtained from VPFIT. The Fe II profile seems to differ in structure from C II and Si II profiles in the sense that almost no Fe II is seen in the central components seen in C II and Si II. The peak Fe II absorption seems to be blue-shifted by $\sim 50$~ km s$^{-1}$ with respect to the peak absorption in C II and Si~ II. 

\subsection{Absorber at $z=5.050$ along the sight line to J1233+0622}
\begin{figure}
\begin{tabular}{l}
\hspace*{-0.27in}
\includegraphics[height=9.5cm, width=10cm]{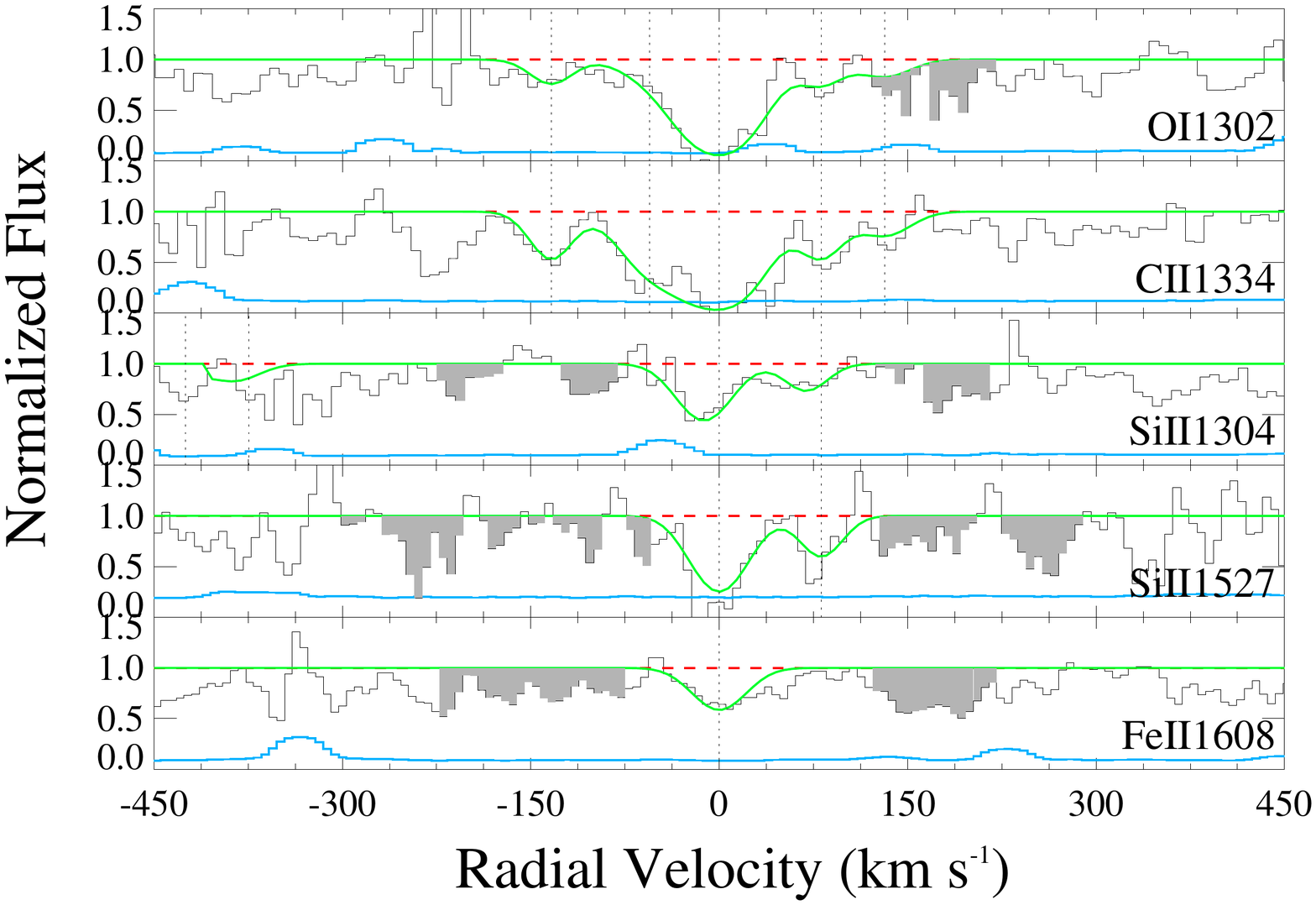}
\end{tabular}
\caption{Same as Fig. \ref{fig:metals_485} but for the absorber at $z=5.050$ in the sight line to J1233+0622. Unrelated absorption features are shaded in grey.}
    \label{fig:metals_505}
\end{figure}

The quasar sight line to J1233+0622 shows a second absorber at $z=5.050$ with log $N_{\rm H\, I} = 20.10\pm0.15$. This absorber is detected in O I $\lambda$1302, C II $\lambda$1334, Si II $\lambda$1304, Si II $\lambda$1527, and Fe II $\lambda$1608. The corresponding Voigt profile fits are shown in Fig. \ref{fig:metals_505}. This is also a metal-rich absorber which shows multiple components spread over a large radial velocity range with [O/H] of $-1.60\pm0.18$. Furthermore, we have analyzed the effect of changing the column density for the strongest component both in O I $\lambda$1302 and C II $\lambda$1334 in steps of 0.1 dex (see Fig. \ref{fig:saturation}). The difference between the data and the profile becomes large, considering the 1 $\sigma$ uncertainty in the normalized flux, if the column density is increased by >0.1 dex for O I and >0.2 dex for C II compared to the adopted column density values. Thus the uncertainties in the column densities are consistent with the values obtained from VPFIT (0.12 dex for O I and 0.16 dex for C II). As Si II $\lambda$1527 is saturated, we used Si II $\lambda$1304 to estimate the abundance for silicon. While oxygen and carbon are detected in multiple components, the absence of some corresponding components in silicon might suggest spatial variation in the dust depletion. The central velocity of the Fe II profile is consistent with the velocities of the main components in O I, C II and Si II profiles.

\subsection{Absorber at $z=4.589$ along the sight line to J1253+1046}
\begin{figure}
\begin{tabular}{l}
\hspace*{-0.27in}
\includegraphics[height=9.5cm, width=10cm]{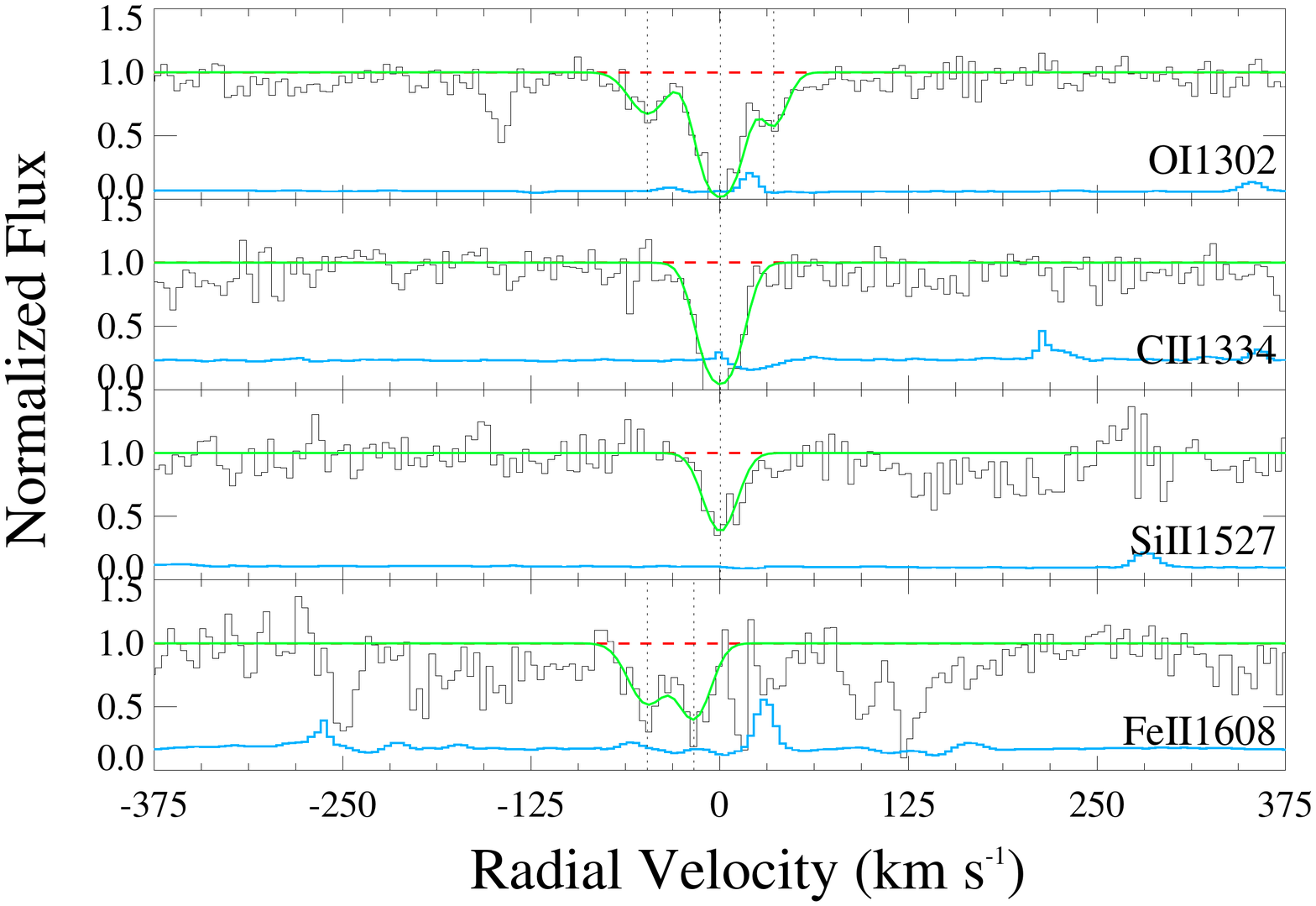}
\end{tabular}
\caption{Velocity plots for metal lines for the absorber at $z=4.589$ in the sight line to J1253+1046 from the MIKE spectrograph. In each panel, the observed data are shown in black and the fitted profiles are shown in green. The blue line at the bottom of each panel shows the $1\sigma$ error in the normalized flux and the placement of continuum is shown by the horizontal dashed red line. The vertical dotted lines show the different velocity components included in the profile fits.}
    \label{fig:metals_4589}
\end{figure}

The sight line to quasar J1253+1046 probes a sub-DLA at a redshift $z=4.589$. Although the Lyman-alpha for this absorber is heavily blended with a nearby absorber, making it difficult to determine the hydrogen column density, we were able to use the less blended Lyman-beta line, and estimate log $N_{\rm H\, I} = 19.75\pm0.15$. The corresponding Voigt profile fits for O I $\lambda$1302, C II $\lambda$1334, Si II $\lambda$1527, and Fe II $\lambda$1608 are shown in Fig. \ref{fig:metals_4589}. O I $\lambda$1302 is saturated, but based on Fig. \ref{fig:saturation}, the O I column density for the strongest component cannot be higher by $>0.2$ dex than the adopted value due to the effect in the line wings. Also, the difference between the data and the profile for C II $\lambda$1334 becomes large if the column density is increased by $>$0.2 dex compared to the adopted value, which is consistent with the uncertainty of 0.22 dex obtained from VPFIT. The [O/H] value of $-1.43\pm0.17$ for this absorber suggests this as a metal-rich sub-DLA. Besides the main component, oxygen is the only element showing multiple components, one at -40 $\rm km s^{-1}$ and another at 43 $\rm km s^{-1}$. Oxygen seems to be enhanced over silicon suggesting possible depletion into dust grains. Surprisingly, one of these  higher velocity components seen in O I but not in C II or Si II is seen in Fe II. Furthermore, as in the case of the $z=4.859$ absorber toward J1233+0622, the main central component seen in O I, C II, and Si II is not seen in Fe II. The center of the Fe II profile seems to be blue-shifted by $\sim 35$ km s$^{-1}$ with respect to the central components in O I, C II, and Si~ II.

\subsection{Absorber at $z=4.600$ along the sight line to J1253+1046}
\begin{figure}
\begin{tabular}{l}
\hspace*{-0.27in}
\includegraphics[height=9.5cm, width=10cm]{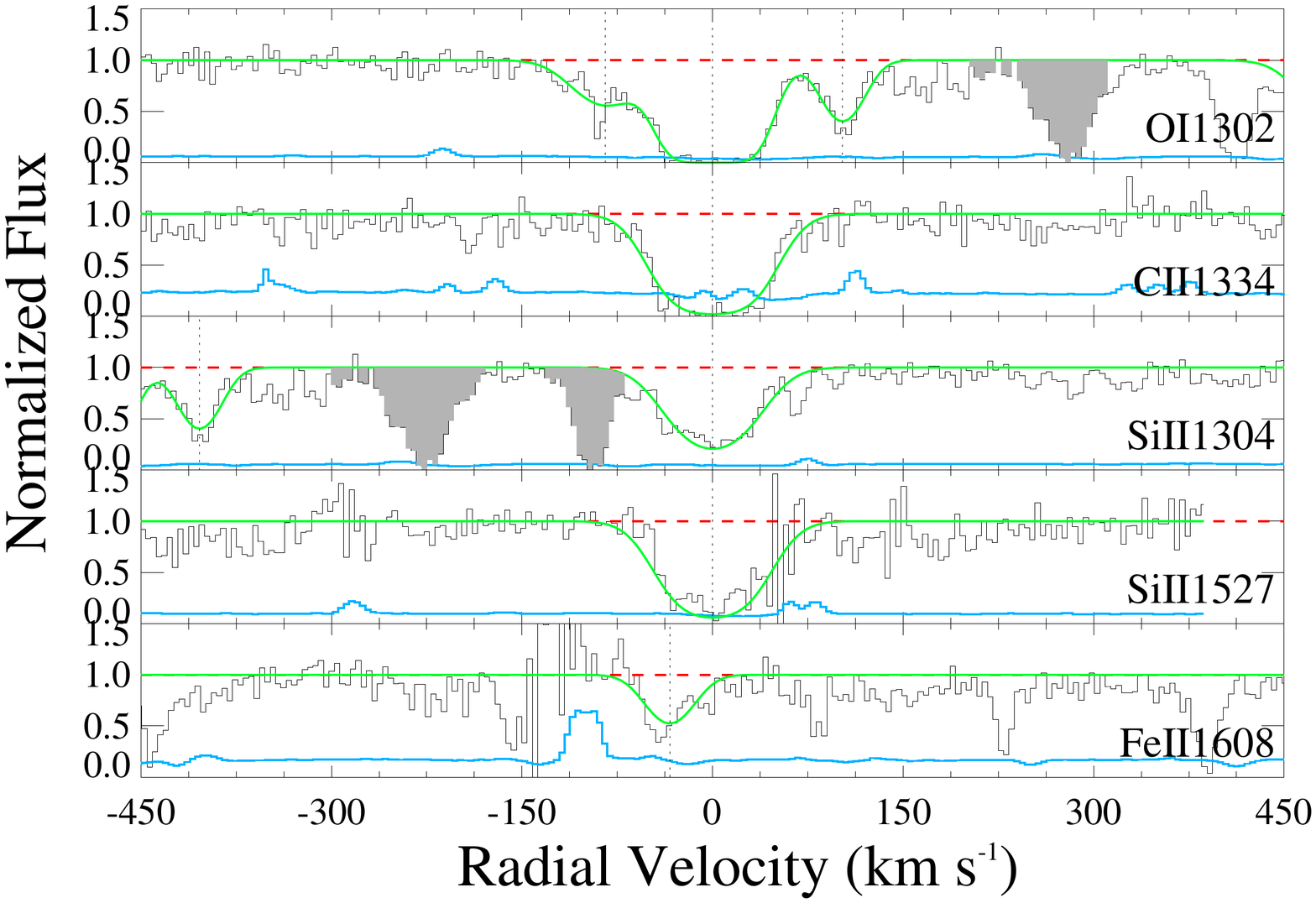}
\end{tabular}
\caption{Same as Fig. \ref{fig:metals_4589} but for the absorber at $z=4.600$ in the sight line to J1253+1046. The shaded regions are unrelated absorption features.}
    \label{fig:metals_460}
\end{figure}

The quasar sight line to J1253+1046 also shows a DLA absorber at $z=4.600$. The column density estimation for Lyman-alpha gives log $N_{\rm H\, I}= 20.35\pm0.15$. O I $\lambda$1302, C II $\lambda$1334,  Si II $\lambda$1304, Si II $\lambda$1527, and Fe II $\lambda$1608 are detected in this absorber and the corresponding Voigt profile fits are shown in Fig. \ref{fig:metals_460}. This is a metal-rich DLA, [O/H] has a lower limit of $-1.46$. Since C II $\lambda$1334 is also saturated, we adopted a lower limit as shown in Table. \ref{tab:rel_abn}. This absorber also shows spatial variation in the relative abundances as two extra components are detected in O I $\lambda$1302 but not in silicon and carbon. Once again, the Fe II profile is quite distinct from the profiles of O I, C II, Si II, and seems to be shifted to the blue by $\sim$ 50~ km s$^{-1}$.

\subsection{Absorber at $z=4.793$ along the sight line to J1253+1046}
\begin{figure}
\begin{tabular}{l}
\hspace*{-0.27in}
\includegraphics[height=9.5cm, width=10cm]{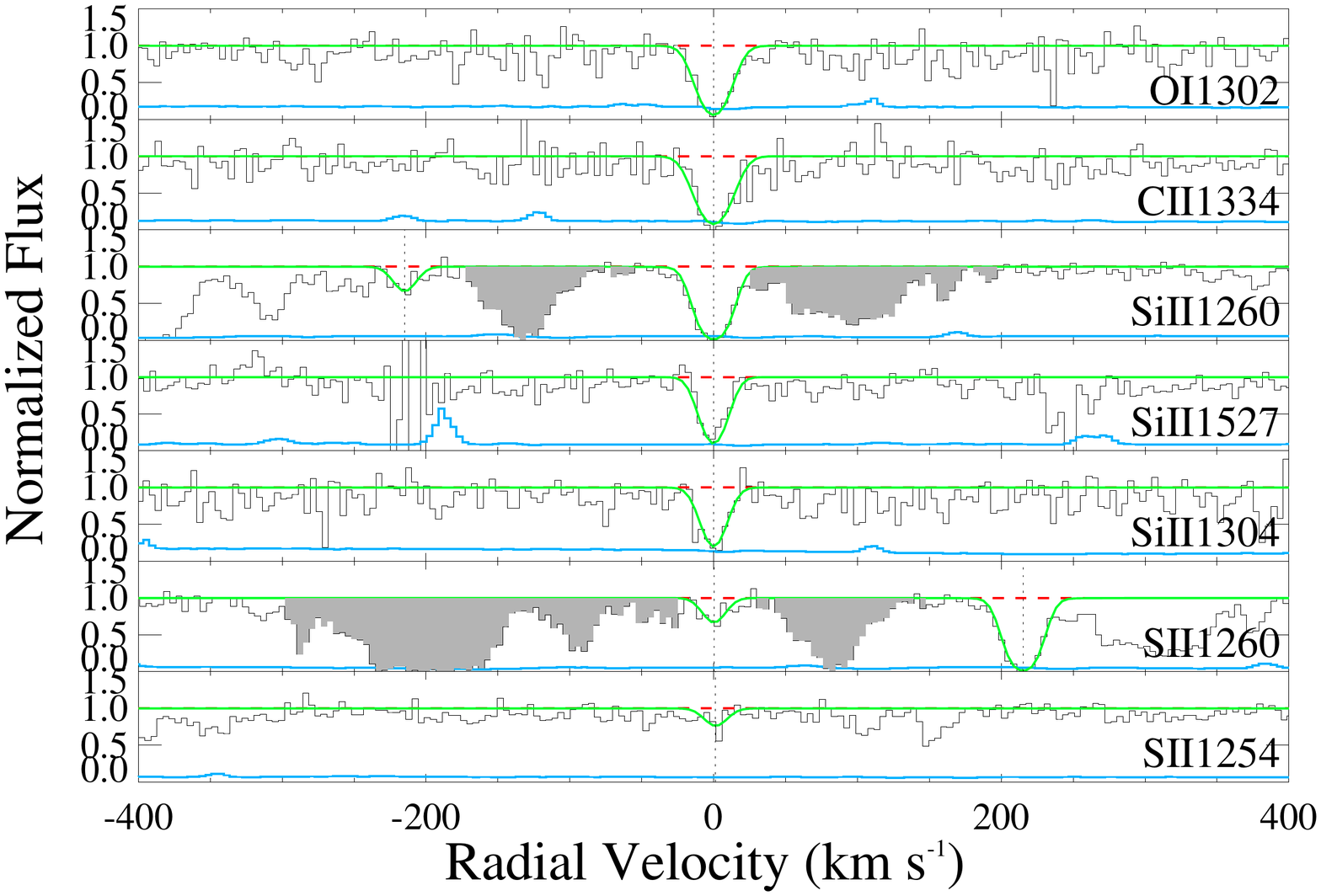}
\end{tabular}
\caption{Same as Fig. \ref{fig:metals_4589} but for the absorber at $z=4.793$ in the sight line to J1253+1046. The shaded regions are unrelated absorption features.}
    \label{fig:metals_4793}
\end{figure}

This is the second sub-DLA detected along the same sight line to J1253+1046 with log $N_{\rm H\, I} = 19.65\pm0.10$. Besides O I $\lambda$1302.2, C II $\lambda$1334.5, Si II $\lambda$1304.4, Si II $\lambda$1260.4, Si II $\lambda$1527, this absorber also shows detection of S II $\lambda\lambda$1254,1260. The corresponding Voigt profile fits are shown in Fig. \ref{fig:metals_4793}. Since O I $\lambda$1302 is saturated, the abundance based on Voigt profile fitting has a lower limit of $-1.63$. However, the difference between the data and the profile for C II $\lambda$1334 based on Fig. \ref{fig:saturation} becomes large if the C II column density is increased by $>$0.2 dex compared to the adopted value, which is consistent with the uncertainty of 0.17 dex obtained from VPFIT. Moreover, the sulphur abundance is significantly higher with a value of [S/H] = $-0.59\pm0.20$. In fact, such a high metallicity is quite rare in DLAs even at z $<$ 2. Also, Si shows a much smaller abundance than S, suggesting strong dust depletion. To date, this absorber has the highest metallicity among $z > 4.5$ sub-DLAs or DLAs.

\subsection{Absorber at $z=4.627$ along the sight line to J1557+1018}
\begin{figure}
\begin{tabular}{l}
\hspace*{-0.27in}
\includegraphics[height=9.5cm, width=10cm]{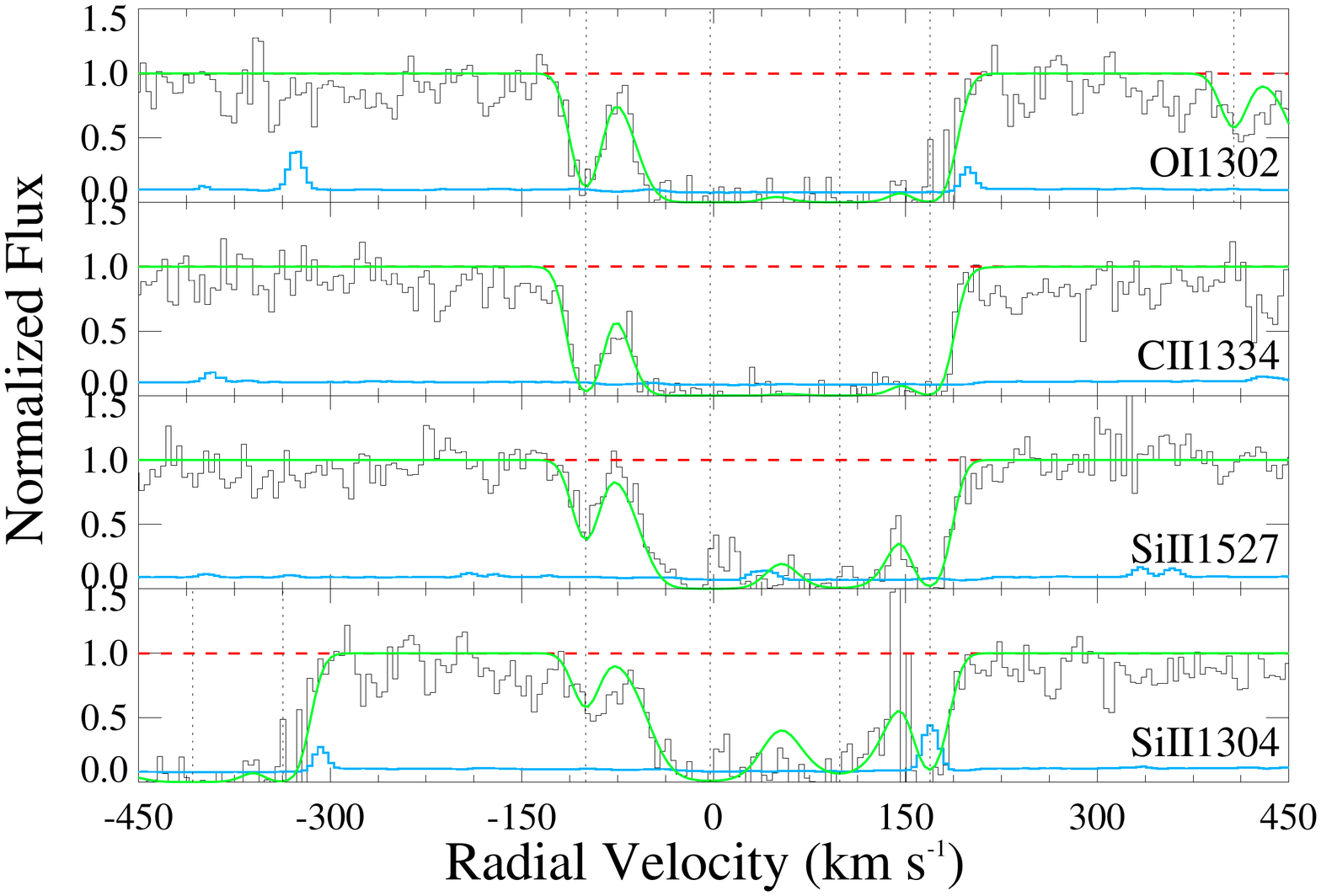}
\end{tabular}
\caption{Same as Fig. \ref{fig:metals_4589} but for the absorber at $z=4.627$ in the sight line to J1557+1018.}
    \label{fig:metals_4628}
\end{figure}

The sight line to J1557+1018 shows a DLA at $z=4.627$ that was estimated by \citet{Noterdaeme et al. 2012} to have log $N_{\rm H\, I} = 21.30$ based on its SDSS spectrum. Based on our higher resolution MIKE spectrum, we get a value of log $N_{\rm H\, I} = 20.75\pm0.15$. However, we note that the S/N at the position of the Lyman-alpha line is low. Higher S/N data would be useful to obtain a more definitive determination of $N_{\rm H\, I}$ in this absorber. Fig. \ref{fig:metals_4628} shows the fitted metal lines for O I $\lambda$1302, C II $\lambda$1334, Si II $\lambda$1304, and Si II $\lambda$1527. \\

All the accessible absorption lines in this absorber are saturated. Thus we can estimate only lower limits on the metal column densities and the corresponding element abundances. For example, based on O I $\lambda$1302, we estimate [O/H] $\ge$ -1.47. This DLA is thus remarkably metal-rich for its high redshift.

\begin{table*}
	\centering
	\caption{Results of Voigt profile fitting for different elements in the absorbers along the sight lines of all quasars in our sample. For the sake of comparison, the total AOD column densities for each of these systems are also given, where available, under the Voigt profile results.}
	\label{tab:voigt_metals}
	\begin{tabular}{ccccccccc} 
	\hline
		 \hline	
		   z  & b$_{\rm eff}$ (km s$^{-1})$  & log $N_{\rm O\, I}$ & log $N_{\rm C\, II}$  & log $N_{\rm Si\, II}$& log $N_{\rm Fe\, II}$ & log $N_{\rm S\, II}$ \\
		  \hline
		  \hline
		   J0306+1853, z$_{abs}=4.987$ &&&&&\\
                 \hline
                 $4.98241\pm0.00004$ & $5.23\pm7.14$ &       ...                   & $13.40\pm0.18$&      ...                  &...&...\\
                 $4.98436\pm0.00001$ & $4.54\pm2.09$ &  $13.86\pm0.12$ &      ...                  & $13.68\pm0.09$&...&...\\	
	        $4.98661\pm0.00004$ & $8.54\pm1.63$ &  $14.28\pm0.16$ & $14.13\pm0.15$& $13.90\pm0.07$&...&...\\		
	        $4.98808\pm0.00001$ & $4.53\pm7.14$ &        ...                  & $13.88\pm0.28$&      ...                  &...&... \\
	        $4.99010\pm0.00002$ & $7.44\pm3.41$ &  $14.14\pm0.06$ &      ...                  & $13.58\pm0.07$&...&...\\
	       	\hline
		Total log N & ... & $14.60\pm0.08$& $14.37\pm0.13$ & $14.22\pm0.05$ &...&...\\
		AOD log N & ... & $14.48\pm0.03$& $14.12\pm0.04$ & $14.05\pm0.04$ &...&...\\

                  J1233+0622, z$_{abs}=4.859$ &&&&&&\\
                 \hline
	        $4.85300\pm0.00003$ & $10.46\pm2.83$ &  ...            & $13.09\pm0.21$&  $13.61\pm0.21$&    ...                 &...\\		
	        $4.85388\pm0.00002$ & $10.46\pm4.11$ &        ...       & $13.19\pm0.18$&    ...                   &$13.13\pm0.20$&...\\
	        $4.85511\pm0.00003$ & $12.23\pm4.19$ &     ...         & ...                        & $13.48\pm0.23$&     ...                 &...\\		
		$4.85644\pm0.00005$ & $8.25\pm1.13$ &      ...         & ...                        & $13.68\pm0.24$&      ...                & ...\\	
		$4.85691\pm0.00006$ & $8.71\pm2.34$ &    ...           & $13.61\pm0.10$ &    ...                    &$13.44\pm0.16$&...\\
		$4.85795\pm0.00004$ & $18.12\pm4.14$& $ >15.30$& $14.42\pm0.07$ & $13.82\pm0.19$&       ...                &...\\
		$4.85908\pm0.00004$ & $7.53\pm2.04$ &    ...           & $14.01\pm0.14$ &  ...                      &        ...               &...\\
		$4.85996\pm0.00005$ & $8.46\pm3.04$ &    ...           & ...                        & $13.59\pm0.24$&        ...               &...\\
		$4.86019\pm0.00003$ & $8.46\pm2.44$ &    ...           & $13.89\pm0.11$ & ...                       &         ...              &...\\
		$4.86605\pm0.00006$ & $8.25\pm2.04$ &    ...           & $13.63\pm0.11$ & ...                       &         ...              &...\\
		\hline
		Total log N &...  & $>15.30$& $14.74\pm0.05$ & $14.35\pm0.10$&$13.61\pm0.13$ &...\\
                 AOD log N &...  & ... & $14.65\pm0.02$ & $14.44\pm1.16$&$13.68\pm0.12$ &...\\

		 J1233+0622, z$_{abs}=5.050$ &&&&&\\
                 \hline
                 $5.04754\pm0.00015$ & $10.11\pm1.12$ &  $13.88\pm0.18$ & $13.89\pm0.19$ &... &...& ...  \\
                 $5.04912\pm0.00002$ & $24.66\pm2.47$ &  $13.79\pm0.28$ & $14.05\pm0.14$ &...  &...&...\\
                 $5.05024\pm0.00001$ & $30.17\pm3.02$ &  $15.11\pm0.12$ & $14.75\pm0.16$& $14.19\pm0.11$& $14.07\pm0.09$&...\\		
	        	$5.05188\pm0.00004$ & $10.71\pm2.09$ &  $13.88\pm0.19$ & $13.84\pm0.19$ & $13.69\pm0.17$&...&...\\		
		$5.05290\pm0.00015$ & $18.58\pm1.89$ &  $13.76\pm0.20$ & $13.49\pm0.23$ &... &... & ...\\	
		\hline
		Total log N &...  & $15.19\pm0.10$& $14.93\pm0.11$ & $14.31\pm0.09$ & $14.07\pm0.09$&...\\	
		AOD log N &...  & $15.11\pm0.41$& $14.80\pm0.05$ & $14.27\pm0.05$ & $14.14\pm0.05$&...\\

		J1253+1046, z$_{abs}=4.589$ &&&&&\\
                 \hline
                 $4.58859\pm0.00001$ & $13.94\pm3.69$ &  $13.84\pm0.08$ &... & ...&$13.98\pm0.14$ &...\\	
                 $4.58916\pm0.00005$ & $11.44\pm7.14$ &... & ...&... &  $13.89\pm0.17$&... \\	
	         $4.58948\pm0.00004$ & $11.44\pm4.73$ &  $14.94\pm0.10$ & $14.34\pm0.22$& $13.67\pm0.09$&...&...\\		
	         $4.59014\pm0.00003$ & $8.38\pm2.36$ &  $13.86\pm0.07$ &...   &... & ...&...\\
		 \hline
		Total log N & ... & $15.00\pm0.08$& $14.34\pm0.22$ & $13.67\pm0.09$ & $14.24\pm0.11$&...\\
		 AOD log N & ... & $14.78\pm0.27$& $14.29\pm0.38$ & $13.88\pm0.04$ & $14.28\pm0.06$&...\\
		
		  J1253+1046, z$_{abs}=4.600$ &&&&&\\
                 \hline
                 $4.59845\pm0.00002$ & $32.33\pm3.41$ &  $14.32\pm0.12$ &...   &...  &...&...\\
                 $4.59940\pm0.00005$ & $23.24\pm7.14$ &... & ...&... &  $14.06\pm0.29$ &...\\	
	         $4.60003\pm0.00003$ & $30.37\pm1.63$ &  $15.53\pm0.08$ & $14.78\pm0.20$& $14.57\pm0.05$&...&...\\		
	         $4.60194\pm0.00003$ & $19.61\pm2.09$ &  $14.32\pm0.09$ &... & ...&...&...\\		
		 \hline
		Total log N & ... & $>15.58$& $>14.78$ & $14.57\pm0.05$ & $14.06\pm0.29$&...\\
                AOD log N & ... & $15.31\pm0.06$& $14.78\pm0.45$ & $14.56\pm0.01$ & $14.26\pm0.05$&...\\

                  J1253+1046, z$_{abs}=4.793$ &&&&&\\
		\hline
		
	        $4.79347\pm0.00001$ & $10.00\pm1.24$ &  $>14.71$ & $14.21\pm0.17$ & $14.15\pm0.18$ &...& $14.18\pm0.17$ \\	
	         AOD log N & ... &  $14.56\pm0.12$ & $14.22\pm0.07$ & $14.08\pm0.07$ &...& $14.25\pm0.04$ \\

	        J1557+1018, z$_{abs}=4.627$ &&&&&\\
	           \hline
	        $4.62512\pm0.00004$ & $10.76\pm1.17$ &  $14.54\pm0.11$ & $14.38\pm0.15$& $13.65\pm0.06$&...&...\\		
	        $4.62694\pm0.00002$ & $35.53\pm3.75$ &  $15.64\pm0.09$ & $15.52\pm0.11$& $15.00\pm0.04$&...&...\\
	        $4.62885\pm0.00003$ & $38.41\pm4.84$ &  $15.55\pm0.19$ & $15.10\pm0.21$& $14.74\pm0.04$&...&...\\		
		$4.63018\pm0.00015$ & $14.29\pm3.43$ &  $15.01\pm0.19$ & $14.53\pm0.19$& $14.35\pm0.09$&...&...\\	
		                     
		\hline
		Total log N  & ... & $>15.97$& $>15.71$ & $>15.26$ &...&...\\	
		 AOD log N & ... & $15.64\pm0.62$& $15.23\pm0.11$ & $15.26\pm0.84$ &...&...\\
	       	\hline
		\hline
		
		\end{tabular}
\end{table*}

\begin{table*}
	\centering
	\caption{Relative abundances and velocity dispersions}
	\label{tab:rel_abn}
	\begin{tabular}{cccccccc} 
		\hline
		\parbox[t]{0.50in}{Relative \par abundances \strut} & \parbox[t]{0.64in}{J0306+1853 \par z$_{abs}=4.987$ \strut} & \parbox[t]{0.64in}{J1233+0622 \par z$_{abs}=4.859$ \strut} & \parbox[t]{0.64in}{J1233+0622 \par z$_{abs}=5.050$ \strut} &	\parbox[t]{0.64in}{J1253+1046 \par z$_{abs}=4.589$ \strut} & 	\parbox[t]{0.64in}{J1253+1046 \par z$_{abs}=4.600$ \strut} &  \parbox[t]{0.64in}{J1253+1046 \par z$_{abs}=4.793$ \strut} & \parbox[t]{0.64in}{J1557+1018 \par z$_{abs}=4.627$ \strut}  \\
		\hline
		$\rm [O/H]$ & $-2.69\pm0.17 $ &      $>-2.14$                  &$-1.60\pm0.18$& $-1.43\pm0.17$  & $    >-1.46       $ & $>-1.63          $  & $>-1.47$          \\
		$\rm [C/H]$ &$ -2.66\pm0.19 $ &$-2.44\pm0.15$ &$-1.60\pm0.19$& $-1.84\pm0.27$  & $    >-2.0       $ & $-1.87\pm0.20$  & $>-1.47$           \\
		$\rm [Si/H]$ &$ -1.89\pm0.16$ &$-1.91\pm0.18$ &$-1.30\pm0.18$& $-1.59\pm0.17$  & $-1.29\pm0.16$ & $-1.01\pm0.21$  & $>-1.00$           \\
		$\rm [S/H]$ &      ...                   &         ...                          &            ...          &       ...                   &           ...             & $-0.59\pm0.20$  &      ...                 \\
		$\rm [Fe/H]$&      ...                  & $-2.64\pm0.20$&$-1.53\pm0.17$& $-1.01\pm0.18$  &  $-1.79\pm0.33$ &    ...                     &     ...                 \\
		$\rm [C/O]$ & $0.03\pm0.15$  &      $<-0.30$                 & $0.00\pm0.15$ & $ -0.41\pm0.24$  &           ...             & $<-0.24$           &     ...                  \\
		$\rm [Si/O]$& $0.79\pm0.09 $ &      $<0.23$                  & $0.30\pm0.14$ & $ -0.16\pm0.12$  &  $<0.17$             & $<0.62$             &     ...                   \\
		$\rm [Fe/O]$&     ...                  &      $<-0.50$               & $0.07\pm0.14$ &  $0.42\pm0.14 $  &   $<-0.33$           &       ...                   &        ...              \\
		$\rm [Si/C]$& $0.76\pm0.13$  &      ...                          &   ...                    &        ...                  &            ...               &         ...                &         ...             \\
		\hline
		\parbox[t]{0.50in}{Velocity \par dispersion \strut}	 &   328.4 km s$^{-1}$	&  $<$322.3 km s$^{-1}$		&  114.0 km s$^{-1}$ &  25.4 km s$^{-1}$ & 88.3 km s$^{-1}$ & 21.0 km s$^{-1}$	& $<$209.3 km s$^{-1}$\\
		\hline		
\end{tabular}
\end{table*}

\begin{figure*}
\centering
  \begin{tabular}{@{}cc@{}}
    \includegraphics[width=.35\textwidth]{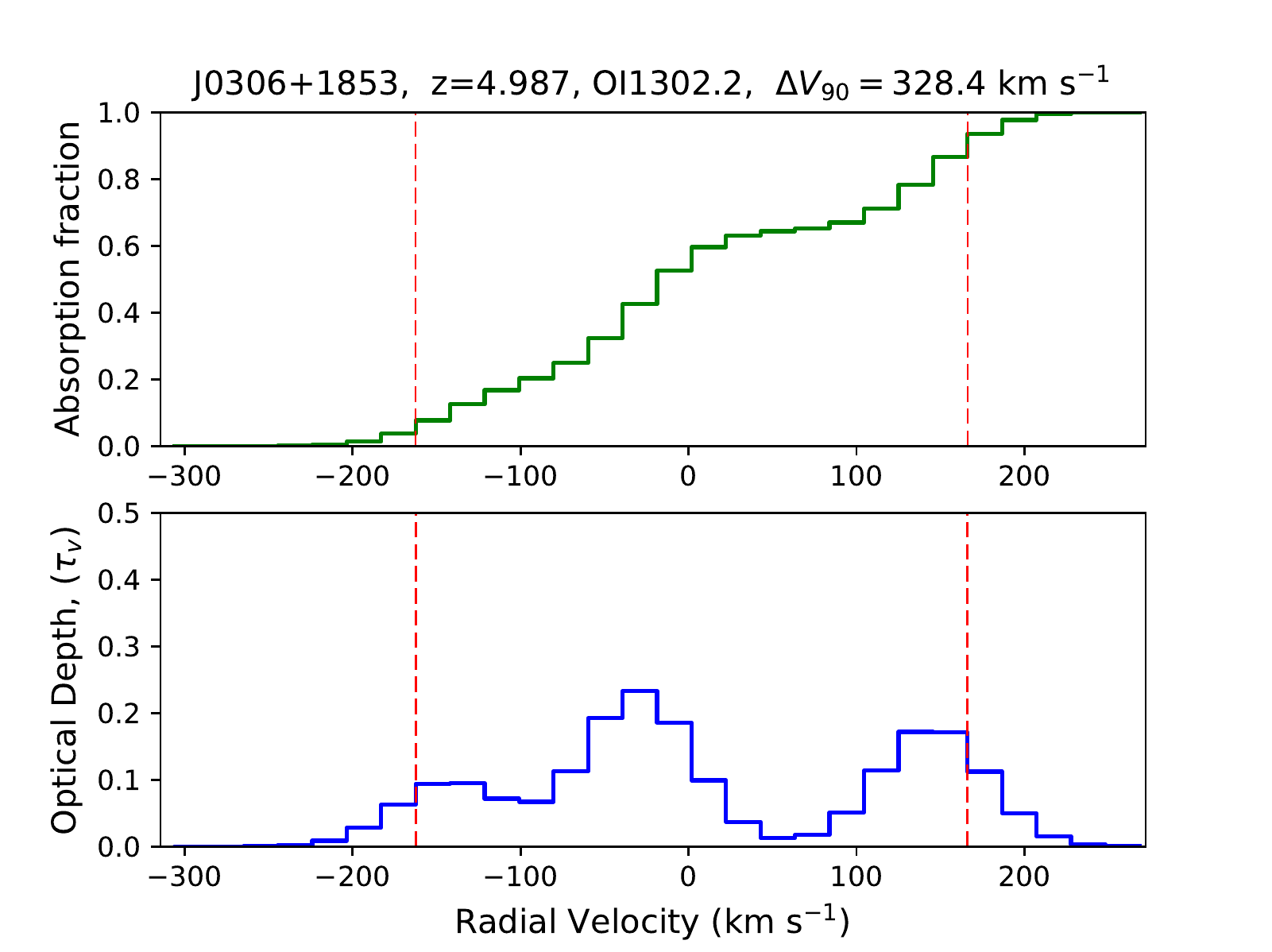} &
    \includegraphics[width=.35\textwidth]{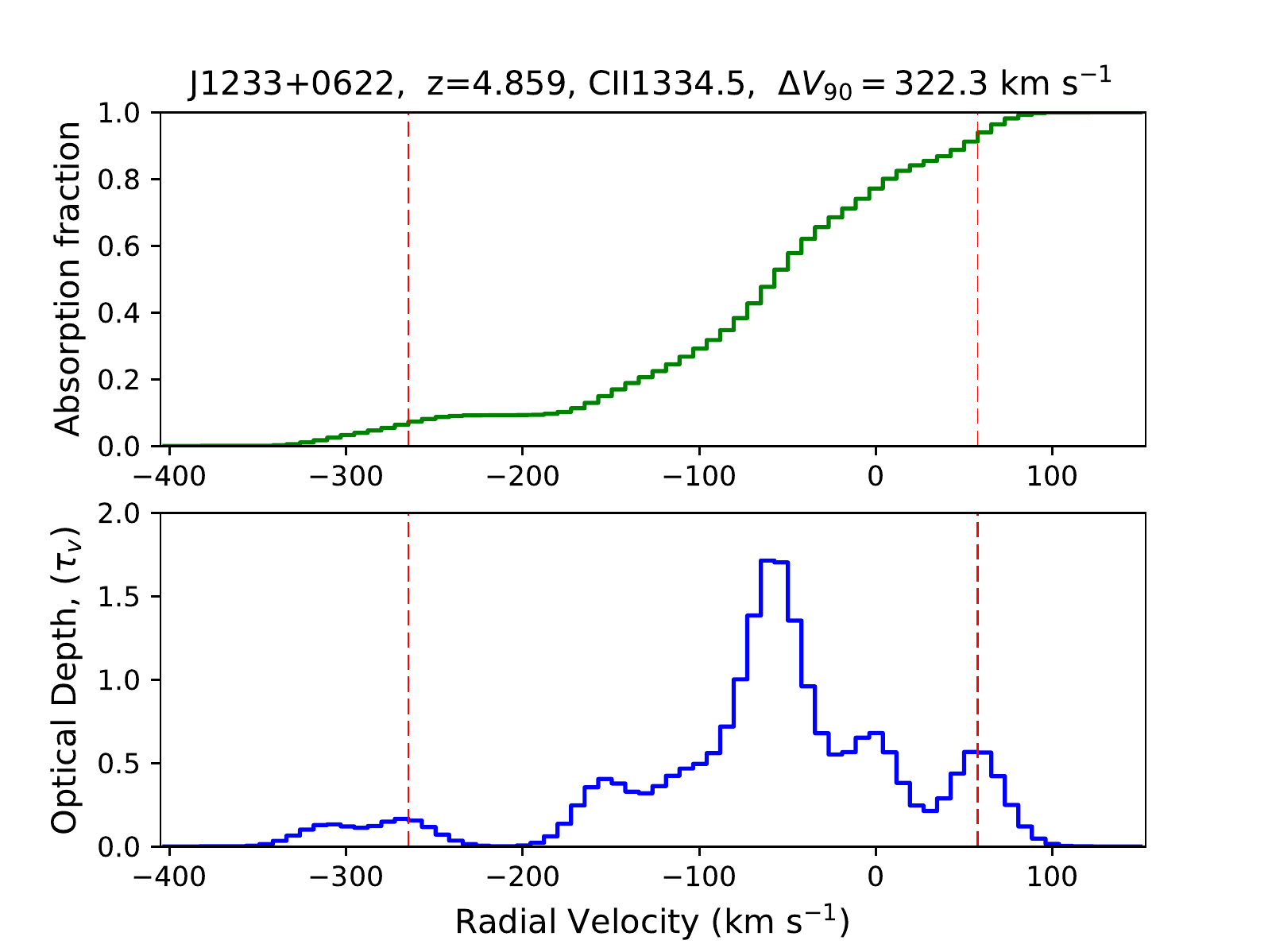} \\  
      \includegraphics[width=.35\textwidth]{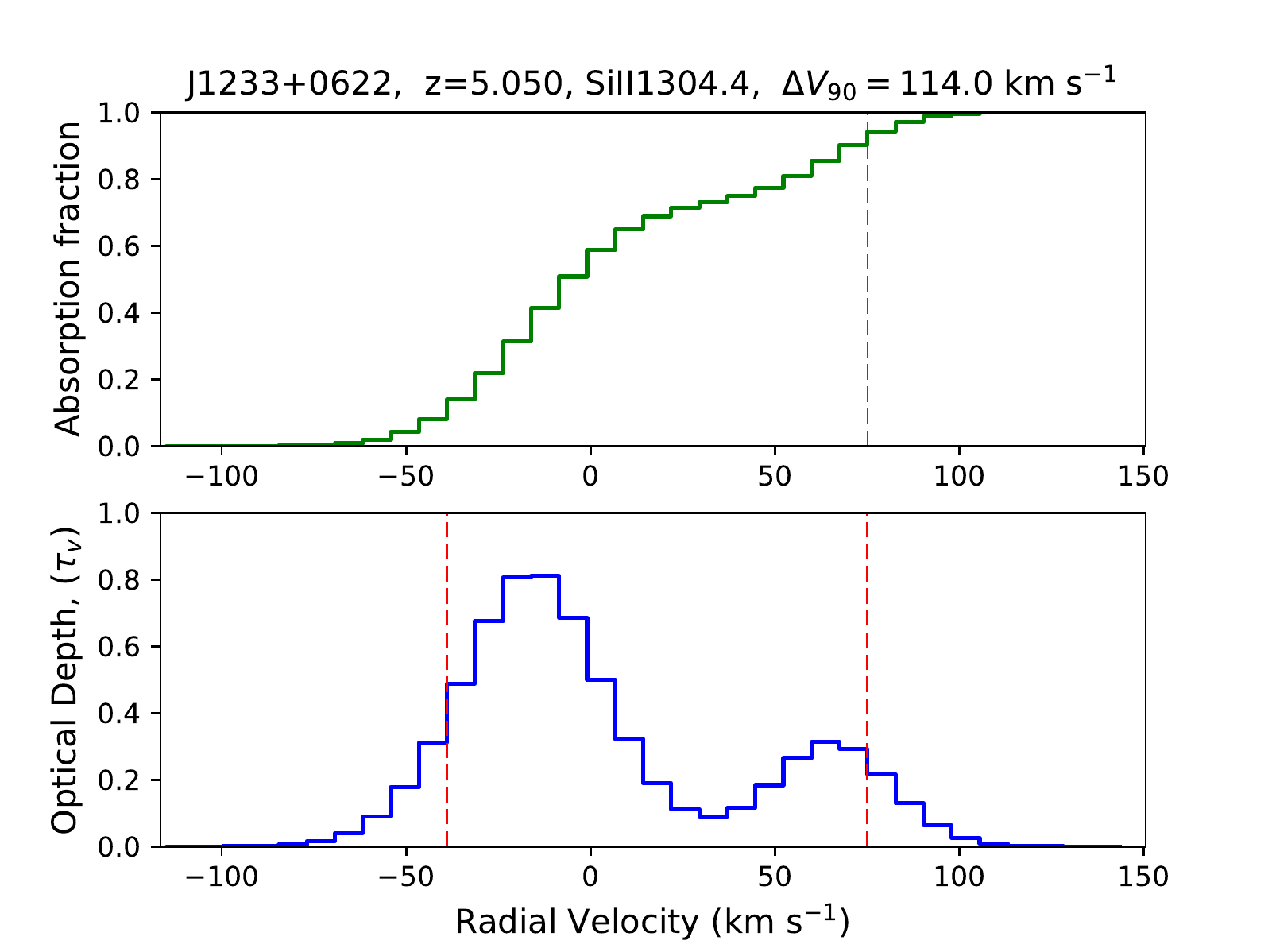} &  
    \includegraphics[width=.35\textwidth]{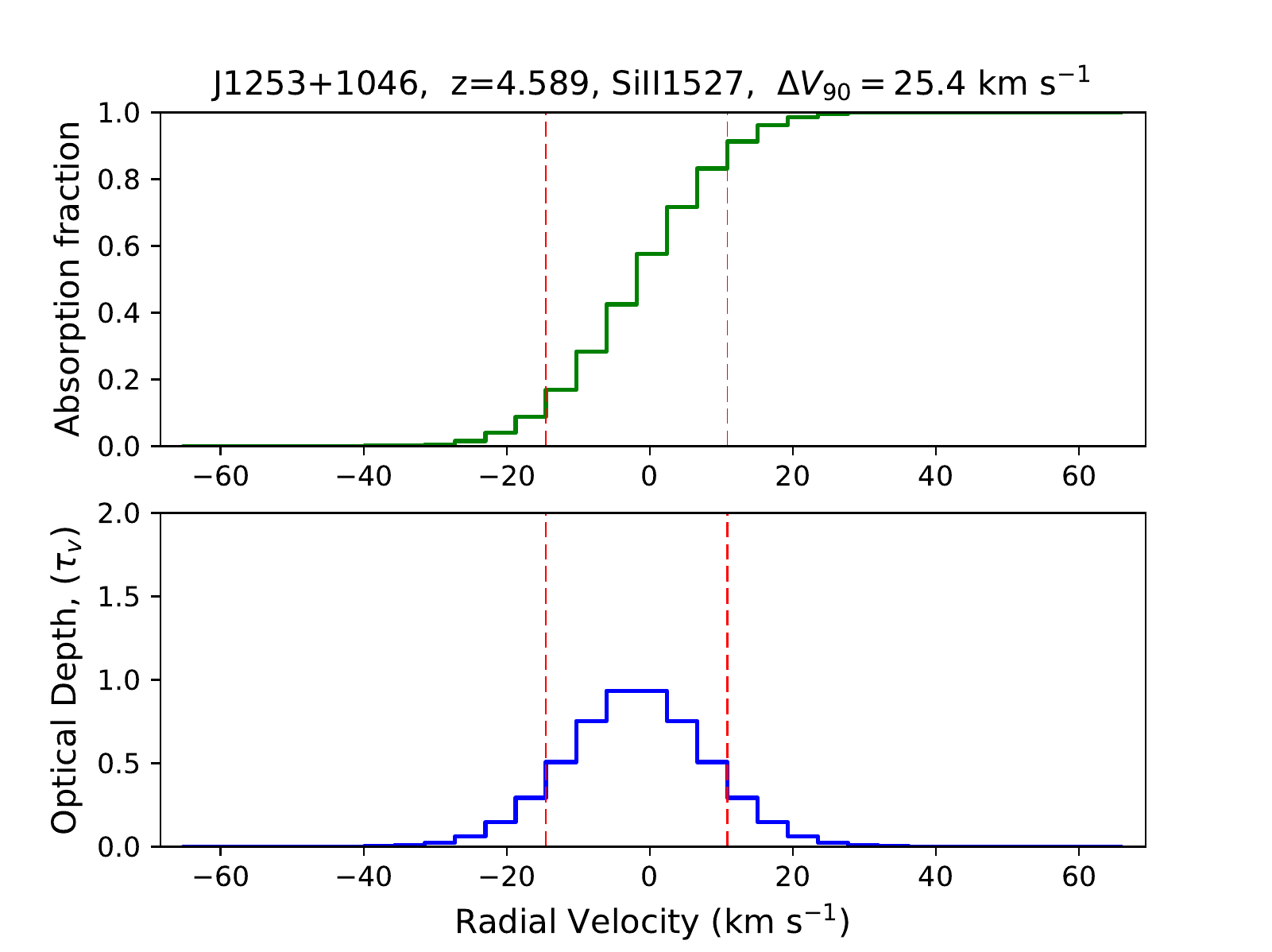} \\
    \includegraphics[width=.35\textwidth]{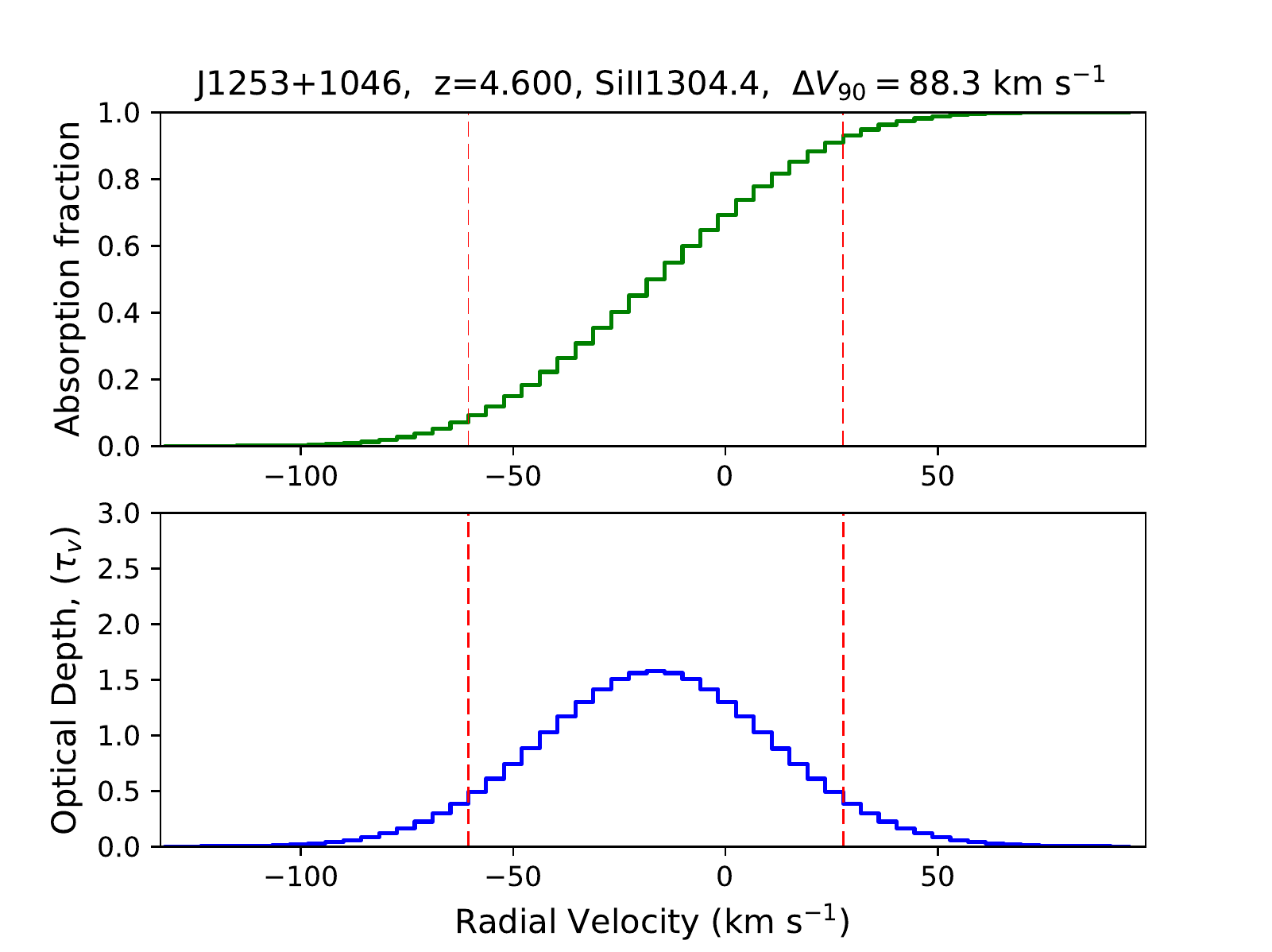} &
    \includegraphics[width=.35\textwidth]{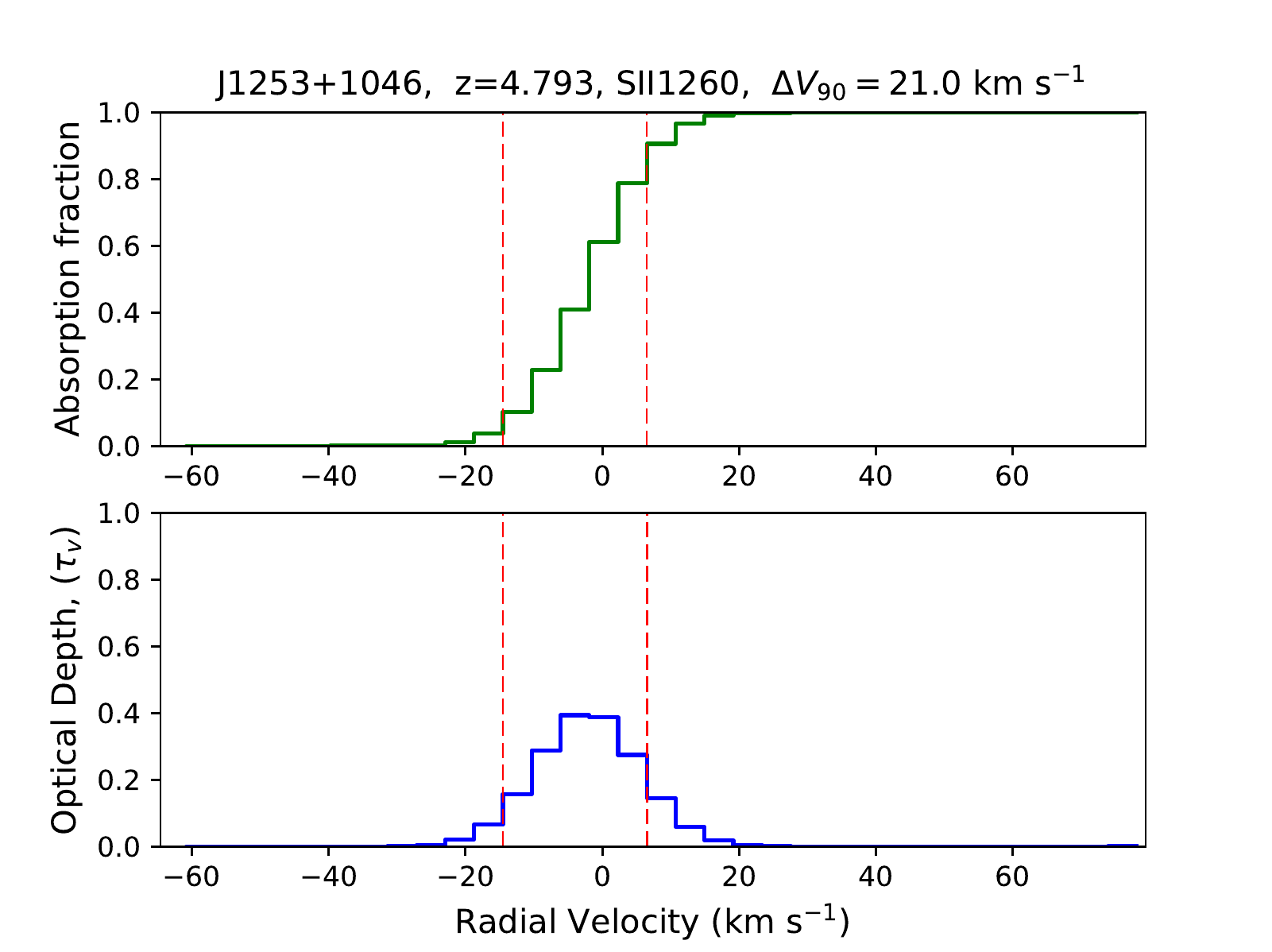}\\
     \multicolumn{2}{c}{\includegraphics[width=.35\textwidth]{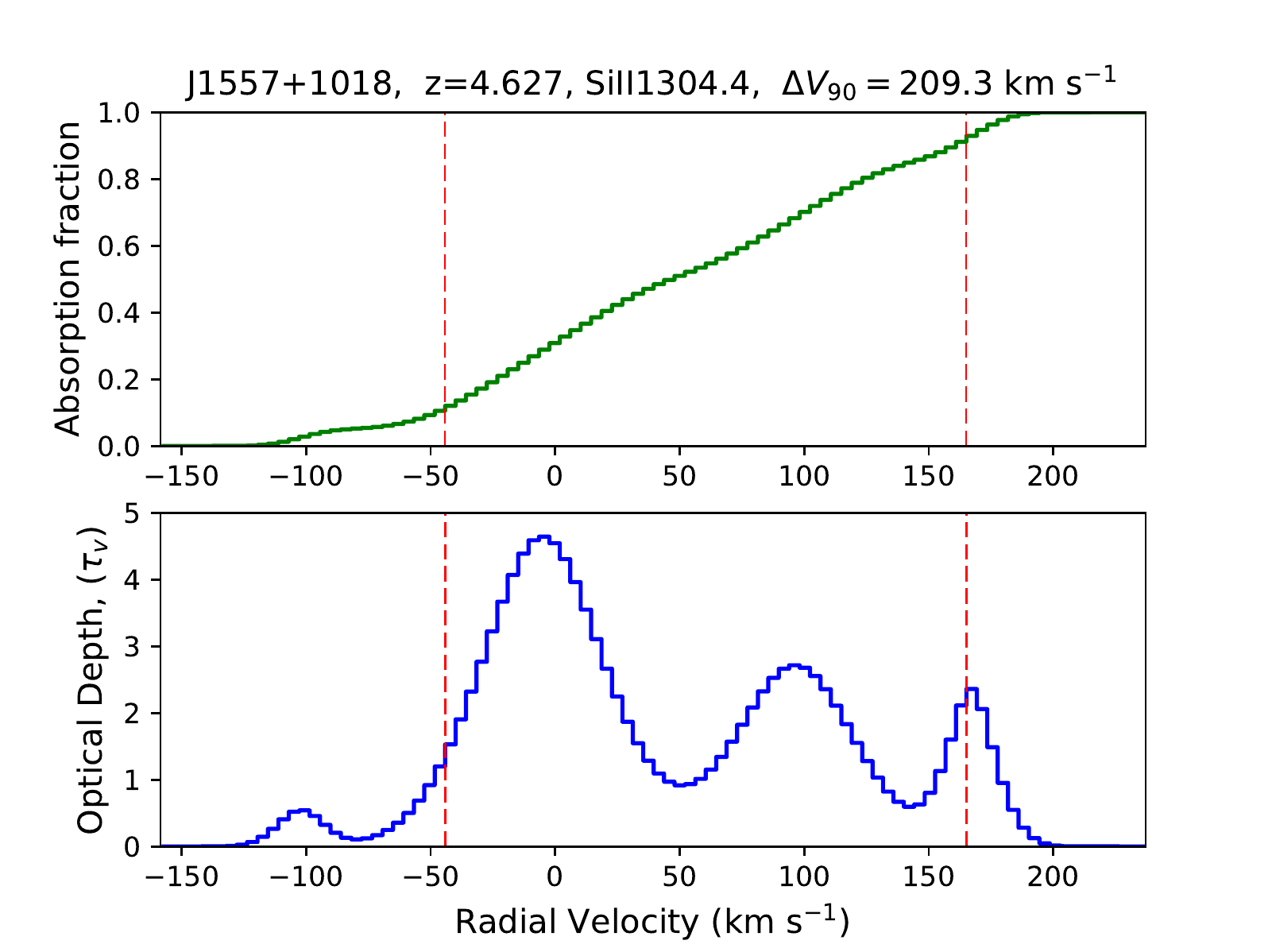}}
  \end{tabular}
  \caption{Plots showing absorption fraction vs. radial velocity (upper panel) and optical depth vs. radial velocity (lower panel) for metal lines from this work. The names of the sight lines, redshifts of the absorbers, names of the metal lines used, and the estimated values of the velocity dispersions are given above the top panel in each figure. The velocity dispersion is the length of the radial velocity within which 90 percent of the cumulative optical depth is contained. The left and right vertical dashed lines in red are drawn corresponding to 5 percent and 95 percent of cumulative optical depth.}
  \label{fig:dispersion}
\end{figure*}

\section{Discussion}
\label{sec:discussion}
Now we combine our measurements with those from the literature and perform a comparative study of metallicity-velocity dispersion, relative abundances, dust depletion, and metallicity evolution.

\subsection{Profile Shapes and the Metallicity vs. Velocity Dispersion Relation}
\begin{figure}
\centering
  \begin{tabular}{@{}c@{}}
\includegraphics[width=\columnwidth]{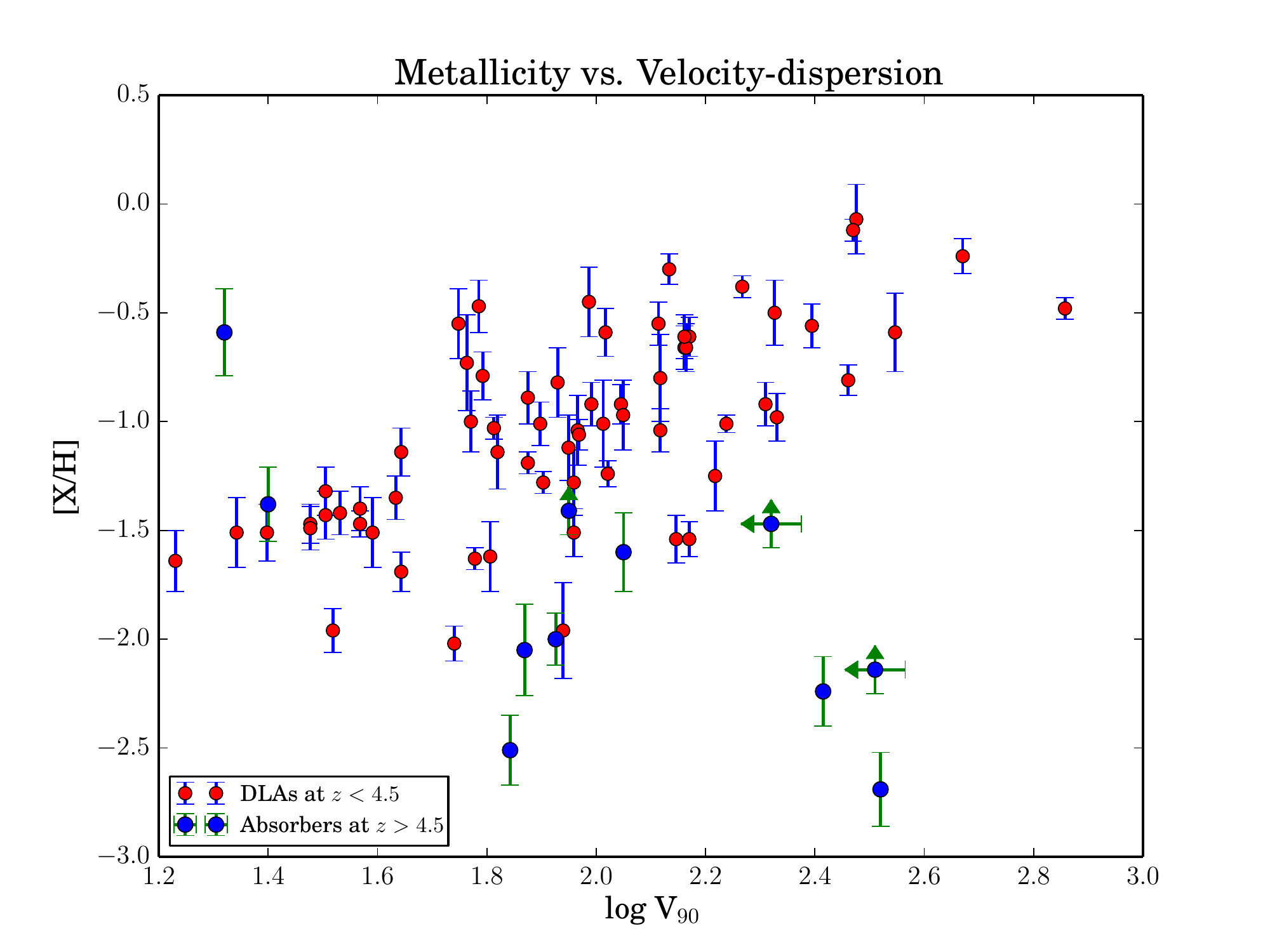}
\end{tabular}
\caption{ Plot showing metallicity vs. velocity-dispersion relation for volatile elements in DLA. Where X=O or S for $z\sim5$ absorbers, and X=O, S or Zn for lower redshift DLAs. The red dots with blue error bars show the measurements for lower redshift DLAs from the literature. The blue dots with green error bars show the measurements for $z>4.5$ absorbers taken from \citet{Poudel et al. 2018}, \citet{Morrison et al. 2016}, and this work. Lower and upper limits are represented by unidirectional arrows.}
    \label{fig:vel_dis}
\end{figure}

The gas kinematics of a galaxy can be understood from the measurement of velocity dispersion of its gas. One measure of the gas velocity dispersion is the quantity $\Delta v_{90}$, i.e., the velocity range that spans 90 percent of the total absorption determined from relatively weak and unsaturated metal lines whenever possible \citep*[e.g.][]{Wolfe et al. 1998}. Fig. \ref{fig:dispersion} shows the measurements of velocity dispersion for all the absorbers in this paper using weaker and unsaturated transitions where available. We treat the $\Delta v_{90}$ derived from the strong, saturated lines as upper limits. To estimate $\Delta v_{90}$, we measure the radial velocity width between 5 percent and 95 percent of the integrated optical depth using the fitted profiles as shown in Fig. \ref{fig:dispersion}. The value of the velocity dispersion may depend on the inflow or outflow of gases within the galaxy and also on the mass of the galaxy. If the velocity dispersion depends primarily on mass, and the role played by gas flows is small, then the velocity dispersion vs. metallicity relation can suggest a relationship between mass and metallicity (MZR) of absorber host galaxies. In lower redshift DLAs and sub-DLAs, a fairly tight correlation exists between the metallicity and the velocity dispersion of the gas \citep*[e.g.][]{Peroux et al. 2003, Ledoux et al. 2006, Moller et al. 2013, Som et al. 2015}. However, this relation is still unknown for high-$z$ absorbers. Based on a small sample, \citet{Poudel et al. 2018} reported a somewhat flatter relation at $z\sim 5$. Fig. \ref{fig:vel_dis} shows the metallicity vs. velocity dispersion relation for the absorbers  presented in this paper, along with those from \citet{Poudel et al. 2018} and other sources from the literature. This larger sample also shows a much flatter metallicity vs. velocity dispersion relation at $z > 4.5$ than for the $z < 4.5$ absorbers. Since we don't find the well established metallicity vs. velocity dispersion trend seen at lower redshifts, it is possible that the strength of the correlation decreases with increasing redshift.\\

A flatter metallicity vs. velocity dispersion relation could be caused by stronger inflows of metal-poor gas at $z > 4.5$. Alternatively, if the velocity dispersion is dependent of the galaxy mass, then we speculate that the flatter trend for $z > 4.5$ absorbers may suggest that these absorbers are more dominated by dark matter and thus have lower stellar mass and lower metallicity.\\

Finally, we note that the velocity profile shapes show some potentially interesting differences. For example, for 3 of the 4 absorbers where we have measurements of Fe II profiles, we see a blueshift in the Fe II profiles with respect to the O I, C II, and Si II profiles. These 3 absorbers are at $z=4.59$, 4.60, and 4.86. Such a shift, if real, would be unusual, compared to lower-redshift DLAs, even the VMP DLAs in Cooke et al. (2017). However, we note that not all of our high-$z$ absorbers show a blueshift in the Fe II profile. For example, such a shift is not seen in the $z=5.05$  absorber toward J1233+0622, or in either of the 2 absorbers in Poudel et al. (2018) at $z=4.81$ or $z=4.83$ where Fe II was observed. While for the $z=4.86$ absorber toward J1233+0622, the shift in the Fe II absorption could arise partly due to a slight offset in the wavelength calibrations of the visible and near-IR parts of the X-Shooter spectra, we do not find any offsets in the wavelength calibration of the MIKE data for the absorbers at $z=$4.59, 4.60 toward J1253+1046. For the absorbers presented here that do show a shift in Fe II profiles, the shift could be indicative of moderate-velocity outflows enriched in Fe II by type Ia supernovae. Alternately, the Fe II profiles may seem shifted if the central components arise in systematically more dusty regions leading to stronger depletion of Fe II in those components. However, the possibility of systematic dust depletion occurring in each case is small. Observations of more high-$z$ DLA/sub-DLAs are essential to assess how common such Fe~ II blueshifts are.

\subsection{Relative Abundances and Nucleosynthesis}
\begin{figure}
\centering
  \begin{tabular}{@{}c@{}}
  \includegraphics[width=\columnwidth]{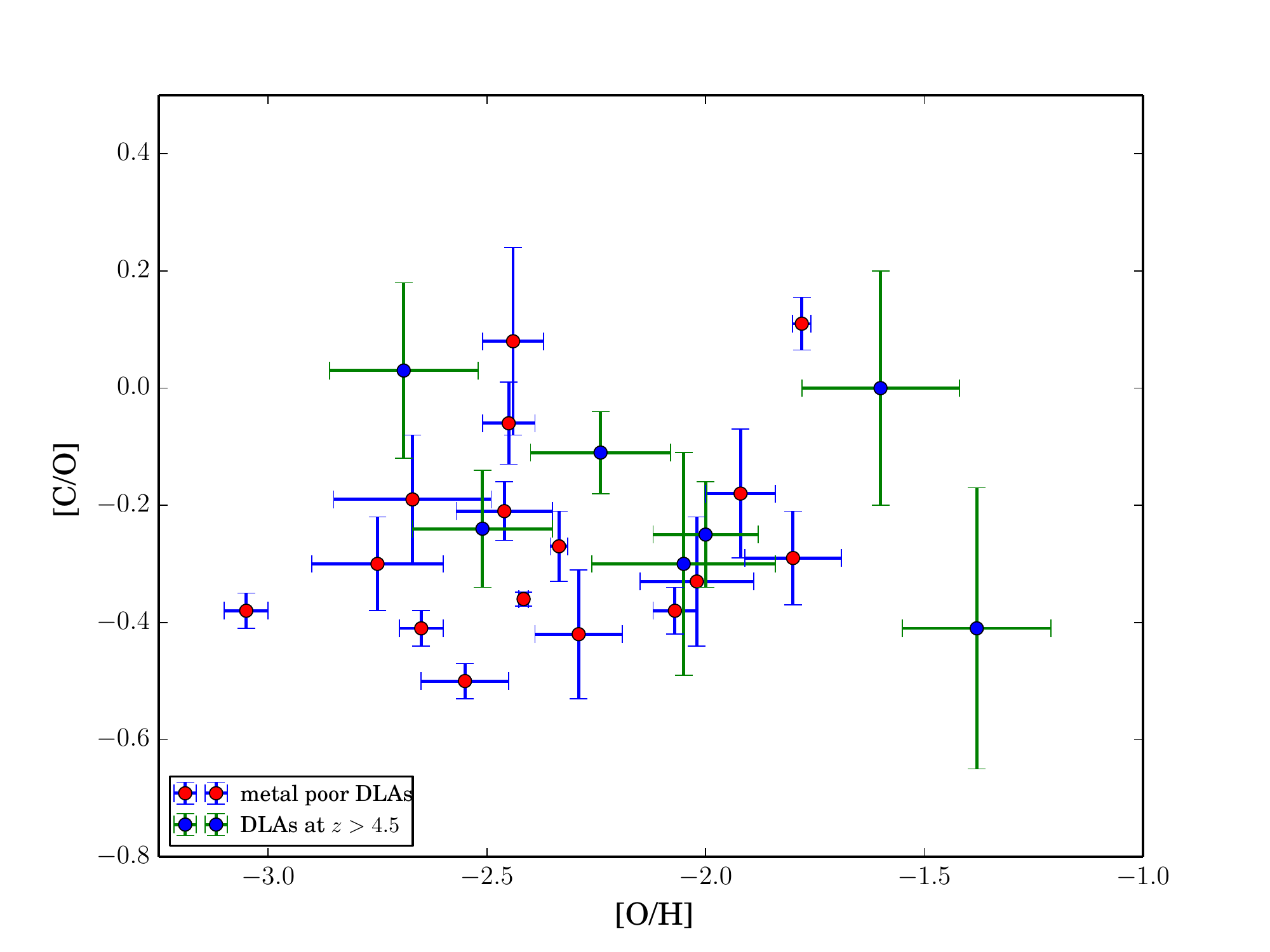}
\end{tabular}
\caption{[C/O] vs. [O/H] for DLAs. The red dots with blue error bars show the measurements for metal-poor DLAs taken from \citet{Cooke et al. 2017}. The blue dots with green error bars are the measurements for $z>4.5$ absorbers which contain the measurements from \citet{Rafelski et al. 2012, Rafelski et al. 2014, Morrison et al. 2016, Poudel et al. 2018}, and this work.} 
    \label{fig:c2o}
\end{figure}

Fig. \ref{fig:c2o} shows a plot of [C/O] vs. [O/H]  for both the DLAs at z$\geqslant$4.5 and the metal-poor DLAs at lower redshifts from the literature. Although the $z>4.5$ DLAs are not always metal-poor, their average [C/O] value of $-0.19\pm0.12$ is consistent with that for the very metal-poor (VMP) DLAs, which have an average [C/O] of $-0.28\pm0.12$ \citep*[e.g.][]{Cooke et al. 2011, Cooke et al. 2017}. \citet{Cooke et al. 2011, Cooke et al. 2017} have suggested that metal-poor DLAs may have been enriched by metal-free progenitor stars. The similar [C/O] ratios between metal-poor DLAs and the $z\sim5$ DLAs suggest that DLAs at $z\sim5$ may also have been enriched by early stars, including perhaps a leftover signature of Population III stars. Thus, it seems reasonable to compare the chemistry of metal-poor DLAs with the $z\sim5$ DLAs to constrain the nature of the progenitor stars, especially their Initial Mass Function (IMF).\\

\begin{figure}
\centering
  \begin{tabular}{@{}c@{}}
  \includegraphics[height=10cm, width=9cm]{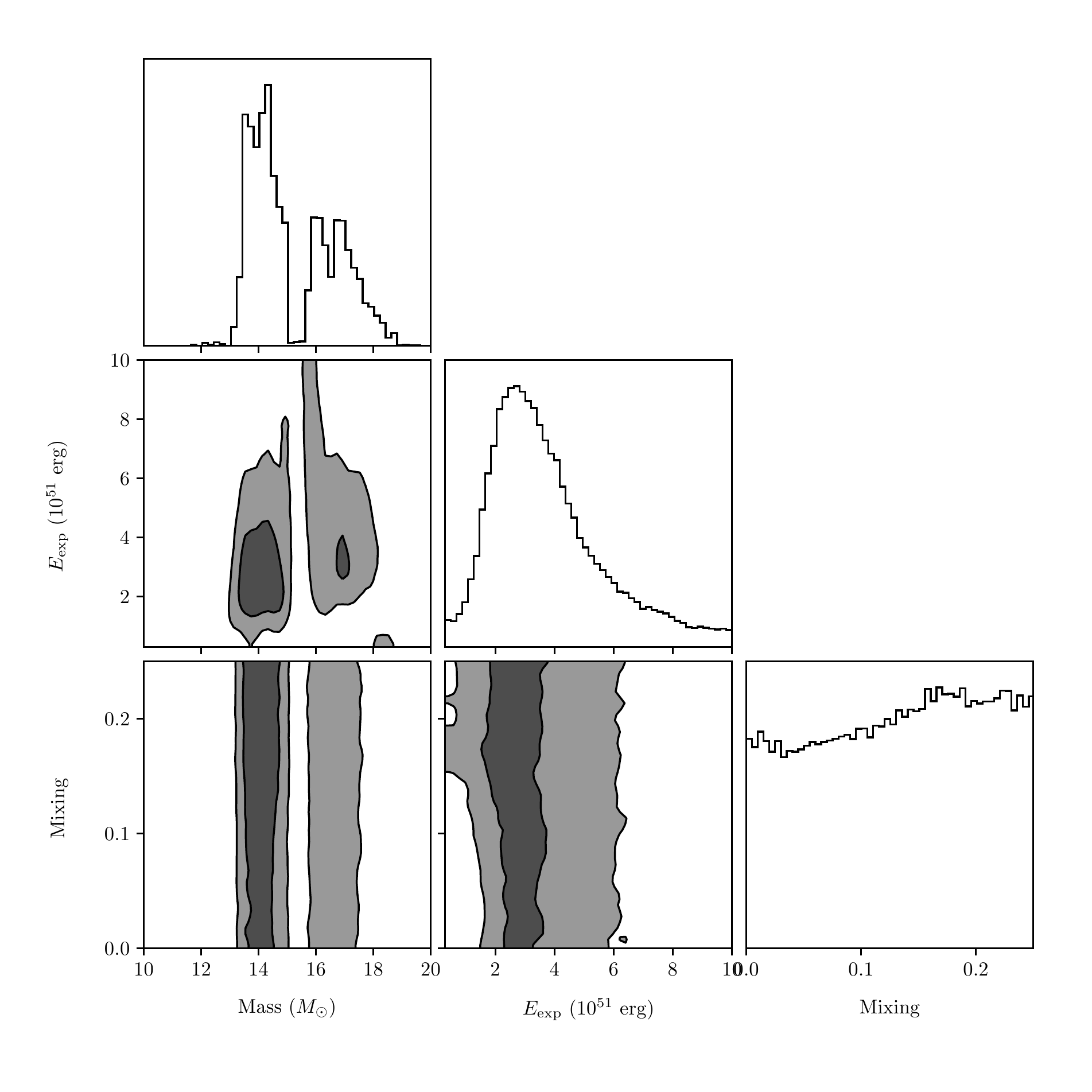}
  \end{tabular}
  \caption{The one- and two-dimensional projections of the posterior probability distributions of progenitor mass, explosion energy, and the stellar mixing parameter of the star that might have enriched the DLA at $z=5.335$ in the sight line to J0231-0728. We combined observed [C/O] and [Si/O] with the \citet{Heger 2010} nucleosynthesis calculations, which comprised 16800 combinations of these three parameters. The marginalised probability of the model parameters on the x-axis are shown in the diagonal panel. The non-diagonal panels with dark and light shades are the two-dimensional projections of the parameters representing the 68 and 95 per cent confidence contours, respectively.}
  \label{fig:heger1}
\end{figure}

\begin{figure}
\centering
  \begin{tabular}{@{}c@{}}
  \includegraphics[height=10cm, width=9cm]{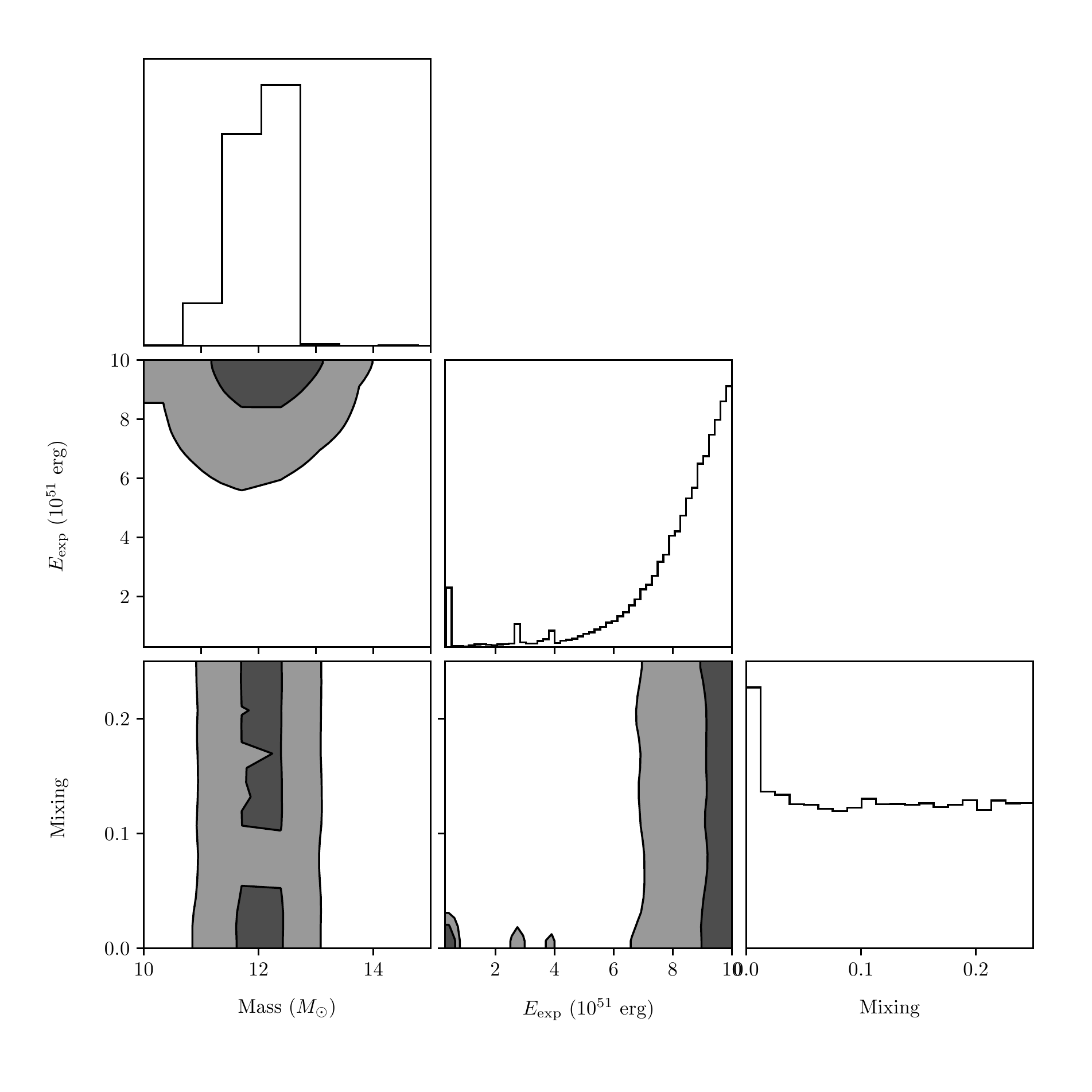}
  \end{tabular}
  \caption{Same as Fig. \ref{fig:heger1} but for the DLA at z=4.987 in the sight line to J0306+1853.}
  \label{fig:heger2}
\end{figure}

\begin{figure}
\centering
  \begin{tabular}{@{}c@{}}
  \includegraphics[height=10cm, width=9cm]{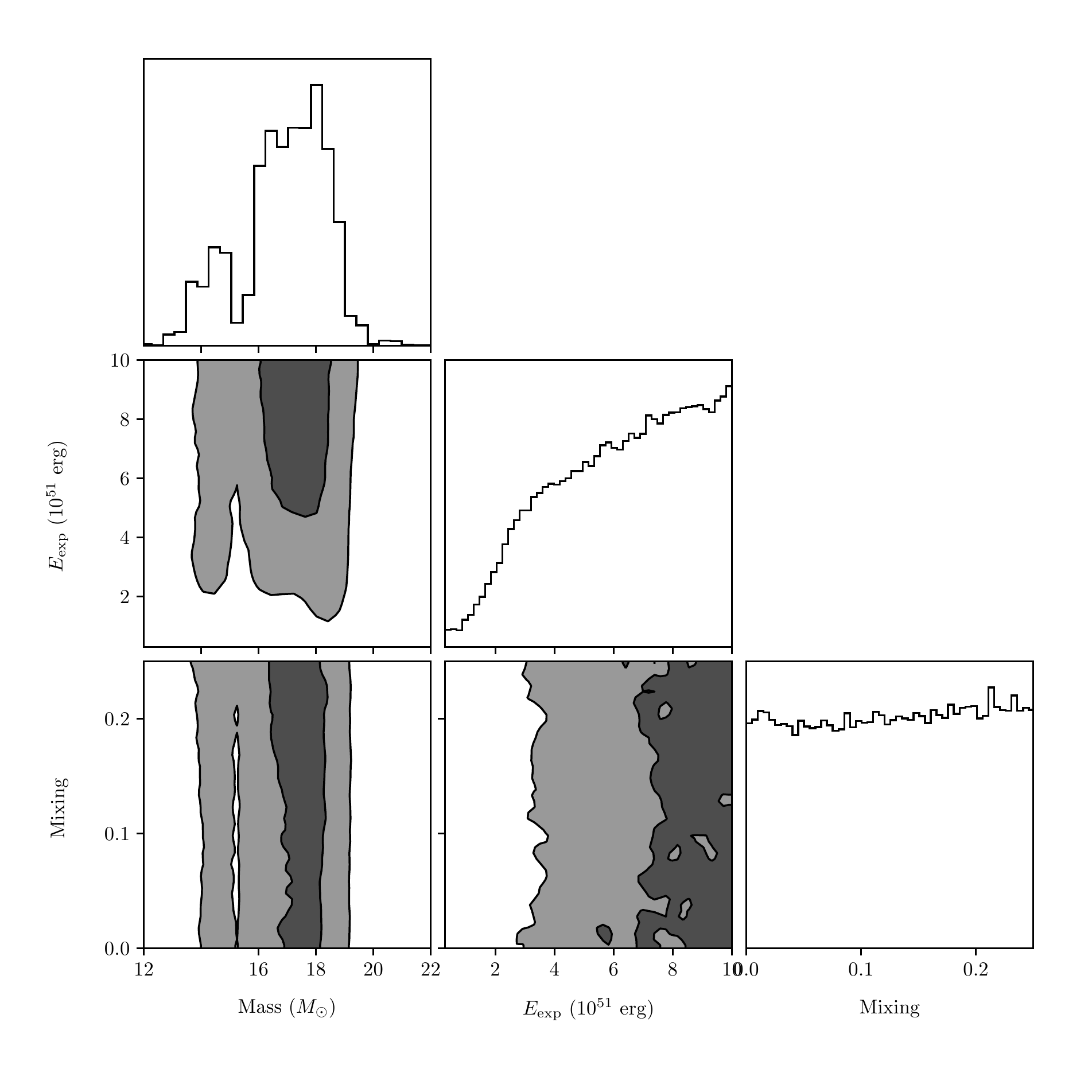}
  \end{tabular}
  \caption{Same as Fig. \ref{fig:heger1} but for the DLA at z=4.809 in the sight line to J0824+1302.}
  \label{fig:heger3}
\end{figure}

To this end, we have selected three $z\sim5$ DLAs which have [O/H]$ < -2.20$ (1 from this work and 2 from \citet{Poudel et al. 2018}) to compare with the nucleosynthesis models of massive metal-free stars from \citet{Heger 2010}.  We compared [C/O] in these 3 DLAs with those predicted by \citet{Heger 2010}, because [C/O] is believed to put a strong constraint on the progenitor mass of Population III stars \citep*[e.g.][]{Woosley 1995, Heger 2010, Cooke et al. 2011, Cooke et al. 2017}. The models of \citet{Heger 2010} include 120 simulated stars spanning progenitor star masses in the range of 10-100 M$_{\odot}$, with a mass resolution of $\rm \delta M > 0.1 M_{\odot}$. To consider the mixing between different stellar layers, these models use grids with 14 different mixing widths. Furthermore, the explosion energy of the supernova ranges from 0.3 to 10 $\times 10^{51}$ erg. The position of the piston just before the explosion is just below the oxygen burning shell where the entropy per baryon is given by $\rm S/k_{B} = 4$, where k$_{B}$ is the Boltzmann constant. We compared the [C/O] and [Si/O] ratios predicted by these various models with the ratios observed for our DLAs of interest, in order to estimate the range of progenitor mass, explosion energy, and mixing width that may have enriched these DLAs. To accomplish this, we linearly interpolated this three-dimensional space and ran a Markov Chain Monte Carlo analysis using the techniques described in \citet{Cooke et al. 2017}. The one- and two-dimensional projections of the samples are plotted in Fig. \ref{fig:heger1}, \ref{fig:heger2} and \ref{fig:heger3} for the DLAs towards J0231-0728, J0306+1853 and J0824+1302 respectively. We were able to put a strong constraint on the progenitor masses for all of these DLAs. The probability distributions of the progenitor masses were centered around 14.6, 12.2, and 17.1 M$_{\odot}$ for the $z=5.335$, 4.987, and 4.809 DLAs towards J0231-0728, J0306+1853, and J0824+1302 respectively. The DLAs in the sight lines to J0231-0728 and J0824+1302 were taken from \citet{Poudel et al. 2018}. The explosion energies were more or less constrained towards the upper end of the given range (E$_{\rm exp} > 5 \times 10^{51}$ erg) for the DLAs towards J0306+1853 and J0824+1302 and 2-4 $\times 10^{51}$ erg for the DLA towards J0231-0728. However, the mixing parameters could not be constrained well for any of our DLAs. Our results are roughly similar to those found by \citet{Cooke et al. 2017}, who obtained a progenitor mass of 20.5 M$_{\odot}$ and explosion energy of 6-8 $\times 10^{51}$ erg for a VMP DLA at $z \approx 3.1$ with [O/H] of -3.05.\\ 

\subsection{Dust Depletion}
To estimate the extent of depletion of elements on dust grains, \citet{Jenkins 2009} proposed a method using four parameters $F_{*}$, $A_{x}$, $B_{x}$, and $Z_{x}$. One of these parameters, $F_{*}$, manifests the strength of dust depletion for the given sight line. Higher value of $F_{*}$ signifies higher depletion level. The other three parameters derived empirically by \citet{Jenkins 2009} are unique constants for each element X. They are related to the gas-phase abundance of element X and  $F_{*}$ by the relation [X$_{gas}$/H] = $B_{x} + A_{x} (F_{*} - Z_{x})$. $F_{*}$, which represents the depletion strength for the given sight line, lies in range 0-1 for neutral gas in the Milky Way but is lower for warm-ionized gas \citep*[e.g. $F_{*} = -0.1$,][]{Draine 2011}. The values for $A_{x}$, $B_{x}$, and $Z_{x}$ are presented in \citet{Jenkins 2009} and \citet{Jenkins 2017}. \citet{Jenkins 2017} suggested that the values in the Small Magellanic Cloud (SMC) would be appropriate for studies of depletion in DLA and sub-DLA absorbers. In a plot of [X/H]$_{\rm obs}$ - $B_{x}$ + $A_{x} Z_{x}$ vs. $A_{x}$, slope represents the $F_{*}$ value and the intercept represents the depletion-corrected metallicity [X/H]$_{\rm intrinsic}$ \citep*[see, e.g.][]{Quiret et al. 2016}. \\

\begin{figure*}
\begin{tabular}{l}
\includegraphics[scale=0.75]{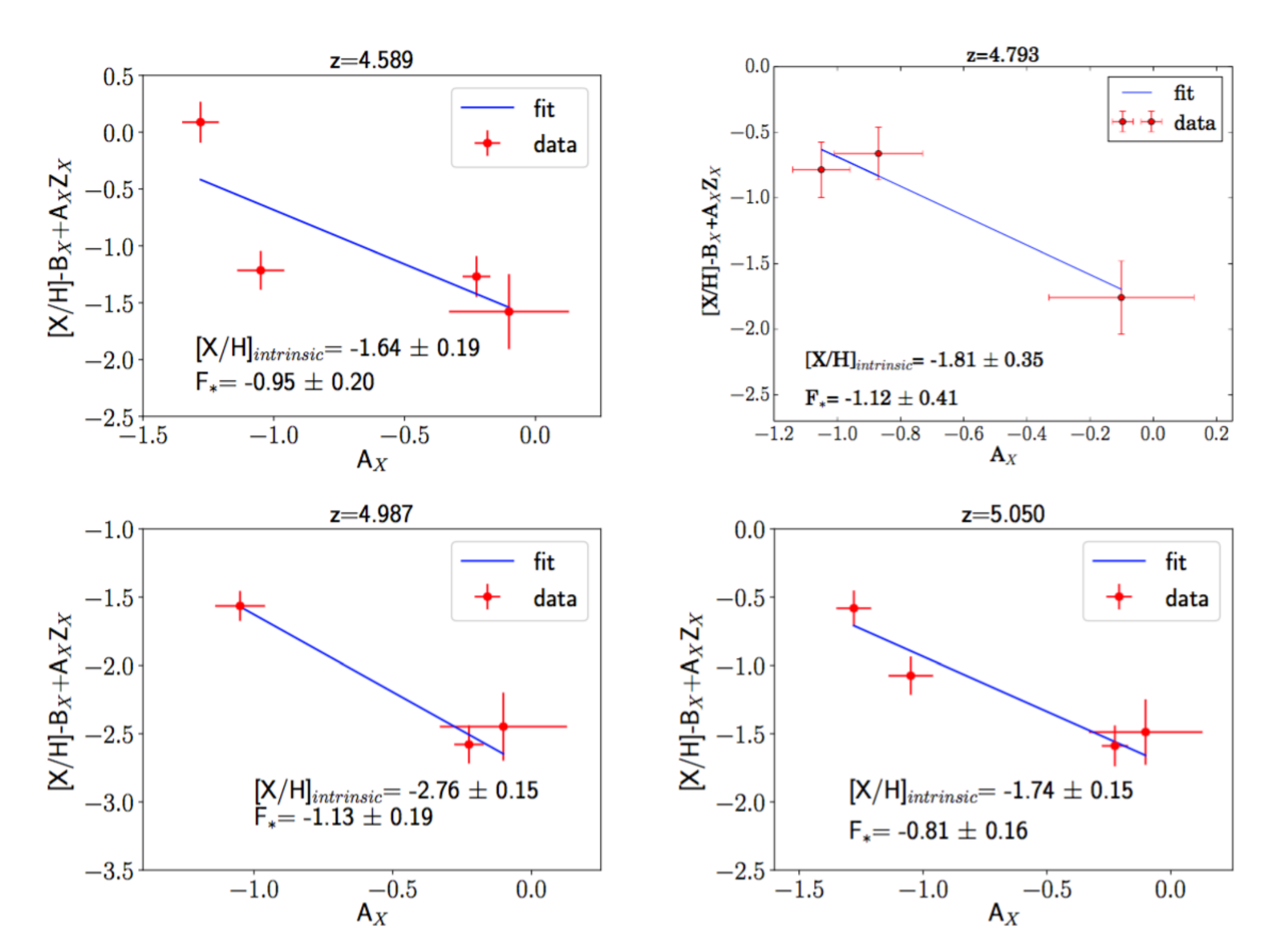}
\end{tabular}
\caption{Plots showing [X/H]$_{\rm obs}$ - $B_{x}$ + $A_{x} Z_{x}$ vs. $A_{x}$ for four absorbers at $z=4.589$, 4.793, 4.987, and 5.050. The $F_{*}$ values and the depletion-corrected metallicity [X/H]$_{\rm intrinsic}$ are shown in the bottom of each panel.}
    \label{fig:dust}
\end{figure*}

Fig. \ref{fig:dust} shows the plots of [X/H]$_{\rm obs}$ - $B_{x}$ + $A_{x} Z_{x}$ vs. $A_{x}$ to calculate the intrinsic metallicities and $F_{*}$ values for four absorbers from this paper for which we have determinations of [O/H] and the associated uncertainties (not just lower limits). We have adopted values of $A_{x}$, $B_{x}$, and $Z_{x}$ for O and C from \citet{Jenkins 2009} and for Si, S, and Fe from \citet{Jenkins 2017}. The $F_{*}$ values for these four absorbers are in the range from $-0.81\pm0.16$ to $-1.13\pm0.19$. These values are closer to the MW's halo gas which shows $F_{*}$ = -0.28 than the MW's cool disc gas, warm disc gas or disc + halo gas which have $F_{*}$ = 0.90, 0.12, and -0.08, respectively. Four other $z>4.5$ absorbers in \citet{Poudel et al. 2018}  and \citet{Morrison et al. 2016} showed $F_{*}$ values comparable to those for the absorbers reported here. Overall, 5 of the 8 absorbers at $z > 4.5$ have $F_{*}$ values equal to or greater than the typical value of $-0.70 \pm 0.06$ found for lower-redshit DLAs by \citet{Quiret et al. 2016}. Finally, the dust-corrected (i.e., intrinsic) metallicities obtained for these absorbers from Jenkins' approach are consistent with the observed [O/H] values within the uncertainties. Furthermore, including the cases with limits, at least 3 out of 12 absorbers have [Si/X]$<$0 and at least 4 out of 9 absorbers have [Fe/X]$<$0, where X denotes O or S. Of the 8 systems with definitive measurements (not just limits) for [Si/X], 2 show [Si/X]~$< $0 at $\ga 2 \sigma$ level. Of the 5 systems with definitive measurements (not just limits) for [Fe/X], 2 show [Fe/X] $<$ 0 at $\ga 2 \sigma$ level. While the samples are still small, it is clear that a significant fraction of $z > 4.5$ absorbers show depletions of Si and Fe. Thus, dust depletion appears to be significant in a substantial fraction of absorbers at $z > 4.5$.


\subsection{Metallicity evolution}
The cosmic mean metallicity of DLAs, also defined as the $N_{\rm H\, I}$-weighted mean metallicity, < $Z$ >  is the ratio of the mean comoving densities of metals and hydrogen, given by < $Z$ > = log $\rm (\Omega_{M}/\Omega_{H})$ -log (M/H)$_{\odot}$ where, $\rm \Omega_{M}$ is the comoving density of metals and $\rm \Omega_{H}$ is the comoving density of neutral hydrogen \citep*[e.g.][]{Lanzetta 1995, Kulkarni 2002}. It is standard practice to use < $Z$ > to investigate the metallicity evolution of DLAs \citep*[e.g.][]{Prochaska et al. 2003a, Kulkarni et al. 2005, Kulkarni et al. 2010, Rafelski et al. 2012, Rafelski et al. 2014, Som et al. 2015, Quiret et al. 2016, Poudel et al. 2018}. In this section, we determine the cosmic mean metallicity of DLAs at redshift $z>4.5$ and compare it with previous measurement at $z>4.5$ as well as $z<4.5$.\\

\begin{figure*}
\begin{tabular}{l}
\includegraphics[scale=0.9]{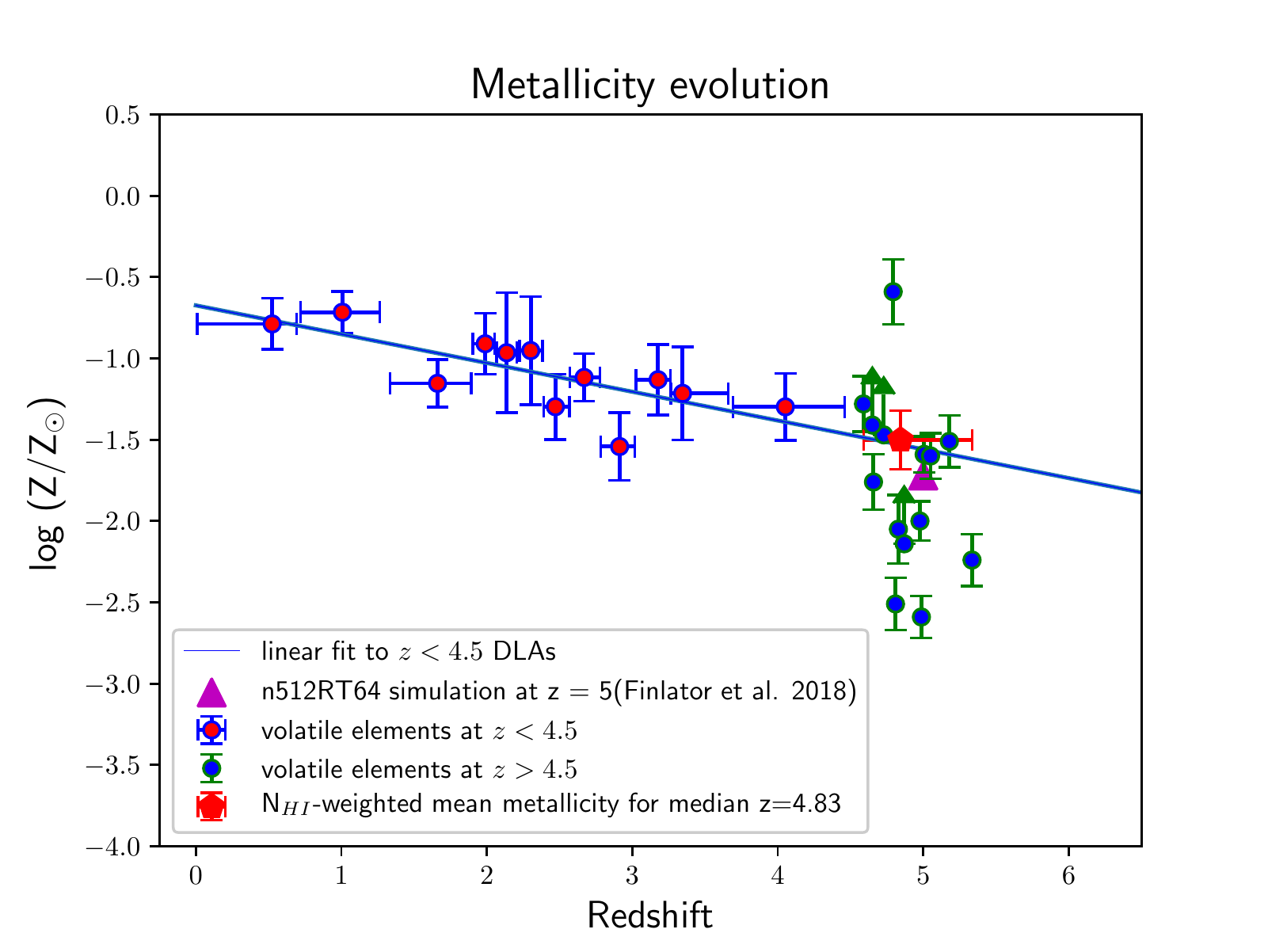}
\end{tabular}
\caption{ Metallicity evolution with redshift. The red dots with blue error bars are the binned $N_{\rm H\,I}$-weighted mean metallicities of DLAs at $z<4.5$ using volatile elements from the literature, with each bin containing 16 or 17 DLAs and the solid blue line showing the corresponding best fit. The blue dots with green error bars are the individual measurements for $z>4.5$ absorbers also using volatile elements. Three of these measurements are taken from \citet{Rafelski et al. 2012, Rafelski et al. 2014}, one from \citet{Morrison et al. 2016}, three from \citet{Poudel et al. 2018}, and seven from this paper. The $N_{\rm H\,I}$-weighted mean metallicity at a median redshift of $z=4.83$ for all absorbers in the $z>4.5$ bin is shown by a pentagon in red. The binned means and the corresponding errors are calculated using survival analysis to handle a mixture of detections and limits. For comparison, the $N_{\rm H\,I}$-weighted mean metallicity of DLAs at $z=5$ for the n512RT64 simulation from \citet{Finlator et al. 2018} is shown by a triangle in magenta.}
    \label{fig:evolution}
\end{figure*}

The metallicity of DLAs decreases by $\approx$0.2 dex per unit increase in redshift as we go from $z=0$ to $z=4.5$ \citep*[e.g.][]{Prochaska et al. 2003a, Kulkarni et al. 2005, Kulkarni et al. 2007, Som et al. 2013, Rafelski et al. 2012, Jorgenson et al. 2013, Som et al. 2015, Quiret et al. 2016}. At $z > 4.7$, a sudden decline of cosmic mean metallicity to a  level of < $Z$ > = -2.03 (6$\sigma$ below the prediction from lower-redshft DLAs) has been claimed \citep*[e.g.][]{Rafelski et al. 2012, Rafelski et al. 2014}. However, this conclusion was based on $z > 4.7$ measurements of primarily Si and Fe, which are refractory elements. Several studies have noted that such refractory elements are more strongly depleted on interstellar dust grains compared to volatile elements such as S, O, and Zn in the Milky Way interstellar medium (ISM) \citep*[e.g.][]{Savage 1996, Jenkins 2009}. Furthermore, a number of studies have noted the prevalence of dust depletion and the difference in depletions of refractory and volatile elements in DLAs \citep*[e.g.][]{Pettini et al. 1997, Kulkarni 1997, Jenkins 2009, De Cia et al. 2016}. Indeed, several recent studies have noted that dust depletion exists in some DLAs even at $z \sim 5$ \citep*[e.g.][]{Morrison et al. 2016, Poudel et al. 2018}. Therefore, to make a robust comparison between the metallicities of the absorbers at $z > 4.5$ and $z < 4.5$, we use the elements O, S, and Zn, which have lower condensation temperatures and thus lower depletion levels. The metallicity of DLAs based on O, S, and Zn is shown in Fig. \ref{fig:evolution}.\\

As seen in Table \ref{tab:rel_abn} or in Fig. \ref{fig:evolution}, for some of the absorbers in our sample, we could only place lower limits on the metallicity. We include both these limits as well as the definitive determinations while calculating the mean metallicity < $Z$ > using survival analysis, which allows estimation of the mean value and the associated error for measurements consisting of a mixture of detections and limits (either upper limits or lower limits). Each bin at $z < 4.5$ included 16 or 17 DLAs and the high-redshift bin (spanning $4.589 < z < 5.335$, with a median redshift of 4.83) included 14 absorbers with measurements of either O or S. The $N_{\rm H\,I}$-weighted mean metallicity < $Z$ > in the highest redshift bin (i.e. for $z > 4.5$) is $-1.51 \pm 0.18$, and is shown in Fig. \ref{fig:evolution} as a red pentagon.  The binned $N_{\rm H\,I}$-weighted mean metallicity of lower redshift systems is indicated as red circles with blue error bars and the corresponding linear best fit is shown as a solid blue line. The predicted metallicity from lower redshift data is consistent with our value in the highest redshift bin. The difference between the predicted and observed values is $< 0.5\sigma$, even if the uncertainty in the predicted value is ignored. \\

A few of the absorbers in our $z > 4.5$ sample are sub-DLAs (6 out of 14) rather than DLAs. It is interesting to ask whether the inclusion of these sub-DLAs is the cause of our cosmic mean metallicity being higher, given that sub-DLAs are found to be more metal-rich than DLAs at $z < 3$ \citep*[e.g.][]{Som et al. 2013, Som et al. 2015, Quiret et al. 2016}. We therefore repeated our calculations of the cosmic mean metallicity at $z > 4.5$ for only the DLAs, and found very little difference. The $N_{\rm H\,I}$-weighted mean metallicity at $z > 4.5$ after excluding the sub-DLAs is $-1.50\pm0.16$. The negligible difference between the values obtained with and without the inclusion of sub-DLAs implies that either < $Z$ > at $z > 4.5$ is dominated by absorbers with higher H I column densities, or that there is much less difference between the metallicity evolution trends for DLAs and sub-DLAs at these high redshifts. To help discriminate between these possibilities, we calculate the $N_{\rm H\,I}$-weighted mean metallicity for just the sub-DLAs at $z > 4.5$, and find to be -1.36$\pm$0.30. Thus, it appears that the difference between the mean metallicities for  DLAs and sub-DLAs is much smaller at $z > 4.5$ than at $z < 3$. This suggests that DLAs and sub-DLAs may have shared a common origin sometime during the second Gyr of cosmic history (i.e., sometime during $3 < z < 4.5$). However, observations of more DLAs and sub-DLAs at $z > 4.5$ are essential to understand how robust the results seen in the small existing samples are.\\

It thus appears that the cosmic mean metallicity at $4.6 \la z \la 5.3$ is consistent with the predictions based on the trend observed for DLAs at $z < 4.5$. This result differs from the conclusion of \citet{Rafelski et al. 2012, Rafelski et al. 2014} that there is a sudden drop in metallicity at $z > 4.7$. The difference may arise from our use of only the volatile elements S and O which have a lower condensation temperature and level of depletion in the MW ISM. This point has also been made by \citet{Morrison et al. 2016} and \citet{De Cia et al. 2018}. Our result confirms the conclusion of \citet{Poudel et al. 2018} that the metallicity at $z > 4.5$ shows a smooth decrease rather than a sudden decline. Indeed, a gradual decrease agrees better with expectations than a sudden decline, in the absence of any evidence of a sudden change in the cosmic star formation history at or shortly before $z \sim 5$.\\

Furthermore, we note that the $N_{\rm H\,I}$-weighted mean metallicity agrees within $\sim 1 \sigma $ with the results from the n512RT64 simulation by \citet{Finlator et al. 2018}, which predicts a value of -1.72 dex for the DLAs at $z=5$.\\

\begin{figure*}
\begin{tabular}{l}
\includegraphics[scale=0.65]{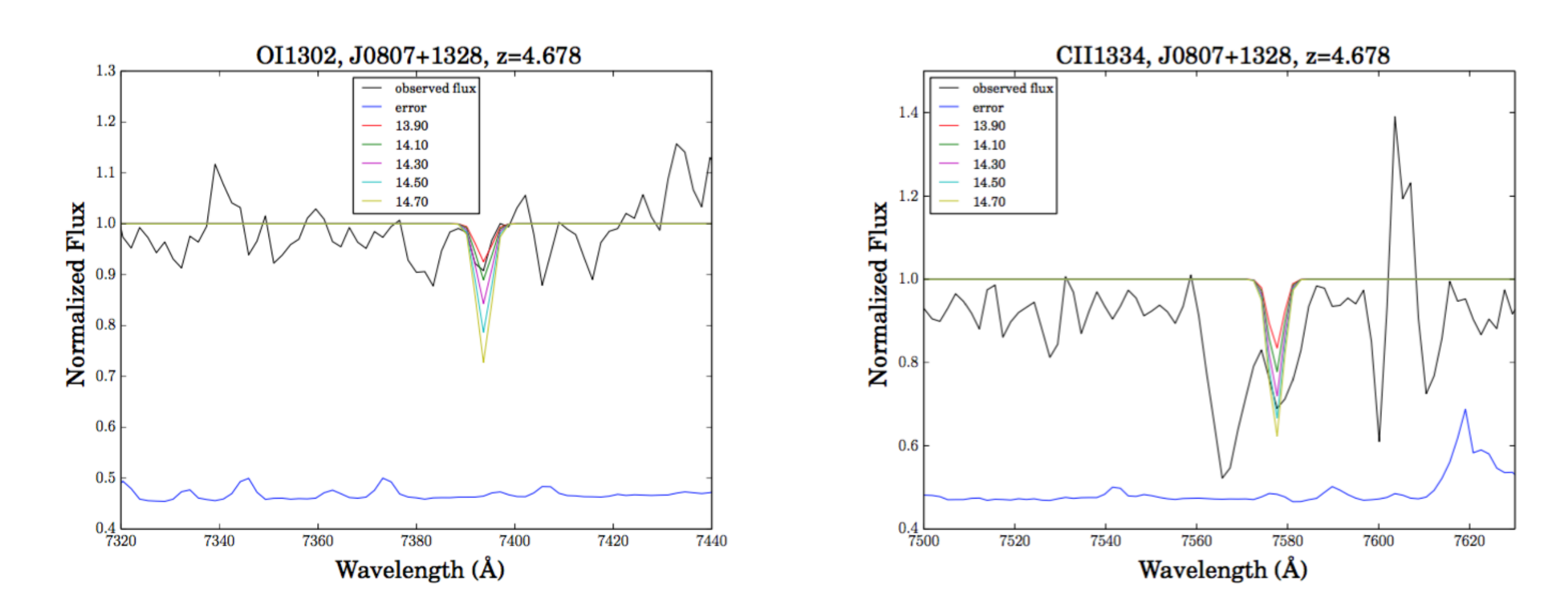}
\end{tabular}
\caption{Overplotting of different column density (log N$_{\rm O I}$ and log N$_{\rm C II}$ ) profiles convolved with SDSS resolution using typical SDSS spectra excluded from our sample to determine the SDSS metallicity floor. In each case, the observed continuum normalized flux is shown in black and the the $1\sigma$ error in the normalized flux is shown in blue at the bottom of each panel and is shifted by +0.4 for the purpose of displaying on the same scale. The metal line, sightline to the quasar, and redshift are given at the top of the figures in each panel.}
    \label{fig:floor}
\end{figure*}

\begin{figure}
\begin{tabular}{l}
\hspace*{-0.27in}
\includegraphics[height=7cm, width=9cm]{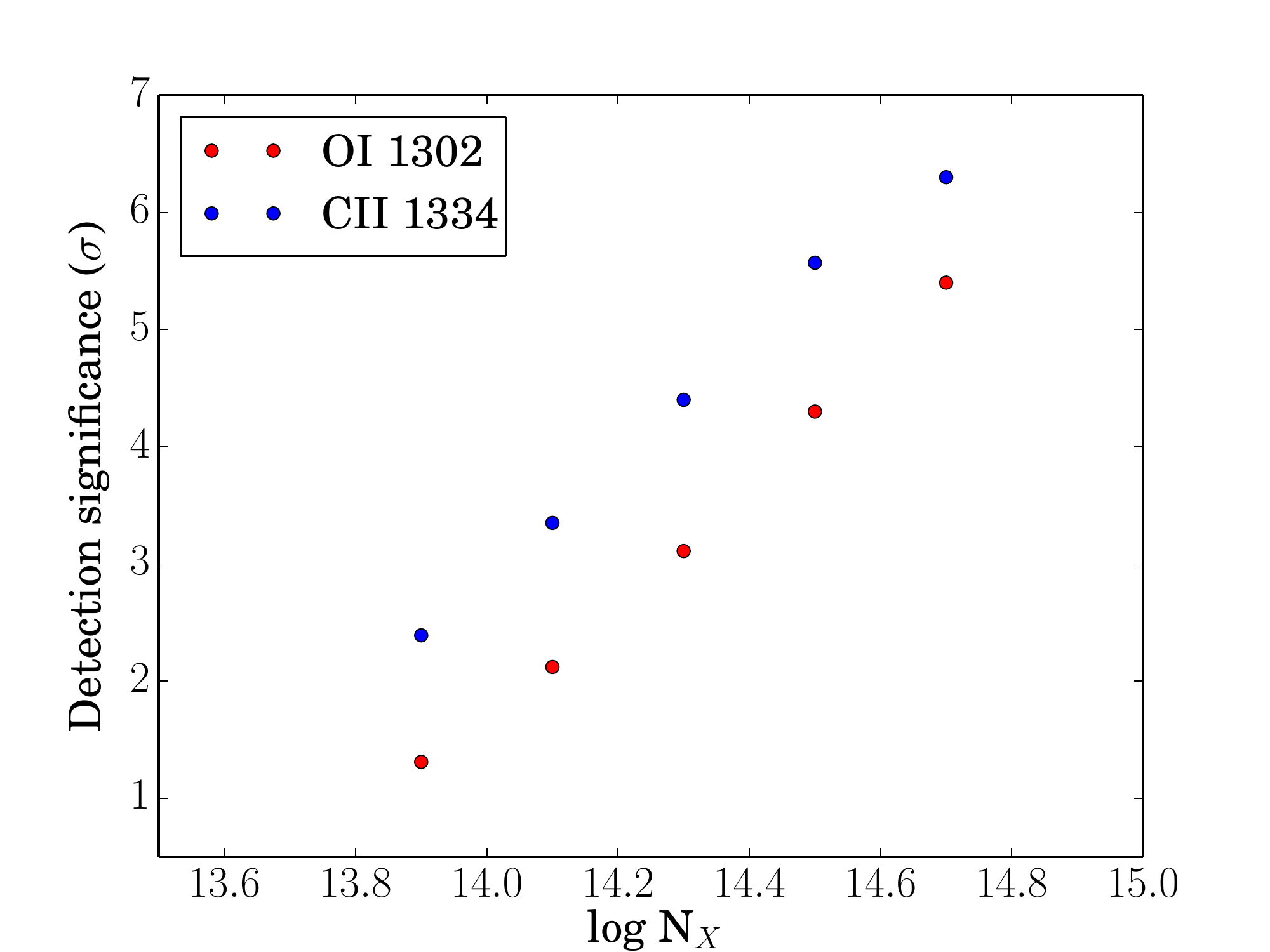}
\end{tabular}
\caption{Detection significance of O I $\lambda$ 1302 and C II $\lambda$ 1334 in the SDSS spectrum of J0807+1328 for different column densities for a hypothetical intervening absorber at z$=$4.678.}
    \label{fig:floor_graph}
\end{figure}

Finally, as discussed in Section \ref{sec:obs}, we emphasize that these absorbers used to estimate the metallicity evolution were selected from the $z>4.5$ SDSS absorbers listed in \citet{Noterdaeme et al. 2012} after checking that their SDSS spectra showed at least 1 metal line (regardless of its strength) at the same redshift as the H I line. The metal lines detected in the SDSS spectra were the lines with the highest oscillator strengths for the the most abundant elements, i.e., O I $\lambda$ 1302 and C II $\lambda$ 1334. A non-detection of either of these lines was conservatively taken as an indication of non-reality of the system, given that identifications based on H I Lyman-alpha in SDSS data have been found to be unreliable for some absorbers \citep*[e.g.][]{Crighton et al. 2015}. We now assess the effect of the sample selection on the metallicity. To do this, we have determined the smallest metal column density that could be detected from the low-resolution and low-S/N SDSS data and the corresponding metal abundances for a typical $N_{\rm H\,I}$ value for our absorbers. To this end, we have performed simulations of O I $\lambda$1302 and C II $\lambda$1334 lines and measured their equivalent widths after convolving with the SDSS instrumental resolution to find the smallest column density that could be detected in the noisy SDSS spectra typical of the objects that were excluded from our sample. The selected quasar, J0807+1328 (RA: 08:07:15, Dec: +13:28:05.2) has an emission redshift of z$=$4.876 and the SDSS i-band magnitude of m$_{i}=$19.39. The absorber redshift (z$=$4.678) was reported by \citet{Noterdaeme et al. 2012} based on its Lyman-alpha measurements. This quasar has a S/N per pixel of $\sim$ 14 near the redshifted O I $\lambda$ 1302 line, which is comparable to the S/N per pixel of $\sim$15 for our sample on average. These simulations show that the equivalent width of a hypothetical O I $\lambda$1302 line for log $N_{\rm O\,I}$=14.30 would be 3 times the 1 $\sigma$ measurement uncertainty estimated from the S/N in the SDSS spectrum, i.e., log $N_{\rm O\,I}$=14.30 would have been detectable at a 3$\sigma$ level (see Fig. \ref{fig:floor}). This limit of log $N_{\rm O\,I}$ = 14.30 would correspond to a limiting metallicity of [O/H]$=$-3.14 dex and -2.79 dex, respectively, at the maximum and median H I column density values for our sample (log $N_{\rm H I}$ = 20.75 and 20.40, respectively). These limiting metallicity values are, respectively, $> 42$ times ($> 9 \sigma$) and $> 19$ times ($> 7 \sigma$) below the $N_{\rm H I}$-weighted mean metallicity that we find for the systems that we do observe. Moreover, the limiting metallicity values of -3.14 and -2.79 dex are far below the $N_{\rm H I}$-weighted mean of Rafelski et al. (2014; $-2.03^{+0.09}_{-0.11}$). Furthermore, the limit of log $N_{\rm C\,II}$ = 14.10 which would be detected at 3 $\sigma$ level, corresponds to a limiting metallicity of [C/H]$=$-3.08 dex and -2.73 dex, respectively, at the maximum and median H I column density values for our sample. We show in Fig. \ref{fig:floor_graph} the detection significance of O I $\lambda$ 1302 and C II $\lambda$ 1334 lines for various column densities. The 3$\sigma$ detection limit corresponds to log $N_{\rm O\,I}$ = 14.30 and log $N_{\rm C\,II}$ = 14.10, far below the column densities detected for our absorbers (see Table \ref{tab:voigt_metals}). Thus, the high mean metallicity we find is not a sample selection effect.

\section{Conclusions}
We have made seven new measurements of metal abundances at $z>4.5$, doubling the existing sample for undepleted elements in gas-rich galaxies, and thus improving the constraints on the first $\sim 1$ billion years of cosmic metal evolution. Our main results are as follows:\\

1. We find a wide spread (factor of $> 100$) in the metallicities, which range from -2.69 dex to -0.59 dex. Our measurements include the highest metallicity observed in a sub-DLA at $z\sim5$, with [S/H]=-0.59$\pm$0.20 for an absorber at $z=4.793$ toward J1253+1046. In the same sight line, we also find a DLA at $z=4.600$ with [O/H]$>$-1.46, which is the most metal-rich DLA known at $z\sim5$.\\

2. Combining our sample with measurements from the literature, we examine the relative abundances in the $z \sim 5$ DLAs, and find their [C/O] ratios to be consistent with those of the VMP DLAs. Furthermore, we estimate the probability distributions of the progenitor masses to be centered around 12 M$_{\odot}$ to 17 M$_{\odot}$, using [C/O] and [Si/O] for three relatively metal-poor $z\sim5$ DLAs.\\

3. In a substantial fraction of absorbers at $z > 4.5$, the extent of dust depletion, as judged by the parameter $F_{*}$, appears to be at least as significant as (and in some cases stronger than)  the typical depletion found in lower-redshift absorbers.\\ 

4. The metallicity vs. velocity dispersion relation for $z \sim 5$ absorbers seems to be different from that for lower redshift DLAs. The flatter trend observed for the $z \sim 5$ absorbers could be explained if these absorbers arise in galaxies with stronger inflows of chemically less enriched gas, or in more dark matter-dominated  galaxies with smaller stellar masses.\\

5. We calculate the $N_{\rm H\,I}$-weighted mean metallicity in the range $4.6 \la z \la 5.3$ and find it to be consistent with the prediction from lower redshifts  DLAs, signifying a smooth decline in DLA metallicity rather than a sudden drop. Furthermore, we demonstrate that this high mean metallicity is not an artifact of sample selection.\\

6. The difference between the mean metallicities for DLAs and sub-DLAs at $z > 4.5$ is small, suggesting that DLAs and sub-DLAs may have been similar at this early epoch. The difference between the DLAs and sub-DLAs seen at $z < 3$ may have thus arisen sometime during $3 < z < 4.5$. \\

7. In 3 of the 4 absorbers where we have detected Fe~ II absorption, the Fe II profile appears to be blue-shifted by $\sim 35-50$ km s$^{-1}$ with respect to the O I, C II, and Si~ II profiles. Such a shift could be indicative of moderate-velocity outflows enriched in Fe II by type Ia supernovae. Alternately, the blue-shifted Fe II profiles may arise from stronger dust depletion of Fe II in the central components. \\

Our results demonstrate the value of obtaining high spectral resolution measurements of volatile elements such as O in absorbers at $z > 4.5$. Clearly, observations of more DLAs and sub-DLAs at $z > 4.5$ are essential to understand how robust the trends seen in the small existing samples are, and thus place more constraints on chemical enrichment of the gas in and around galaxies by early stars.

\label{sec:conclusion}

\section*{Acknowledgements:}

We thank an anonymous referee for helpful comments that have improved this manuscript. SP, VPK, and FHC acknowledge partial support from NASA grants NNX14AG74G and NNX17AJ26G and NASA/STScI support for HST programs GO-12536, 13801 (PI Kulkarni). The authors would like to thank Dr. Xiaohui Fan for making his MagE data for J0306+1853 available to us after publication. The authors would also like to thank Dr. Ryan Cooke for providing his MCMC code to calculate progenitor star parameters from relative element abundances. CP thanks the Alexander von Humboldt Foundation for the granting of a Bessel Research Award held at MPA.
\\







\appendix

\section{Assessment of Metal Line Saturation}
\begin{figure*}
\centering
  \begin{tabular}{@{}cc@{}}
   \includegraphics[width=.45\textwidth]{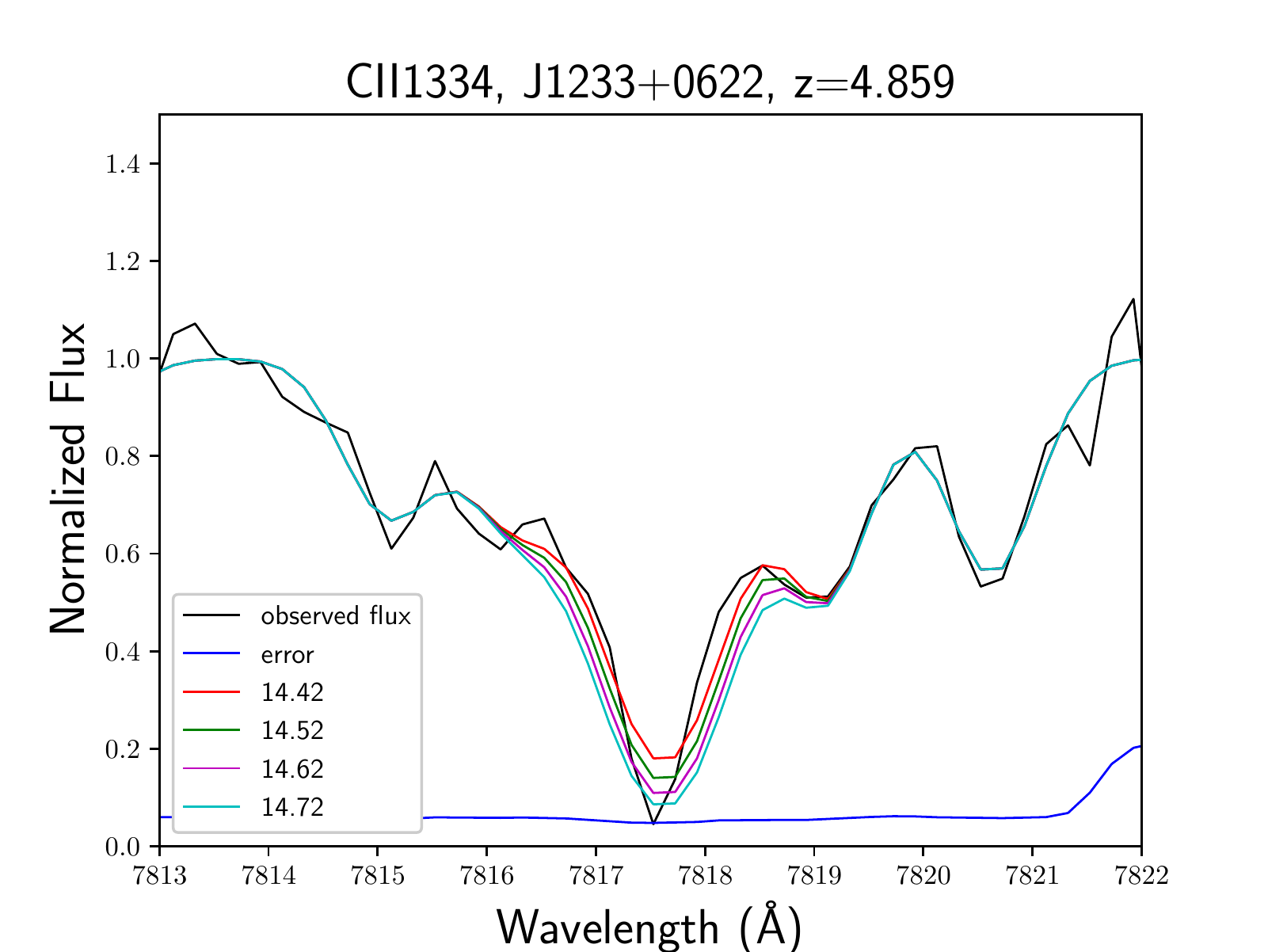} &
    \includegraphics[width=.45\textwidth]{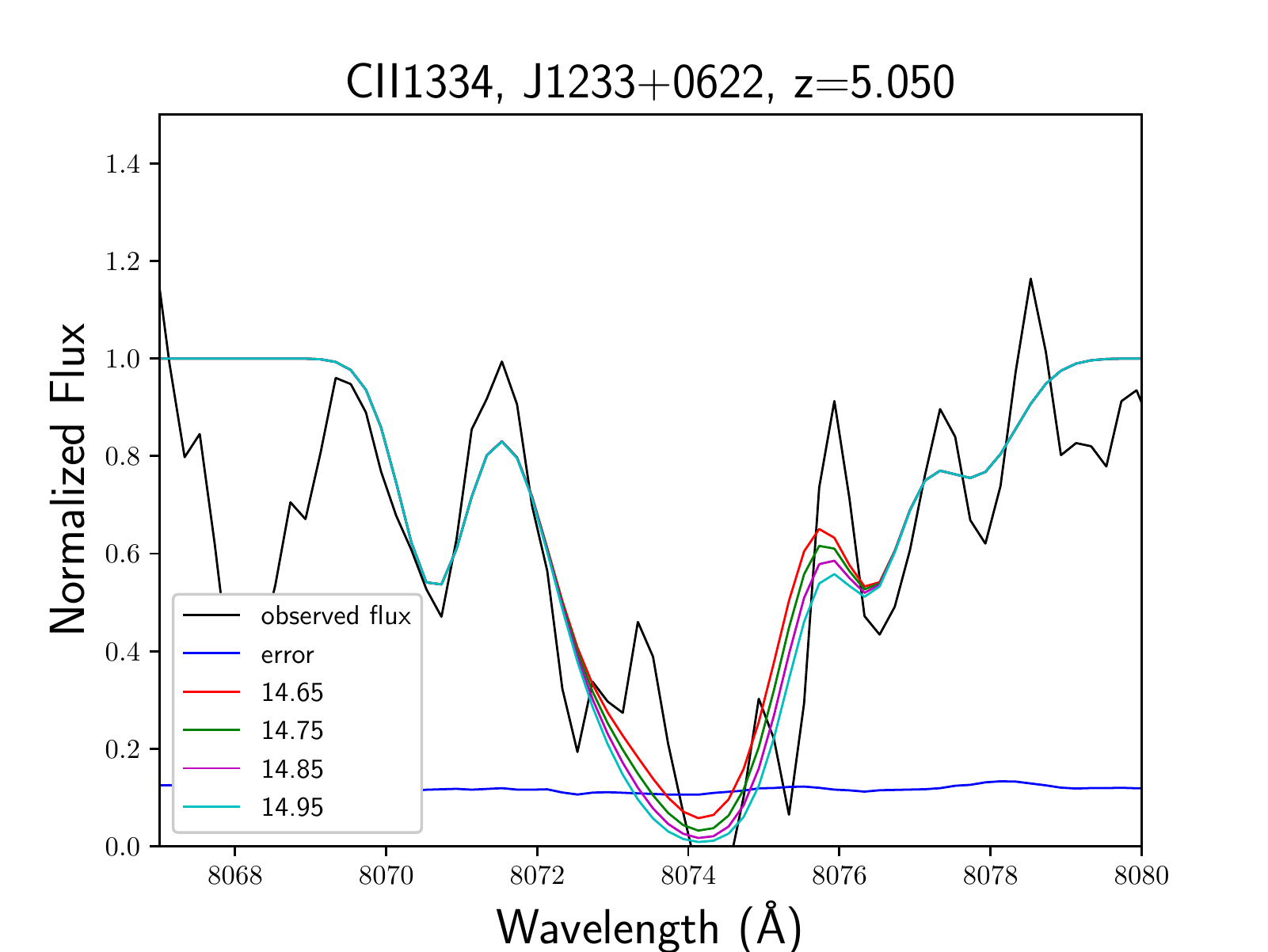} \\  
     \includegraphics[width=.45\textwidth]{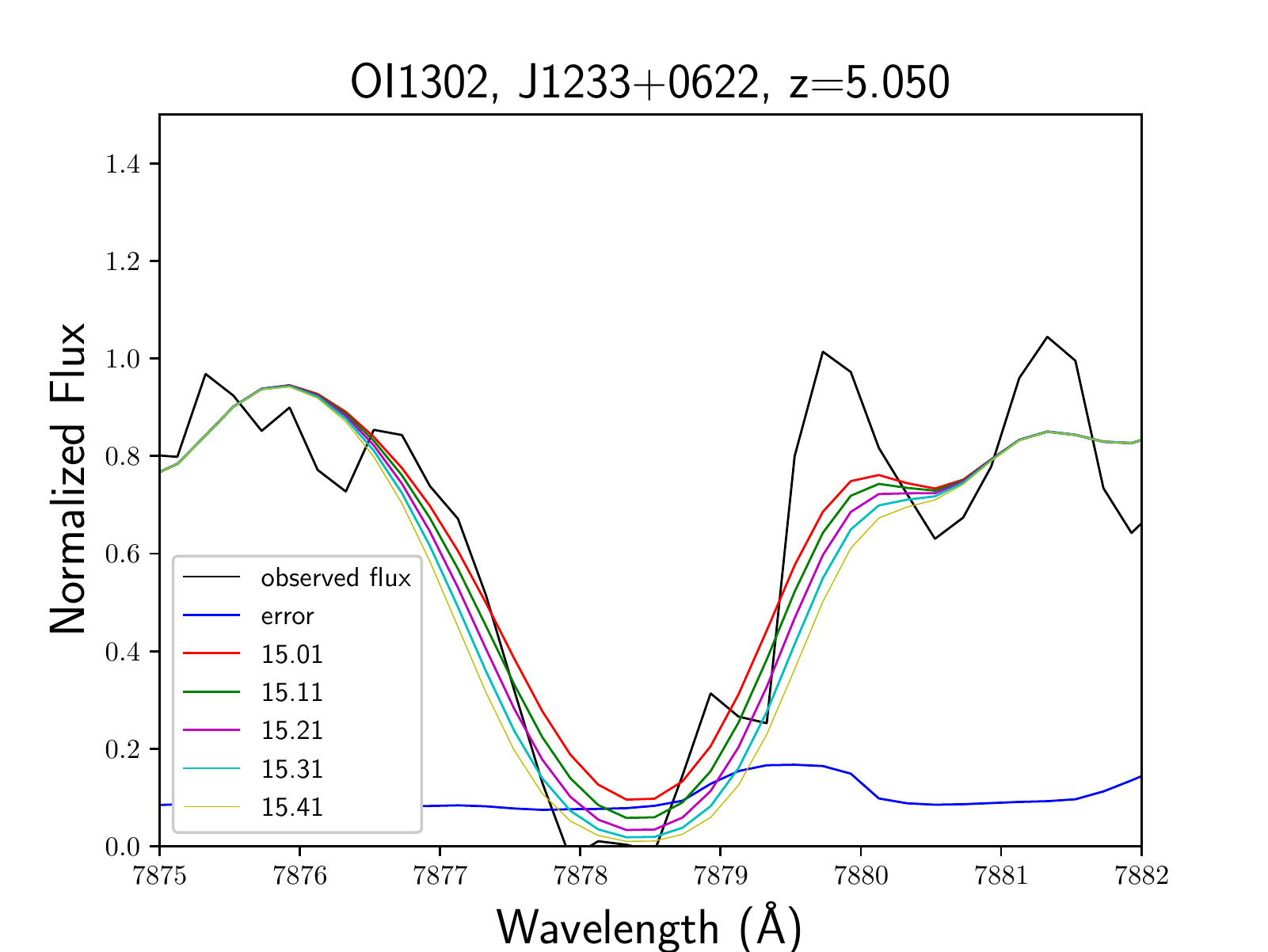} &  
    \includegraphics[width=.45\textwidth]{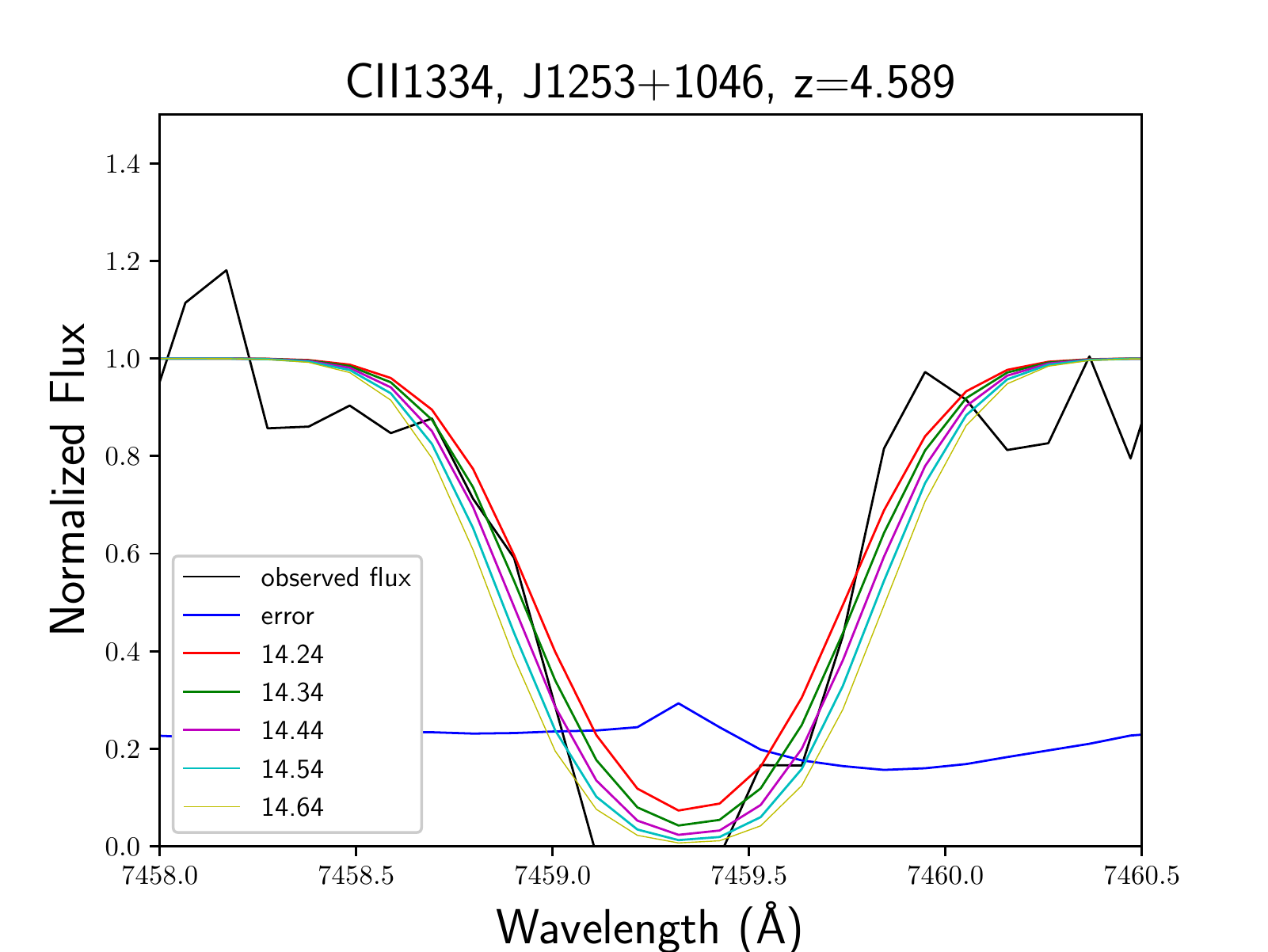} \\
    \includegraphics[width=.45\textwidth]{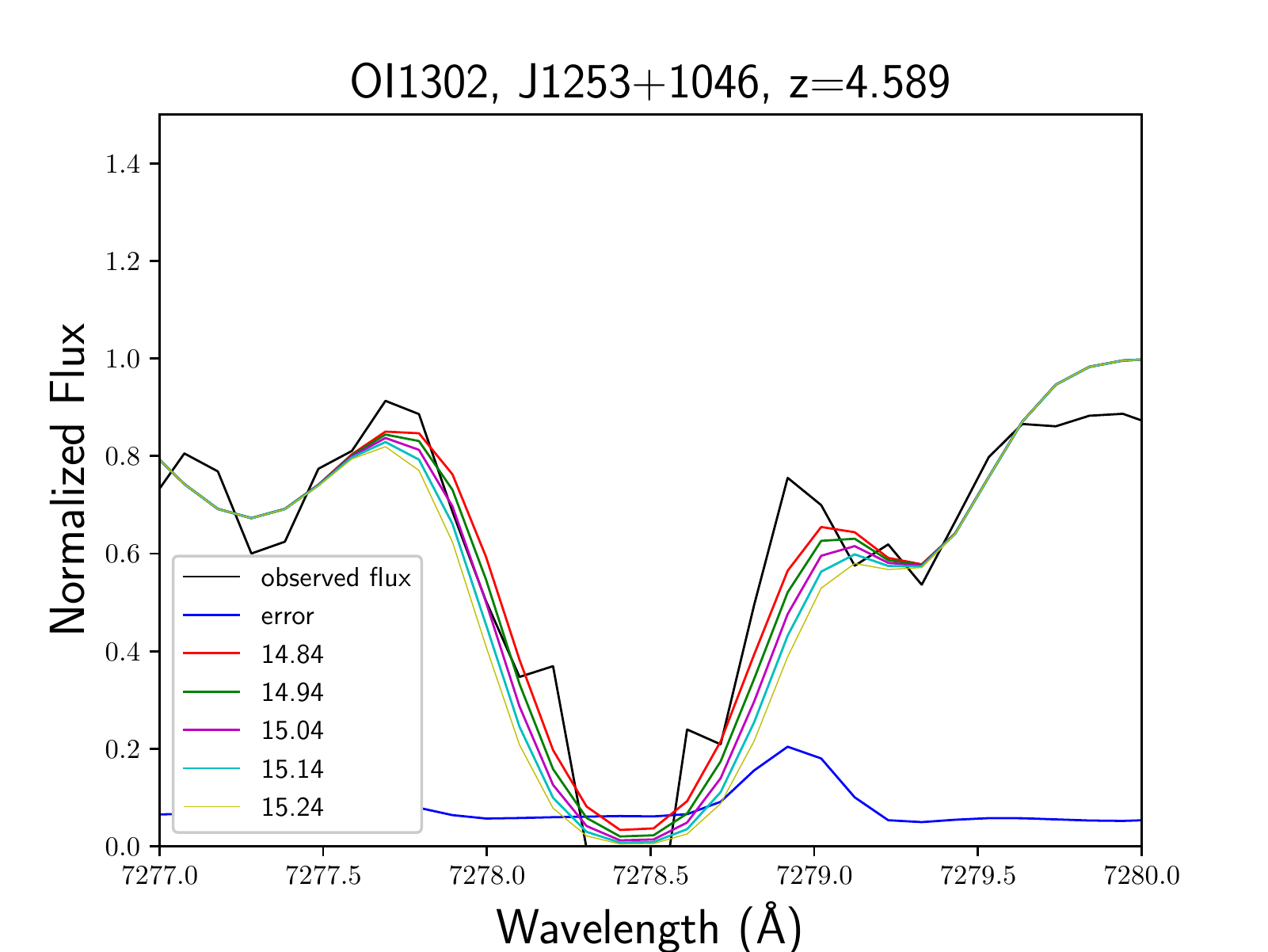} &
    \includegraphics[width=.45\textwidth]{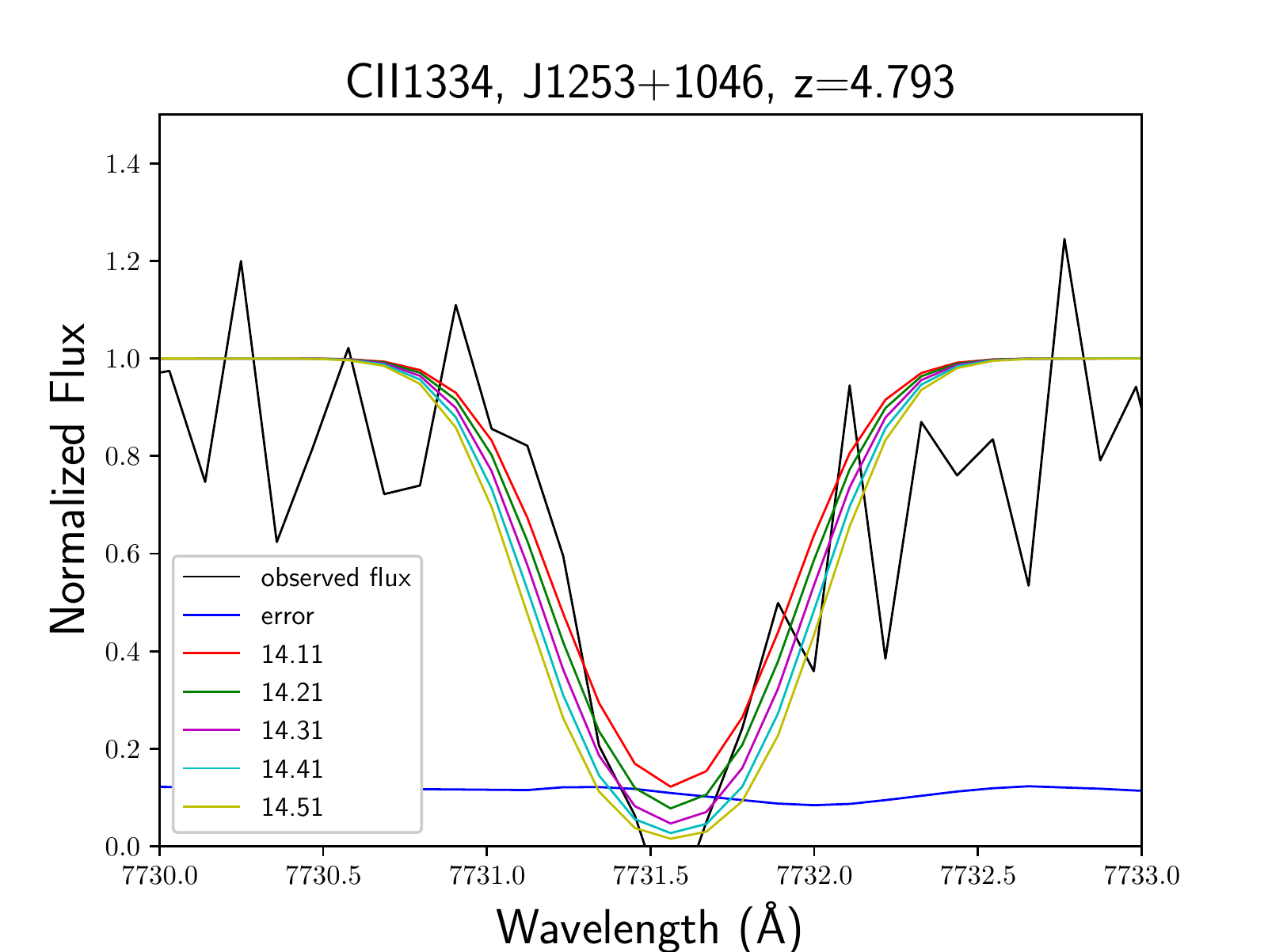}
  \end{tabular}
  \caption{Overplotting of different column density (log N$_{\rm X}$, where, X$=$OI or CII) profiles to estimate the extent of saturation. In each case, the observed continuum normalized flux is shown in black and the $1\sigma$ error in the normalized flux is shown in blue at the bottom of each panel. The metal line, sightline to the quasar, and redshift are given at the top of the figures in each panel.}
  \label{fig:saturation}
\end{figure*}

Here we discuss the estimates of the limits on O I and  C II column densities in cases where 
the lines are strong. Fig. \ref{fig:saturation} shows overplotting  of different column density (log N$_{\rm X}$, where, X$=$OI or CII) profiles to estimate the extent of saturation.

\section{Velocity plots for Lyman-series lines}
Examples of velocity plots of Lyman-series lines for our systems are shown below. It is clear from these plots that in these cases, Lyman lines beyond Lyman-alpha are not useful for estimation of H I column densities.

\begin{figure*}
\centering
  \begin{tabular}{@{}cc@{}}
   \includegraphics[width=.37\textwidth]{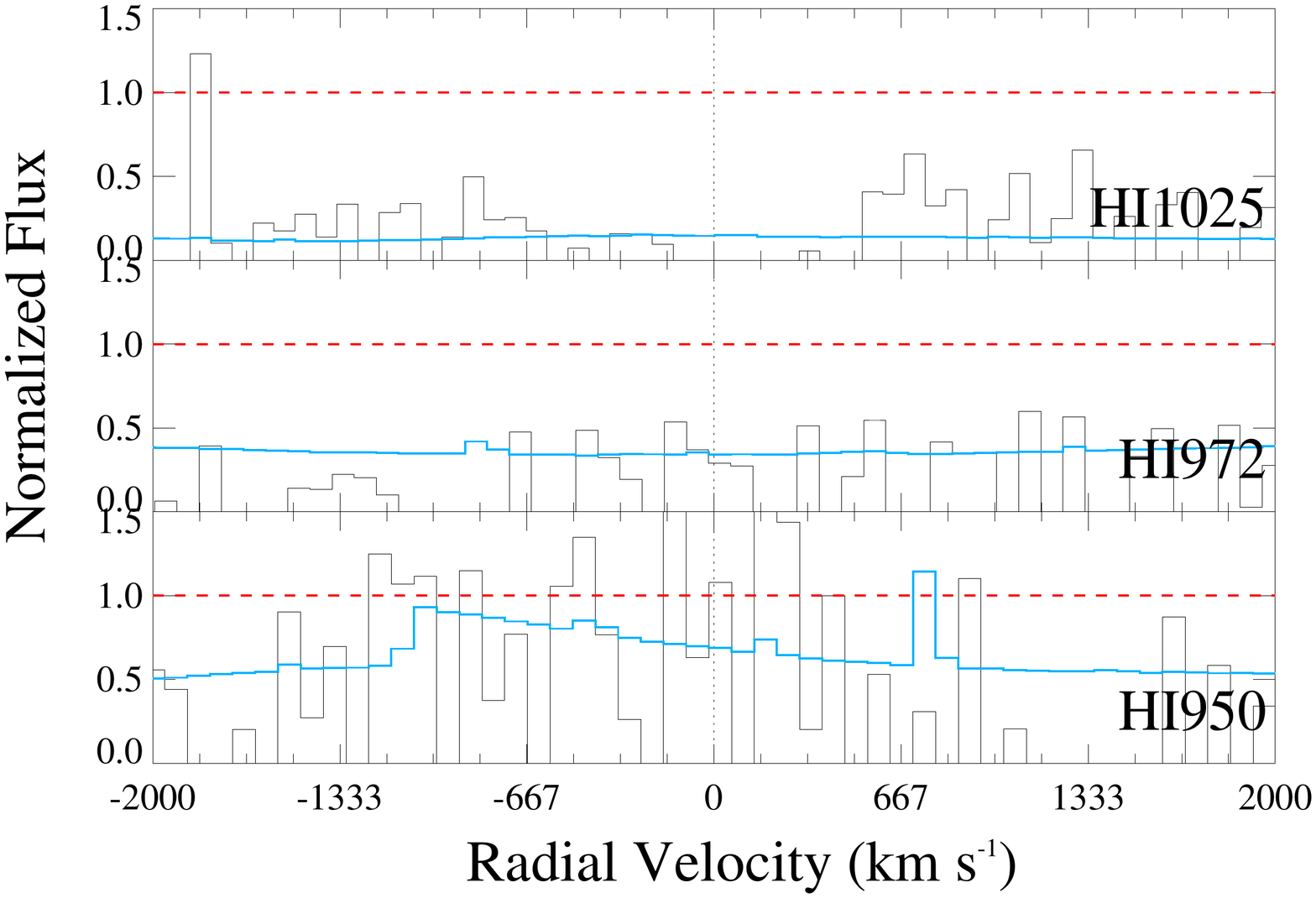} &
    \includegraphics[width=.37\textwidth]{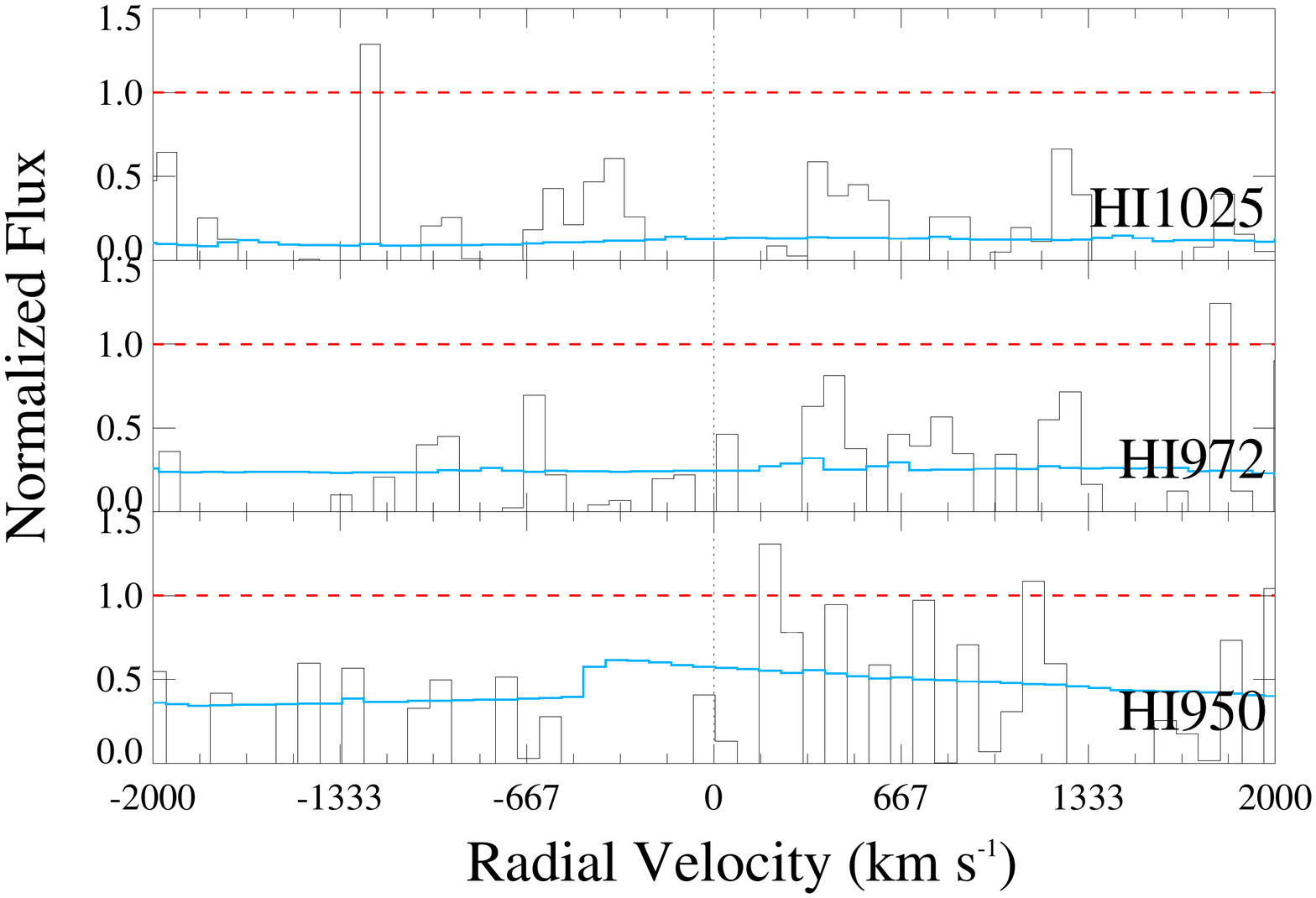} \\  
  \end{tabular}
  \caption{Velocity plots for Lyman-beta, Lyman-gamma, and Lyman-delta for the absorbers at z$=$4.859 (Left) and z$=$5.050 (Right) towards J1233+0622. The corresponding Lyman-alpha lines are shown in Fig. \ref{fig:lyman}. In each panel, the vertical dotted line is at the center of the Lyman-alpha line of the corresponding absorber, the continuum level is shown by horizontal dashed line in red, the observed data are shown in black, and the blue line at the bottom shows the 1$\sigma$ error in the normalized flux.}
  \label{fig:lyman_series1}
\end{figure*}

\begin{figure*}
\centering
  \begin{tabular}{@{}cc@{}}
   \includegraphics[width=.37\textwidth]{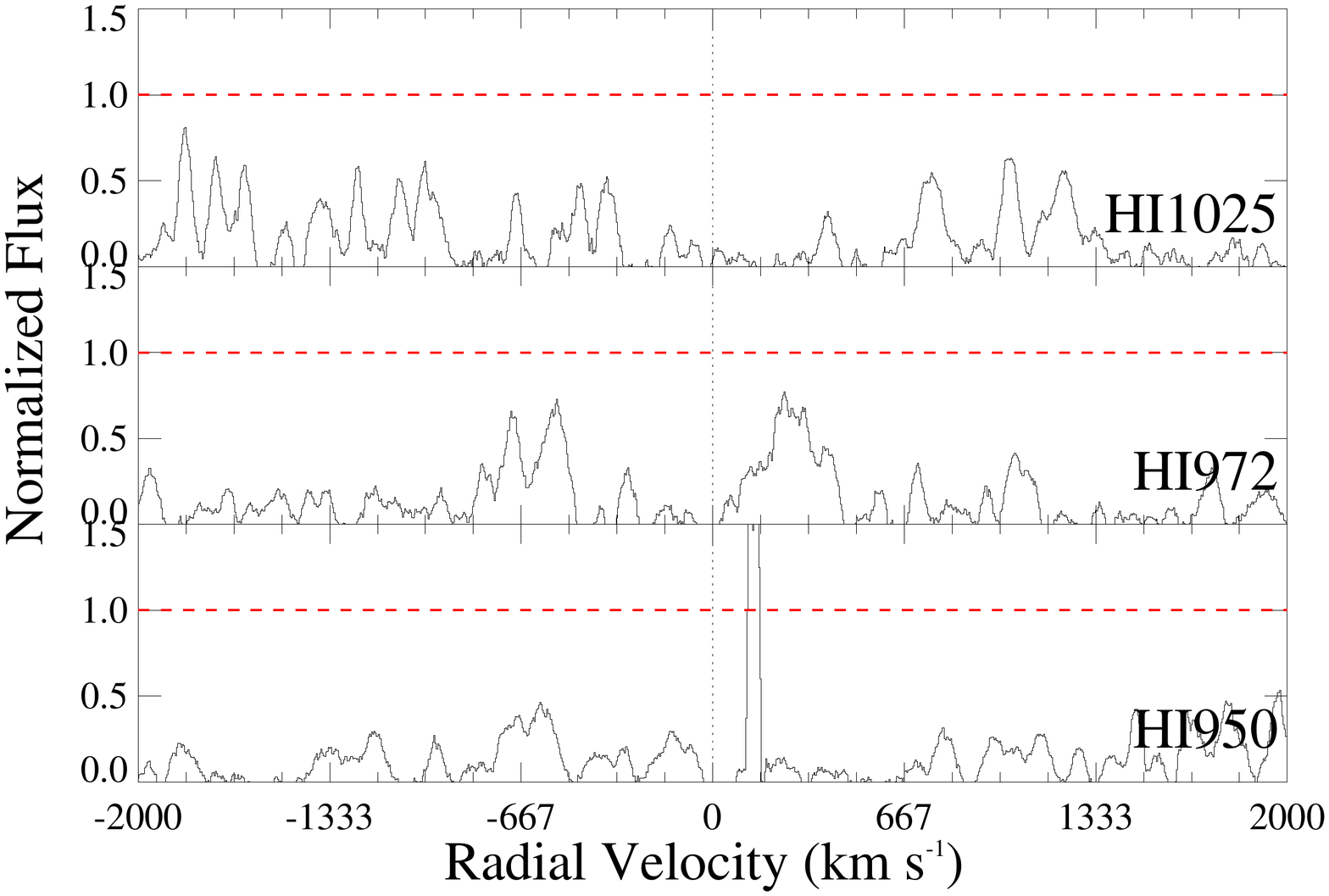} &
    \includegraphics[width=.37\textwidth]{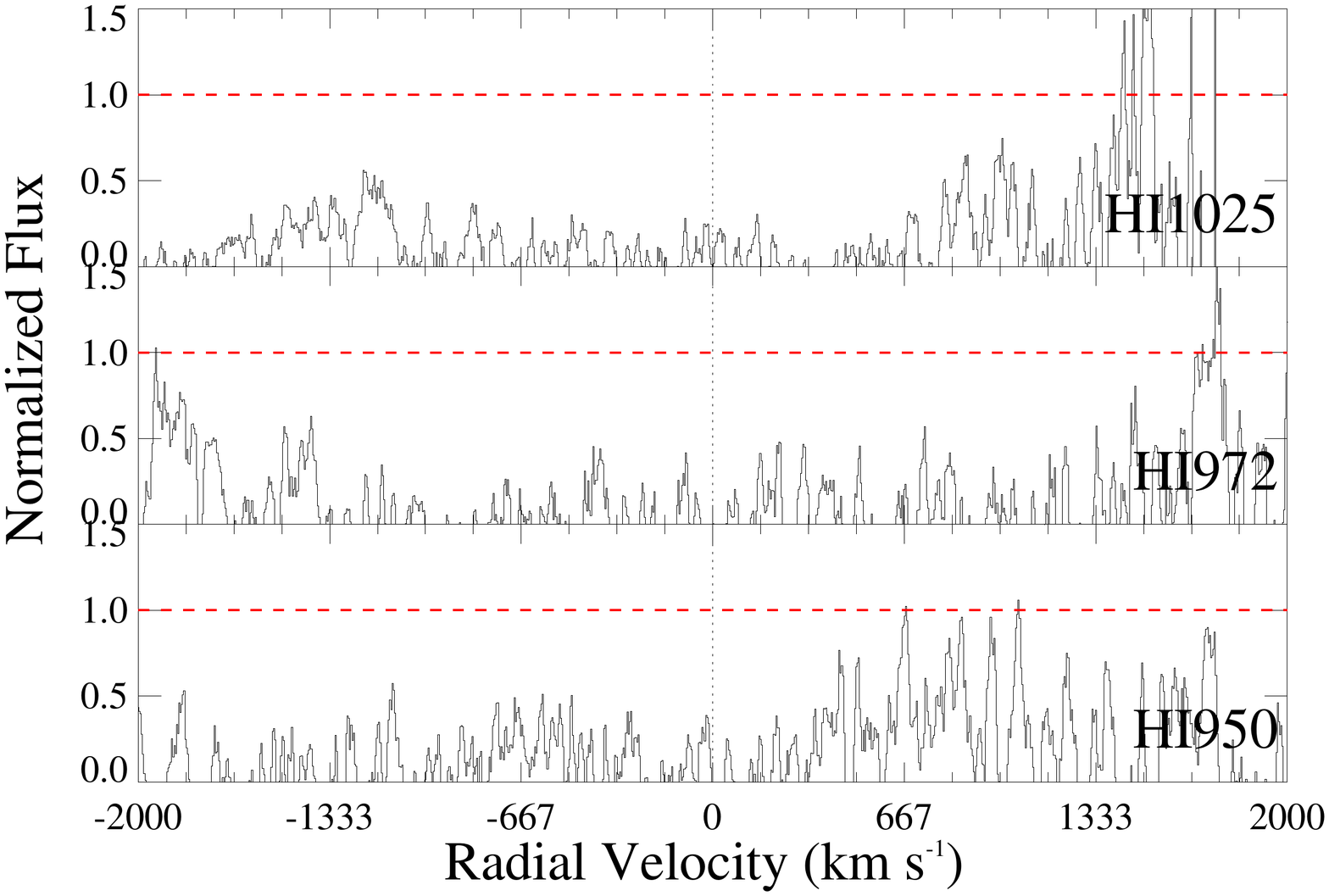} \\  
  \end{tabular}
  \caption{Velocity plots for Lyman-beta, Lyman-gamma, and Lyman-delta for the absorbers at z$=$4.793 (Left) and z$=$4.627 (Right) towards J1253+1046 and J1557+1018 respectively. In each panel, the vertical dotted line is at the center of the Lyman-alpha line of the corresponding absorber, the continuum level is shown by horizontal dashed line in red, the observed data are shown in black.}
  \label{fig:lyman_series2}
\end{figure*}



\bsp	
\label{lastpage}
\end{document}